\documentclass[fleqn,usenatbib]{mnras}

\usepackage{newtxtext,newtxmath}

\usepackage[T1]{fontenc}

\usepackage{orcidlink}

\DeclareRobustCommand{\VAN}[3]{#2}
\let\VANthebibliography\thebibliography
\def\thebibliography{\DeclareRobustCommand{\VAN}[3]{##3}\VANthebibliography}

\usepackage{graphicx}	
\usepackage{amsmath}	

\newcommand{\kms}{km s$^{-1}$}

\newcommand{\orcid}[1]{\href{orcid.org/#1}{\textcolor[HTML]{A6CE39}{\aiOrcid}}}

\def\13co{$^{13}$CO}
\def\aj{AJ}

\def\aap{A\&A}
\def\apj{ApJ}
\def\apjs{ApJS}

\def\c18o{C$^{18}$O}

\def\cm2{cm$^{-2}$}

\def\hi{H{\sc i}}
\def\FCNM{$f_\mathrm{CNM}$}
\def\RHI{$\mathcal{R_{\mathrm{H{\sc I}}}}$}
\def\h2{H$_2$}
\def\kms{km s$^{-1}$}
\def\k0{$\kappa_{353}$}

\def\n2h{N$_2$H$^+$}
\def\NHI{$N_\mathrm{H{\sc I}}$}

\def\nhiUnit{$\times\ 10^{20}$ cm$^{-2}$}
\def\NHIthin{$N^{*}_\mathrm{H{\textsc{I}}}$}

\def\ourmod{\textit{TPCNet}}
\def\ourmods{\textit{TPCNets}}

\def\s{s$^{-1}$}
\def\s353{$\sigma_{353}$}

\def\taupeak{$\tau_{\mathrm{peak}}$}

\def\TBpeak{$T_{\mathrm{b,peak}}$}

\def\Tk{$T_{\mathrm{k}}$}
\def\Tkmax{$T_{\mathrm{k,max}}$}
\def\Ts{$T_{\mathrm{s}}$}
\def\t353{$\tau_{353}$}
\def\xco{$X_\mathrm{CO}$}

\title[\ourmod: Representation learning for \hi\ mapping]{\ourmod: Representation learning for \hi\ mapping}

\author[Nguyen \& Tang et al.]{Hiep Nguyen\orcidlink{https://orcid.org/0000-0002-2712-4156}$^{1}$\thanks{Corresponding author: vanhiep.nguyen@anu.edu.au},
Haiyang Tang$^{2}$\thanks{Co-first author},
Matthew Alger$^{3}$,
Antoine Marchal\orcidlink{https://orcid.org/0000-0002-5501-232X}$^{1}$,
Eric G. M. Muller\orcidlink{https://orcid.org/0000-0001-5621-1577}$^{1}$, \newauthor
Cheng Soon Ong\orcidlink{https://orcid.org/0000-0002-2302-9733}$^{2,4}$,
N. M. McClure-Griffiths\orcidlink{https://orcid.org/0000-0003-2730-957X}$^{1}$
\\
$^{1}$Research School of Astronomy and Astrophysics, The Australian National University, Canberra, ACT 2611, Australia\\
$^{2}$School of Computing, The Australian National University, Canberra, ACT 2601, Australia\\
$^{3}$Google, Pirrama Road, Pyrmont, NSW 2009, Australia\\
$^{4}$Data61, CSIRO, Clunies Ross Street, Acton, ACT 2601, Australia
}

\date{Accepted XXX. Received YYY; in original form ZZZ}

\pubyear{2023}

\begin{document}
\label{firstpage}
\pagerange{\pageref{firstpage}--\pageref{lastpage}}
\maketitle

\begin{abstract}
We introduce \ourmod, a neural network predictor that combines Convolutional and Transformer architectures with Positional encodings, for neutral atomic hydrogen (\hi) spectral analysis. Trained on synthetic datasets, our models predict cold neutral gas fraction (\FCNM) and \hi\ opacity correction factor (\RHI) from emission spectra based on the learned relationships between the desired output parameters and observables (optically-thin column density and peak brightness).
As a follow-up to Murray et al. (2020)’s shallow Convolutional Neural Network (CNN), we construct deep CNN models and compare them to \ourmod\ models. \ourmod\ outperforms deep CNNs, achieving a 10\% average increase in testing accuracy, algorithmic (training) stability, and convergence speed. Our findings highlight the robustness of the proposed model with sinusoidal positional encoding applied directly to the spectral input, addressing perturbations in training dataset shuffling and convolutional network weight initializations. Higher spectral resolutions with increased spectral channels offer advantages, albeit with increased training time. Diverse synthetic datasets enhance model performance and generalization, as demonstrated by producing \FCNM\ and \RHI\ values consistent with evaluation ground truths. Applications of \ourmod\ to observed emission data reveal strong agreement between the predictions and Gaussian decomposition-based estimates (from emission and absorption surveys), emphasizing its potential in \hi\ spectral analysis.
\end{abstract}




\section{Introduction}
\label{sec:intro}
    Neutral atomic hydrogen (\hi), the most abundant gas in the interstellar medium (ISM), plays a crucial role in the evolution of the ISM in galaxies. It provides the initial material for the formation of stars, serves as the basic element of molecular clouds, influences the dynamics of the ISM, acts as a cooling agent and source of radiation shielding, and participates in the feedback mechanism that regulates star formation. In the gas-to-star process, optically-thin warm \hi\ gas ($\sim$8,000 K) must cool down to become optically thick cold gas ($\sim$100 K) before transitioning into molecular gas ($\sim$10 K). The mass fraction of cold atomic gas along a line of sight is a key indicator of this atomic-to-molecular transition. While this fraction can be estimated when both \hi\ emission and absorption are measured, it cannot be directly determined from \hi\ emission alone. In this work, we tackle the challenge of using machine learning to predict properties of optically-thick cold \hi, and demonstrate that our approach is useful on observed emission data.

Within the ISM, \hi\ gas exists in multiple phases with varying temperatures and densities primarily due to thermal instability in stable pressure equilibrium \citep{Field1965,Wolfire2003,Hennebelle2000,Audit2005}. There are two thermally stable phases present: the cold neutral medium (CNM) and the warm neutral medium (WNM), with kinetic temperatures and densities of ($T_\mathrm{k}$, n) = (25--250 K, 10--100 cm$^{-3}$) and ($T_\mathrm{k}$, n) = (4000--8000 K, 0.1--1 cm$^{-3}$), respectively \citep{Field1969,McKee1977,Wolfire2003}. A thermally unstable phase (or unstable neutral medium, UNM) with intermediate temperature and density is present as well \citep{Heiles2003a,Kim2013,Murray2018,Hill2018,Nguyen2019, Marchal2019}. From observations of the Galactic ISM, \hi\ is distributed roughly as follows: $\sim$32--64\% WNM, $\sim$28--40\% CNM, and $\sim$20--28\% UNM \citep[e.g.,][]{McClure-Griffiths2023}.

As the optically-thick CNM is the key building block for molecular hydrogen \h2\ \citep[e.g.,][]{Krumholz2009}, constraining the distributions of CNM fraction and true \hi\ column density across the Galactic ISM is essential for understanding the formation of molecular clouds and subsequent star formation. Nevertheless, precisely obtaining its crucial parameters, such as the fraction of cold neutral medium (\FCNM) and opacity correction to \hi\ column density (\RHI), from neutral hydrogen emissions remains challenging.

In practice, along with \hi\ emission, measuring \hi\ absorption towards a continuum radio source is necessary to estimate the physical properties of the atomic ISM. Such observational pairs offer the most direct method for inferring optical depths, temperatures, and column densities, which are key for observationally distinguishing the different \hi\ phases, as well as constraining their mass fractions \citep{Heiles2003a,Strasser2007,Dickey2003,Dickey2009}. The column density under the optically-thin assumption ($N^{*}_\mathrm{HI}$) is proportional to the brightness \hi\ temperature, hence, can be readily obtained from observed emission profiles. However, this assumption may miss a significant amount of gas mass (about 10\%) for high Galactic latitude line of sights, $b > 20^{\circ}$, \citep[e.g.,][]{Murray2015,Murray2018,Murray2018a,Murray2021} and at least 20\% for sightlines towards the Galactic Plane or molecular clouds \citep[e.g.,][]{Heiles2003b,Stanimirovic2014,Lee2015,Nguyen2019} because the emission includes not only contributions from warm, optically-thin gas, but also from cold, optically-thick gas. In the case when emission/absorption pairs are not available, we alternatively have to apply some kind of opacity correction to the available emission data.

The limited availability of continuum radio sources for measuring absorption is a significant obstacle to acquiring optically-thick \hi\ properties over large sky areas. The current approaches for extracting large-scale \hi\ properties from emission data only, based on spectral line width and amplitude, such as Gaussian decomposition \citep{Mebold1982,Haud2007,Kalberla2018,Marchal2019} and the Fourier Transform approach \citep{Marchal2024}, are limited in their effectiveness due to subjectivity, complex implementation, and high sensitivity to noise. Recently, \citet{Murray2020} (M20, henceforth) proposed an important approach by constructing a two-layer convolutional neural network (CNN) model to extract information from 1D spectral data. While their method represents an outstanding starting/reference point, there is potential for improvement. The simplicity of their two-layer CNN models may lead to limitations in terms of accuracy (as pointed out by \cite{Marchal2024} in their application of the Fourier-transformed method to Galactic intermediate velocity clouds), interpretability, and stable mathematical properties, which are crucial for interpretation. This suggests that there is room for new machine methods capable of extracting information from \hi\ emission with higher accuracy while reducing implementation complexity and computation time.

Building on the work by \cite{Murray2020}, we employ a supervised machine learning technique (regression) to estimate the values of \FCNM\ and \RHI\ from \hi\ emission (without the continuum background sources). The success of regression approaches relies on a meaningful embedding of the raw spectral data into a vector of features. Recently, representation learning has shown to be a successful approach that uses neural networks to learn an appropriate embedding from data \citep[e.g.,][]{Goodfellow2016,Gulati2020,Rozanski2023}. Of particular interest is the implementation of Transformer neural networks (an architecture designed to handle sequence-to-sequence tasks and long-range dependencies), which have achieved remarkable results in natural language processing \citep{Vaswani2017,Devlin2018,Bubeck2023} and image analysis benchmarks \citep[and references therein]{Dosovitskiy2020,Carion2020,Khan2021}. As an extension of their applicability, we aim to assess Transformer neural networks developed specifically for spectral analysis. To pursue this endeavor, we design ``\ourmod'', which combines Convolutional and Transformer architectures (together with Positional encodings) to strategically synergize the strengths of CNNs with the self-attention mechanism of the Transformer \citep{Vaswani2017}. While the CNN excels in capturing short-range features within spectra, the Transformer's multi-head self-attention (MHSA) specializes in revealing the intricate long-range correlations between frequency channels \citep[e.g.,][]{Pan2022, Rozanski2023}. Specifically, the CNN extracts compact features from \hi\ emission spectral data, and these features are then fed into the Transformer, serving as a predictor, to predict the \hi\ parameters of interest (\FCNM\ and \RHI).
Moreover, we will examine various data representations of emission spectra to determine the most effective positional encodings.

Our models are trained on a synthetic training database derived from a range of hydrodynamic (HD, \citealt{Saury2014}) and magnetohydrodynamic (MHD, \citealt{Seta2022}) simulations. Through our training processes, the \ourmod\ models have learned intricate relationships between the desired output parameters and key observables (e.g., optically-thin column density and peak brightness temperature). These relationships, extracted from the synthetic training datasets derived from various simulations, seem to form the foundation of our models' predictive capabilities.

Once trained, we apply the models to observed emission data collected from both emission and absorption surveys to validate their performances across different observational contexts. In this study, we conduct detailed comparisons between the predictions generated by \ourmods\ and those derived from Gaussian decomposition-based methods, such as ROHSA (for emission observations) and emission-absorption Gaussian fitting (for absorption surveys). Additionally, we compare our results with the lower limit of cold gas obtained from Fourier-transformed methods. We also construct deep CNN models in addition to \ourmod\ for comparison purposes. We evaluate the performance of these two model designs and compare them with the shallow CNN from \citet{Murray2020}. These comparisons will demonstrate that \ourmods, trained on synthetic datasets generated by numerical simulations, generalize effectively to make meaningful predictions about cold \hi\ properties using emission data observed by both emission and absorption surveys.

In the context of 21-cm radiative transfer, CNM clouds are prominently encoded within \hi\ emission spectra as narrow features (often located near the emission peaks), with typical linewidths of $\Delta V_\mathrm{CNM} \lesssim 3$ \kms\ \citep{Dickey1978,Dickey1990,Heiles2003a,Stanimirovic2014,Murray2015,Murray2018,Murray2021,Nguyen2019,Nguyen2024}. The current study therefore targets the local Galactic ISM due to the need for high spectral resolutions ($\sim$1 \kms) to accurately extract CNM properties from emission data. Furthermore, the parameters of the simulations, used to generate our training database, closely resemble the physical conditions of the Solar neighborhood. Besides, the Galactic \hi\ absorption surveys (also at high spectral resolutions), provide the most direct estimates of CNM fraction and opacity corrections, making them suitable for validating our methodology. \ourmods\ (and its in-progress Bayesian extension using Bayesian neural networks) yet hold the potential for broader applications, including extragalactic \hi\ studies, such as those targeting nearby galaxies with the Local Group L-Band Survey (LGLBS), which offers a spectral resolution of 0.4 \kms. This approach may also extend to higher redshifts if adequate spectral resolutions can be achieved, particularly in the era of the Square Kilometre Array (SKA). To adapt \ourmods\ for the local-Universe or higher-redshift \hi, we plan to train models on synthetic databases derived from large-scale cosmological simulations, such as Illustris \citep{Genel2014,Vogelsberger2014}, Illustris-TNG \citep{Springel2018}, EAGLE \citep{Schaye2015,Crain2015} and SIMBA \citep{Dave2019}, which account for varying conditions and physical scales.

In this paper, Section \ref{sec:data} introduces our synthetic training database, highlighting key trends in the training sets, as well as the observed emission/absorption data used for model evaluation. Section \ref{sec:prediction} describes \ourmod\ and deep CNN model designs, training processes, and experiments with CNNs and \ourmods\ on synthetic data. Section \ref{sec:pes} discusses the impact of positional encodings in spectral data analysis and a \ourmod\ vs CNN ablation study. We then validate our \ourmod\ model in Section \ref{sec:evaluation} using synthetic evaluation set. Section \ref{sec:compare_with_absorption} offers a comparison between the \ourmod\ predictions and absorption-based results inferred by the most direct methodology (emission-absorption Gaussian decomposition). In Section \ref{sec:application_to_emission}, we represent an application of \ourmods\ to observed emission cube towards an intermediate velocity region (the Low-Latitude Intermediate-Velocity Arch 1) with detailed comparisons with ROHSA Gaussian decomposition, Fourier-transformed method, and a previously published CNN model. Finally, we summarize our findings and outline our future plans in Section \ref{sec:conclusions}.

In the course of this research, we have developed a Python package called ``\textit{tano signal}''\footnote{https://doi.org/10.5281/zenodo.14183481}. This package allows users to utilize trained models for predicting the \FCNM\ and \RHI\ from emission data.

\section{Data}
\label{sec:data}
    \subsection{Synthetic training database}
\label{subsec:training_datasets}

Our training spectral datasets are generated based on Hydrodynamic and Magnetohydrodynamic simulations from \cite{Saury2014} (resolution $\sim$ 0.04 pc) and \cite{Seta2022} (resolution $\sim$ 0.4 pc), respectively. \cite{Saury2014} conducted a numerical study of thermally bi-stable CNM and WNM structures considering the effect of turbulence and thermal instability. \cite{Seta2022}'s simulations investigated the behavior of the non-isothermal turbulent dynamo within a two-phase medium to study the influence of atomic gas phases on the properties of the turbulent dynamo and the magnetic field it amplifies.

We generate synthetic \hi\ observations considering the full radiative transfer of the 21-cm line (described in \citealt{Marchal2019}) by using density, temperature, and velocity fields produced by simulations. Here, we account for thermal and turbulent motions to accurately simulate emission line broadenings. The spectral line widths and shapes arise directly from the intrinsic physical properties of the simulated medium. We do not decompose the synthetic spectra into Gaussian components; instead, we estimate their full widths at half maximum (FWHM) using second-moment analysis. The resulting line widths in our training dataset span from approximately 0.5 \kms\ to 44 \kms. Additionally, in order to replicate real observations, we introduce random noise at levels of 0.5 -- 0.8 K to synthetic spectra and smooth the synthetic data cubes with a Gaussian beam size of 45 arcsec. We then compute the signal-to-noise ratio (S/N) based on the typical peak intensities of the spectra, which generally fall between 3 K and 100 K. This results in an estimated S/N ratio ranging from $\sim$5:1 to 150:1, depending on both the specific line strength and the level of added noise.

Our final training database comprises five position-position-velocity (PPV) cubes, each with dimensions of (512 $\times$ 512 $\times$ \textit{N}), where \textit{N} indicates the number of velocity channels. Using two velocity resolutions: 0.3125 and 0.8 \kms, corresponding to 256 and 101 velocity channels, respectively, we can evaluate the impact of spectral resolutions on our predictive outcomes. Every position pixel in these cubes is associated with an emission spectrum (brightness temperature $T_\mathrm{b}(v)$ in Kelvin) and an absorption spectrum (optical depth $\tau(v)$), both as a function of Doppler velocity $v$ in km\ s$^{-1}$.

The primary aim of our models is to predict the amount of optically-thick cold \hi\ gas based solely on emission across large sky areas. 
We therefore exclude the synthetic optical depth data from the model training and reserve it for comparison purposes with the results obtained from absorption observations. Observationally, \hi\ optical depth can be directly measured only in the directions of strong radio continuum sources, which are sparsely distributed and not always available. While optical depth provides valuable information, its limited availability restricts its application in large-scale predictions. However, with instruments like SKA and Australian SKA Pathfinder (ASKAP) telescopes, absorption measurements can be much more densely sampled, with $\sim$10 measurements per square degree \citep[e.g.,][]{McClureGriffiths2015aska,Nguyen2024}. This higher measurement density would enable future models to incorporate both emission and absorption data, potentially leading to more accurate predictions across larger sky areas.

The distribution and structures of optically-thick \hi\ are governed by the physical processes embedded in the simulations. The diversity of the training dataset, and thus the model’s generalization ability, is intrinsically linked to the variety of simulations used to generate the spectra. The HD/MHD simulations encompass the multi-phase characteristics of neutral \hi\ gas, enabling the synthetic observations to closely resemble realistic atomic scenarios. It is yet important to note that the simulations employ parameters (such as metallicity, radiation field strength, thermal pressure, and heating and cooling prescriptions) within ranges comparable to those only observed in the Galactic Solar neighborhood, and they do not include the atomic-to-molecular transition. Nonetheless, by combining two HD/MHD simulations, we can investigate how different simulation types with a range of astrophysical processes affect the model performances.

\begin{figure}
\centering
\includegraphics[width=0.47\textwidth]{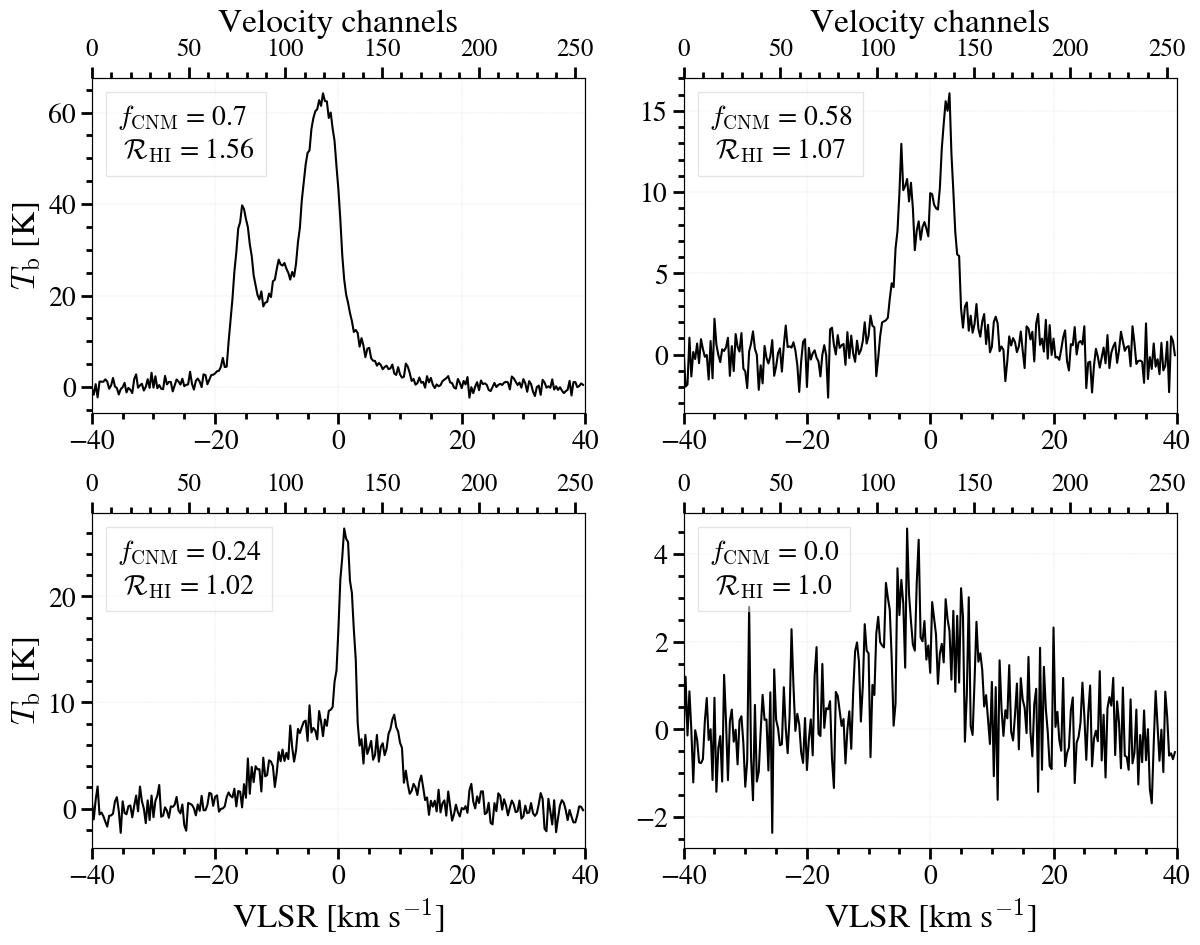}
\caption{Samples of synthetic \hi\ spectra in training database derived from \citet{Seta2022} (left panels) and \citet{Saury2014} (right panels), each paired with their associated CNM fraction \FCNM\ (with $T_\text{k, CNM} < 500$ K) and opacity correction to optically-thin estimate \RHI.  The velocity channel bin width is 0.3125 \kms.}
\label{fig:eg_spectra}
\end{figure}

The integrated \hi\ properties along lines of sight, including column densities, gas phase fractions, and optically-thick \hi, can be estimated using the brightness $T_\text{b}(v)$, optical depth $\tau_\text{HI}(v)$, and gas temperatures. The total \hi\ column density with optical depth and spin (excitation) temperature $T_{\text{s}}$ is given by:

\begin{equation}
\frac{N_\mathrm{HI}}{[\mathrm{cm^{-2}}]} = 1.8224 \times 10^{18} \int \tau_\text{HI}(v) \frac{T_\mathrm{s}}{[\mathrm{K}]}~\frac{dv}{[\mathrm{km~s^{-1}}]}
\label{eq:nhitot}
\end{equation}

The column density for the non-absorbing emission gas under the optically-thin limit ($\tau_\text{HI} \ll 1$) is calculated directly from the emission brightness:

\begin{equation}
\frac{N^{*}_\mathrm{HI}}{[\mathrm{cm^{-2}}]} = 1.8224 \times 10^{18} \int \frac{T_\mathrm{b}}{[\mathrm{K}]}~\frac{dv}{[\mathrm{km~s^{-1}}]}
\label{eq:nhithin}
\end{equation}

In accordance with \cite{Heiles2003a}'s separation of \hi\ gas phases (CNM, UNM, and WNM) based on their kinetic temperatures, we consistently define throughout this paper ``CNM'' as a \hi\ gas parcel characterized by a kinetic temperature $T_\mathrm{k} < 500$ K (with our future work aims to extend these predictions to include the fraction of UNM gas within 500--5000 K). To quantify the cold gas fraction along a line of sight, we compute \FCNM\ as the ratio between CNM column density ($N_\text{HI,CNM}$) and total \hi\ column density ($N_\text{HI}$).
Likewise, to assess the contribution of optically-thick \hi, we determine the opacity correction factor $\mathcal{R}_\text{HI}$ as the ratio between total \hi\ column density and optically-thin column density ($N^{*}_\text{HI}$).

Following the analysis by \cite{Murray2020}, we pair each spectrum with its ground-truth values for \FCNM\ and \RHI. A single PPV cube contains 262,144 (512 $\times$ 512) emission spectra. With five synthetic cubes from simulations, our dataset boasts a total of 1,310,720 spectra. We allocate four cubes, with 1,048,576 spectra, for training the models (\textit{training sets}), while the remaining cube is reserved for evaluation (\textit{evaluation set}). It is pertinent to mention that, within the training sets, the emission signal is centralized over the spectral channels, with weak signal at both ends (mostly representing noise) and strong signal in the central range. Figure \ref{fig:eg_spectra} provides a visual representation of four sample spectra along with their corresponding \FCNM\ and \RHI\ values.

\begin{figure}
\centering
\includegraphics[width=0.45\textwidth]{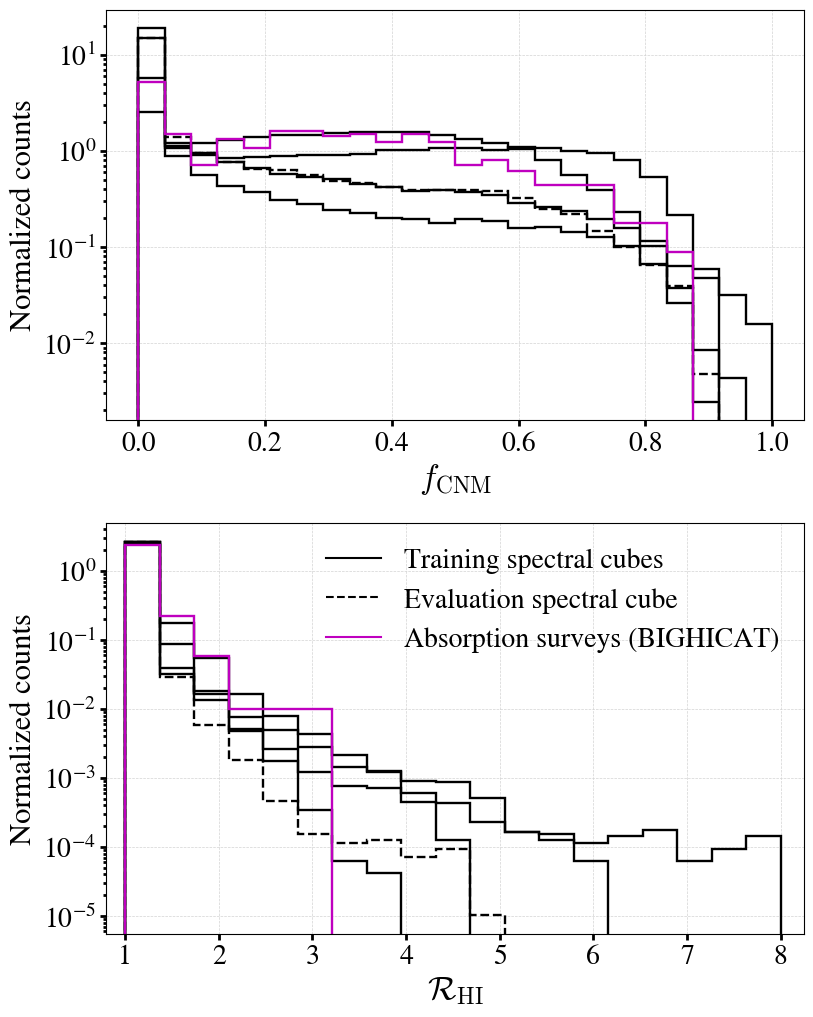}
\caption{Histograms of cold \hi\ gas fraction \FCNM\ ($T_\text{k, CNM} < 500$ K) (upper panel) and \RHI\ (lower panel) from training spectral cubes (solid, black), evaluation cube (dashed, black) and absorption observations (solid, magenta). The opacity correction values are capped at \RHI\ $=$ 8 for visualization purposes (higher \RHI\ are not shown).}
\label{fig:fcnm_rhi_hist}
\end{figure}

\begin{figure}
\centering
\includegraphics[width=0.5\textwidth]{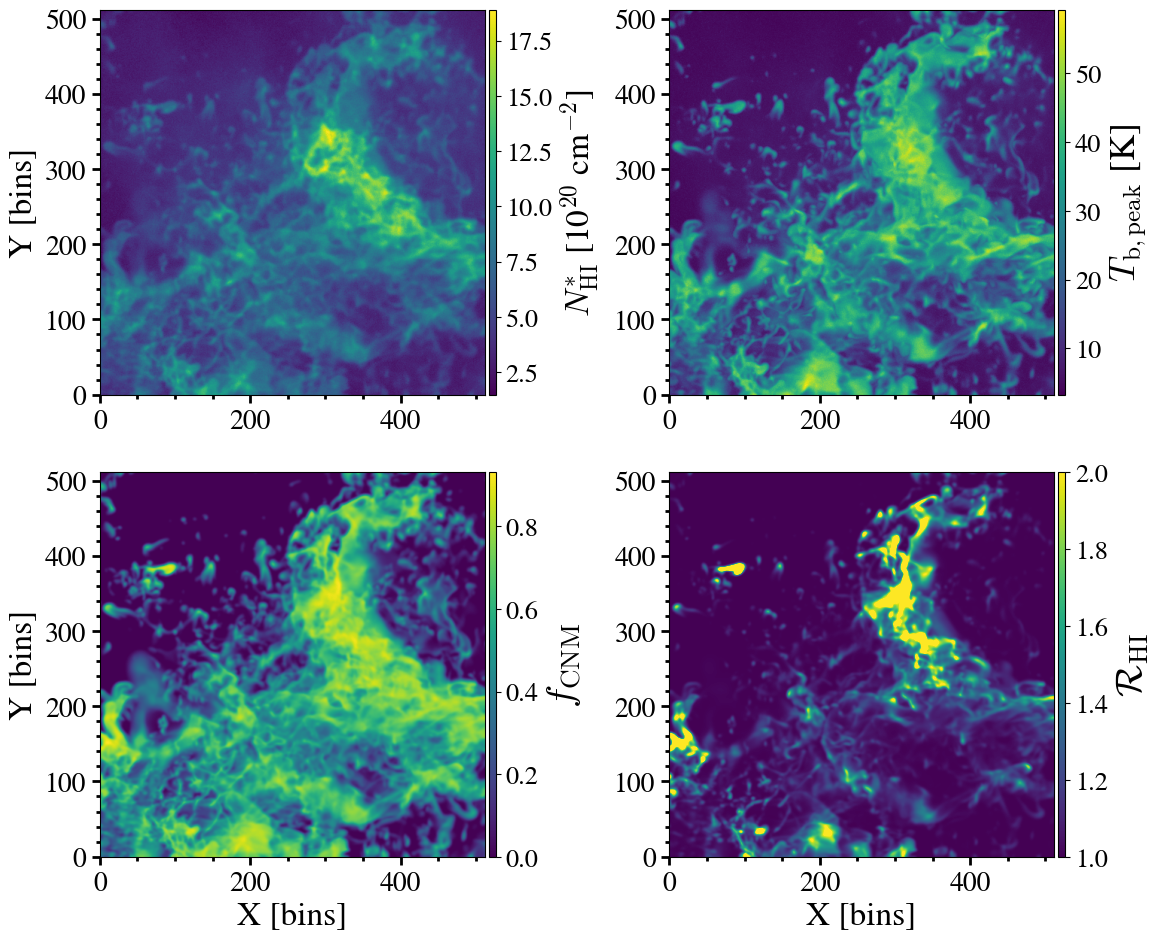}
\caption{\hi\ maps computed from a training spectral cube generated by \citet{Saury2014}'s simulation with a resolution (cell size) of 0.04 pc: Optically-thin \hi\ column density \NHIthin\ (upper left), peak brightness temperature \TBpeak\ (upper right), cold gas mass fraction \FCNM\ with $T_\text{k, CNM} < 500$ K (lower left), and \hi\ opacity correction factor \RHI\ (lower right).}
\label{fig:nhi_tbmax_fcnm_rhi}
\end{figure}

Figure \ref{fig:fcnm_rhi_hist} depicts histograms of \FCNM\ and \RHI\ from both training/evaluation sets (black solid and dashed lines, respectively) and absorption observations (magenta lines), compiled by \cite{McClure-Griffiths2023} under the ``BIGHICAT'' meta-catalog. In training/evaluation datasets, synthetic \FCNM\ spans 0\% to 100\%, while the absorption-based \FCNM\ ranges from 0\% to 88\% (with $T_\text{k, CNM} < 500$ K). \RHI\ values in the synthetic database vary 1 to $\sim$15, whereas estimated \RHI\ from absorption surveys extends from 0 to 3. Synthetic \FCNM\ and \RHI\ are apparently higher than those observed by absorption surveys, indicating potential limitations of the simulations in producing molecular gas and consequently yielding an excess of optically thick \hi\ (refer to Section \ref{sec:compare_with_absorption} for a detailed comparison). This discrepancy may introduce biases during model training, potentially leading models trained from this database to overpredict \FCNM\ and \RHI. To address this concern, we plan to incorporate more recent MHD numerical studies \citep[e.g.,][]{Kim2013,KimOstriker2017,Hu2023,Vijayan2023,Kim2023} in future work. Nevertheless, to ensure effective learning and generalization, it is essential that training sets have diverse output values, allowing training to encounter a broad range of scenarios.

In Figure \ref{fig:nhi_tbmax_fcnm_rhi}, the upper-left panel shows the optically-thin \hi\ column density map of a training cube in the unit of $10^{20}$ cm$^{-2}$. The upper-right panel displays the peak brightness temperature (\TBpeak). The lower-left and lower-right panels depict the maps of \FCNM\ and \RHI, respectively. Substantial variations in \FCNM\ and \RHI\ can be observed across different map areas. These variations are influenced by the distribution of cold neutral hydrogen gas and the structural characteristics of the \hi\ gas phases.

Visualizations of the remaining data cubes are provided in publicly available analysis notebooks at \href{https://doi.org/10.5281/zenodo.14183481}{\textit{DOI 10.5281}}.

\subsection{Trends in training sets}
\label{subsec:training_set_trends}

\begin{figure*}
\centering
\includegraphics[width=\textwidth]{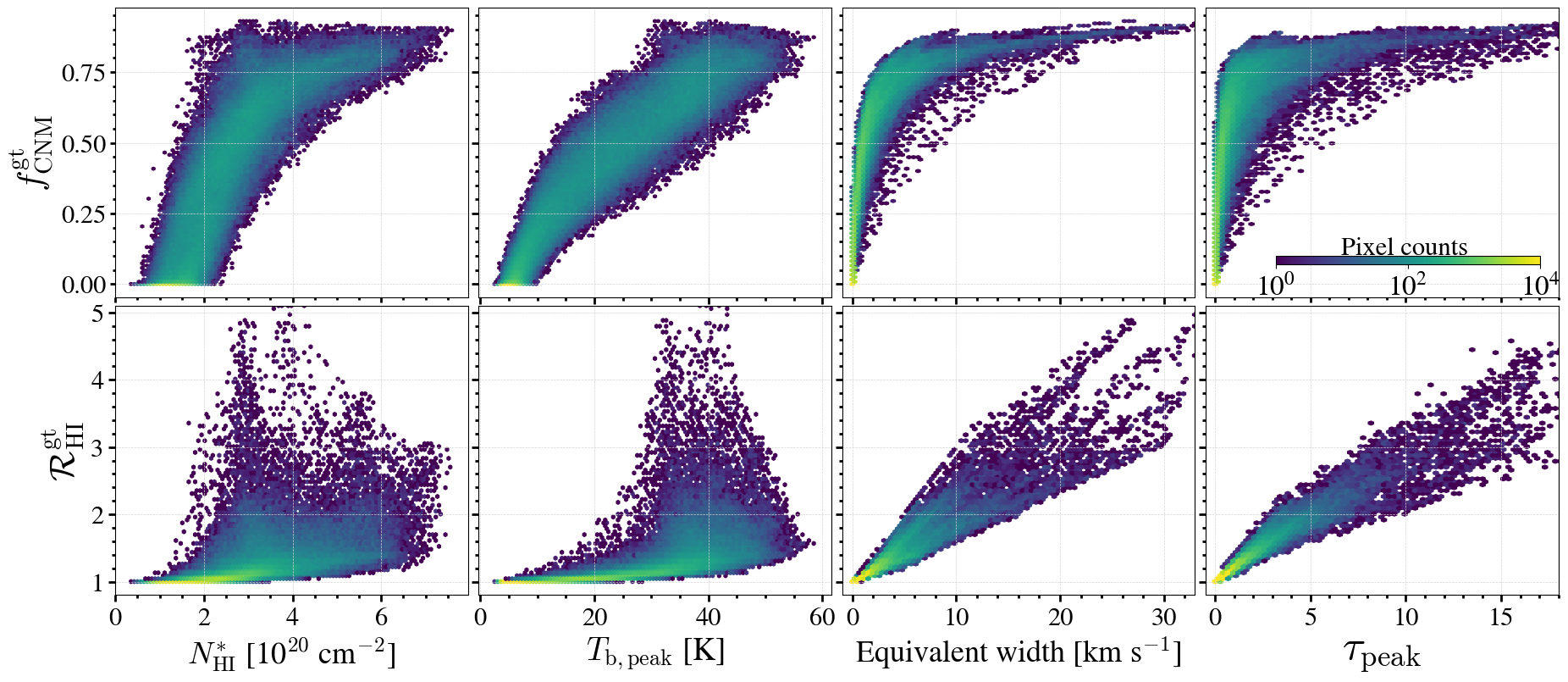}
\caption{The relationships between ground truth (``gt'') values of CNM mass fraction \FCNM, opacity correction factor \RHI\ and optically-thin \hi\ column density \NHIthin\ (first column), maximum brightness temperature $T_\mathrm{b,peak}$ (second column), equivalent width $\text{EW} = \int$$\tau dv$ [km s$^{-1}$] (third column), and \hi\ peak optical depth $\tau_\mathrm{peak}$ (last column) in a synthetic position-position-velocity spectral data cube.}
\label{fig:trends}
\end{figure*}

Examining the training sets obtained from simulations reveals distinct trends in the relationships between synthetic \FCNM\ and \RHI\ with two key (observable) quantities from emission: \TBpeak\ and \NHIthin, as well as with absorption-based quantities: peak optical depth $\tau_\text{peak}$ and equivalent width $\text{EW} = \int \tau (v)dv$ (Figure \ref{fig:trends}). We note that synthetic \FCNM\ and \RHI\ increase in tandem with \TBpeak, \NHI, \taupeak\ and equivalent width increase, showing strong positive correlations among these parameters. Higher values of \TBpeak, \NHI, \taupeak\ and equivalent width are associated with higher \FCNM\ and \RHI. Synthetic \RHI, nevertheless,
show distinctive fluctuations with occasional abnormally high values at certain range of \TBpeak\ and \NHI. These fluctuations in \RHI\ may suggest that its behavior is more complex and potentially influenced by additional factors or interactions in the simulations, which require further investigation.

Similar trends have been observed between \FCNM, \RHI, and column density, equivalent width from absorption analyses \citep[e.g.,][]{Heiles2003b,Lee2015,Murray2015,Nguyen2018, Nguyen2019}. Within optically-thin regime (\NHI\ $\sim$ \NHIthin\ $\lesssim$ 5 \nhiUnit), the absorption-based \FCNM\ reaches its peak ($\sim$0.6) and does not exhibit saturation afterward at higher \NHIthin; also, the observed opacity correction is not significant (\RHI\ $\sim$ 1) in this range of column density. However, within the training database, synthetic \FCNM\ saturates (at $\sim$0.8), and synthetic optically-thick \hi\ becomes significant (\RHI\ $\sim$ 2) at very low column densities (\NHIthin\ $\sim$ 2 \nhiUnit) and low EW ($\sim$ 0.5 \kms).

Despite the differences observed between the synthetic and observational databases, the increasing trends of \FCNM\ and \RHI\ over \TBpeak\ and \NHIthin\ may serve as vital factors for the models to learn from, enabling them to extract comprehensive patterns encoded in the training datasets. Since our models are designed to predict \FCNM\ and \RHI\ from emission only, synthetic optical depth $\tau(v)$ is not used in the model training.

In addition to \TBpeak\ and \NHIthin, we also investigate potential patterns in the relationship between the two parameters of interest (\FCNM\ and \RHI) and the FWHM line widths estimated from second moments (as described in Section \ref{subsec:training_datasets}). Our analysis does not reveal strong correlations between \FCNM\ or \RHI\ and the line widths. Specifically, the Pearson and Spearman coefficients for \FCNM\ and FWHM line widths are $R_\mathrm{Pearson}$ = $-$0.08 and $\rho_\mathrm{Spearman}$ = $-$0.11, respectively, both $p-$values = 0. Similarly, the Pearson and Spearman coefficients for \RHI\ and FWHM line widths are $R_\mathrm{Pearson}$ = $-$0.06 and $\rho_\mathrm{Spearman}$ = $-$0.02, both $p-$values = 0. This suggests that line widths do not substantially affect the predicted values.

\begin{figure}
\centering
\includegraphics[width=0.485\textwidth]{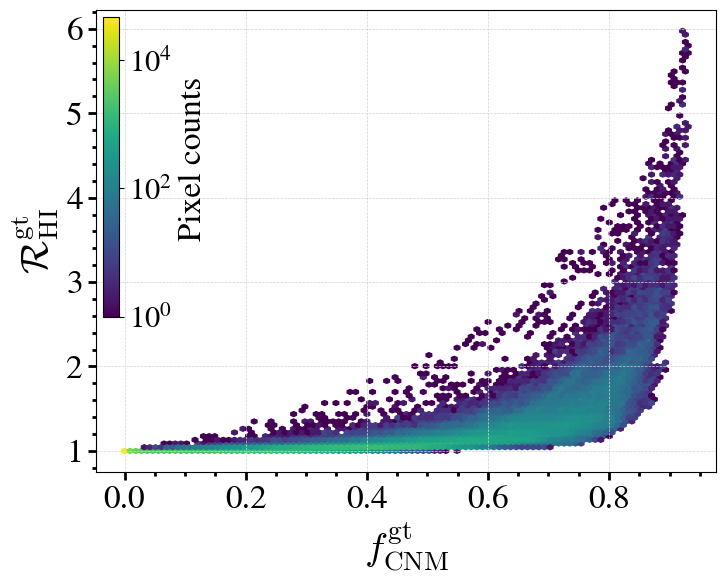}
\caption{The relationships between ground truth (``gt'') values of CNM mass fraction \FCNM\ and opacity correction factor \RHI\ in the same synthetic position-position-velocity spectral data cube as in Figure \ref{fig:trends}.}
\label{fig:fcnm_rhi_trends}
\end{figure}

\subsection{Absorption surveys}
\label{subsec:abs_data}

We evaluate the performance of our \ourmod\ models by comparing their predictions to the findings from absorption surveys in the ``BIGHICAT'' catalog compiled by \citet{McClure-Griffiths2023} (see Section \ref{sec:compare_with_absorption}). This absorption catalog is sourced from various observational studies \citep{Heiles2003a,Mohan2004,Roy2013,Roy2013b,Stanimirovic2014,Murray2015,Denes2018,Murray2018,Murray2018a,Nguyen2019,Murray2021} and provides a large sample (373 unique lines of sight) of \FCNM\ and \RHI\ estimates across different regions of the Galactic ISM  over the velocity range ($-90, -90$) \kms. We focus on surveys that target random sightlines, such as those in the 21-SPONGE (56 sightlines) and Millennium (57 sightlines) surveys, as well as the MACH survey (44 sightlines) which covers high galactic latitudes. These lines of sight are directed away from giant molecular clouds and are assumed to be dominated by atomic gas. The \FCNM\ and \RHI\ along these sightlines have been determined in the most direct approach using Gaussian decompositions of emission-absorption spectral pairs. These selected absorption surveys are heterogeneous in terms of opacity sensitivity ($\sim$5 $\times$ 10$^{-4}$ per 1 \kms\ channel width) and they all employed the Gaussian fitting techniques based on the methodology developed by \cite{Heiles2003a} to model the data. 

Given our definition of the CNM in this study, we recalculated the \FCNM\ along all selected sightlines, considering all identified CNM components with \Ts\ < 500 K. Whereas, the opacity correction factors \RHI\ remain unaffected by the CNM temperature threshold.

We would like to point out that the methodology for estimating emission spectra from absorption surveys differs from that for synthetic emission spectra. For instance, the Mach survey employed 40 EBHIS emission spectra \citep[from the Effelsberg-Bonn All-Northern Sky Survey,][]{Winkel2010,Kerp2011,Winkel2016} around each radio continuum source for Gaussian fitting, whereas other surveys estimated the expected emission spectra (the emission profiles in the absence of a continuum source) from off-source measurements to account for fluctuations in emission around a radio source. In contrast, synthetic emission spectra were derived from the radiative transfer of the 21-cm line using number density, temperature, and velocity fields generated by simulations.

\subsection{Observed emission data}
\label{subsec:observed_set}

\begin{figure*}
\centering
\includegraphics[width=\textwidth]{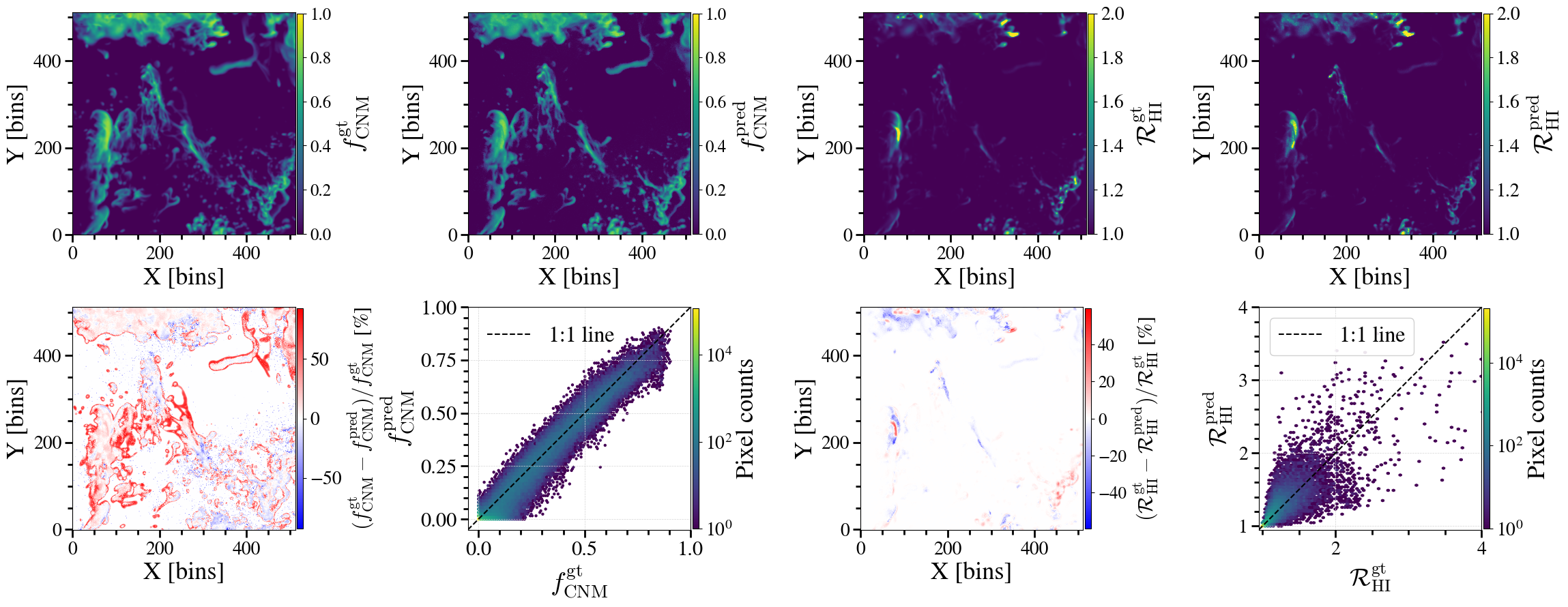}
\caption{Predictions of \FCNM\ (two first columns) and \RHI\ (two last columns) by \ourmods\ using the evaluation cube with a cell size of 0.04 pc. Upper left: ground-truth (``gt''); upper right: predictions (``pred''); lower left: relative difference between ground truths and predictions, with a mean (standard deviation) of 12\% (21\%) for \FCNM\ and 1\% (3\%) for \FCNM\; lower right: one-to-one comparison.}.
\label{fig:ctrans_fcnm_map_c4_compare}
\end{figure*}

We also validate our \ourmod\ models' performances using emission data from the Green Bank Telescope \hi\ Intermediate Galactic Latitude Survey \citep[GHIGLS,][]{Martin2015}. The emission data cube features a spatial resolution of $\sim$9.4 arcmin (pixel size of 3.5 arcmin), a velocity resolution (channel spacing) of 1.0 (0.8) \kms, and an rms noise level of $\lesssim$ 100 mK per 0.8 \kms\ channel. Our analysis focuses on the Low-latitude intermediate-velocity Arch 1 (LLIV1), a 7$^{\circ}$.5 (128 pixels with a pixel size of 3.5 arcmin) square subregion of the North Celestial Pole Loop (NCPL) mosaic centered at ($l,b$) = ($+$144$^{\circ}$, $+$39$^{\circ}$), and spanning a velocity range of ($-81, -27$) \kms. The distance to this subregion is estimated to be between 0.9 and 1.8 kpc (see Table 2 in \citealt{Wakker2001}), implying that each pixel has a physical linear size of $\sim$1$-$2 pc. We will apply \ourmod\ models to predict \FCNM\ and \RHI\ in LLIV1 and then compare these predictions with publicly available results of the cold \hi\ gas properties obtained through Gaussian decomposition \citep[][]{Vujeva2023}, Fourier transform \citep{Marchal2024}, and M20 CNN model (see Section \ref{sec:application_to_emission}).

A subregion toward a high Galactic latitude Intermediate Velocity Cloud (IVC), centered at Galactic coordinates ($l,b$) = ($+$135$^{\circ}$, $+$55$^{\circ}$) in the HI4PI all-sky Survey \citep{hi4pi2016}, is also utilized as an auxiliary set. This IVC is a bright cloud with a brightness temperature $T_\mathrm{b}$ $\approx$ 26 K at velocity $v$ $\approx$ $-50$ \kms. \citet{Marchal2024} investigated this IVC subregion in detail (see their Section 8), and compared their lower limit on CNM mass fraction, derived by Fourier transform of 21-cm emission spectra, with M20 CNN predictions. We include this IVC in our analysis solely for comparing \ourmods\ with the M20 CNN.

In applying machine learning techniques to real observational data, it is crucial to account for potential artefacts and quality issues that can degrade model performance. Common observational artefacts, such as radio frequency interference (RFI), stray radiation, and poor spectral baselines, can introduce biases that affect the accuracy of predicted physical parameters. The emission and absorption datasets used in this work were thoroughly calibrated and corrected for major observational artifacts (e.g., RFI, stray radiation). The observing and data reduction techniques were specifically designed to produce high-quality spectral line data cubes \citep[e.g., see][]{Heiles2003a,McClureGriffiths2009,Martin2015,Stanimirovic2014,Murray2015,Peek2011,Peek2018}.

\section{Prediction using machine learning}
\label{sec:prediction}
    Our supervised machine-learning task aims to quantify optically-thick cold \hi\ gas (\FCNM\ and \RHI) from emission spectra using deep learning models. Models are trained on synthetic spectral cubes with known ground truth labels. During training, models optimize their parameters -- weights and biases -- by minimizing prediction errors, with the Root Mean Square Error (RMSE) chosen as the evaluation metric. We tune hyperparameters and configurations for effective generalization. Once trained and validated, models make predictions on unseen data.

We have designed deep CNNs and Transformer-based \ourmods\ for this regression task. CNNs use convolutional layers to learn patterns within the spectral input, while \ourmods\ combine CNN and Transformer, leveraging multi-head attention for long-range interactions between different parts of emission spectra. Refer to Appendix \ref{sec:models} for details of the two deep learning architectures. Moreover, full descriptions of the experimental setup, training procedures, and optimized hyperparameters (e.g., number of layers, learning rates) for both models are available in Appendix \ref{sec:hyperparams}. After testing various convolutional configurations and learning rates, we determined that an 8-layer convolutional architecture and a learning rate of $5 \times 10^{-3}$ for both CNNs and \ourmods\ provided optimal performances. These settings are consistently used in all subsequent experiments.

To evaluate the role of positional encodings, we experimented with different data representations by incorporating seven positional encoding techniques into the emission spectra as models' inputs. Taking into account the sensitivity to convolutional weight initialization and shuffling, as well as fluctuations in testing RMSEs, we observed that \ourmod\ regularly outperforms deep CNNs, with an average improvement of 10\% in testing accuracy, convergence speed, and model training stability. Among the positional encoding techniques, we identified the \ourmod\ with \textit{``add sinusoidal''} positional encoding (adding a sinusoidal positional function to the original spectrum) as the most robust model. This model is therefore selected for further analyses. See Appendix \ref{sec:pes} for details, where the impact of kernel sizes, spectral resolutions, and noise levels on model performance are also discussed.

\section{Verification on synthetic spectra}
\label{sec:evaluation}
    In this section, we evaluate the performance of the \ourmod\ and outline the outcomes of its application to predicting \FCNM\ and \RHI\ maps using the evaluation spectral cube. In this case, we deploy the \ourmod\ model along with the most robust positional encoding method, specifically the \textit{adding sinusoidal} approach (as pointed out in Section \ref{subsec:comparison}). Figure \ref{fig:ctrans_fcnm_map_c4_compare} shows the predicted \FCNM\ and \RHI\ maps, respectively, as well as a comparison against the ground truths. In each figure, the upper-right panel shows the true values (``gt''), the upper-left panel presents the predictions (``pred''), the lower-left panel displays the relative difference between the predictions and ground truths, and the lower-right panel provides a scatter plot for a one-to-one comparison.

Our hybrid convolutional-Transformer model generally succeeds in predicting \FCNM\ and \RHI\ compared to the ground-truth maps, effectively capturing the spatial distributions of CNM gas and optically-thick \hi\ from the evaluation spectral cube. In both cases of \FCNM\ and \RHI, the model can reproduce the structure of cold \hi\ gas projected onto 2D maps, although it operates on a spectrum-by-spectrum basis and does not incorporate information about spatial coherence. Quantitatively, the predicted \FCNM\ closely aligns with the ground truths, with a mean absolute difference and standard deviation of 0.02 and 0.03 respectively. Their relative difference ranges from $-$250\% to 92\% (values below $-$92\% are clipped in Figure \ref{fig:ctrans_fcnm_map_c4_compare}). The RMSE between the prediction and ground truths is $3.5\%$. All pixels with large relative errors have very small ground truths (\FCNM\ $\sim$ 0). This occurs because when the difference between ground truths and predictions is divided by a small ground truth value, even a small difference results in a large relative error. Indeed, pixels with a relative difference less than $-$50\% or greater than $+$50\% have ground truths \FCNM\ $<$ 0.2.

Likewise, the predicted \RHI\ also exhibits a linear correlation with the true values. Their absolute difference has a mean of 0.01 and a standard deviation of 0.05. The relative difference spans from $-$140\% to $\sim$60\%, with values below $-$60\% clipped in Figure \ref{fig:ctrans_fcnm_map_c4_compare}. The RMSE between the predictions and ground truths is 0.05. At \RHI\ < 2, \ourmod\ adeptly predicts \hi\ opacity correction factors. However, challenges arise at high \RHI, where the \RHI\ values (\RHI\ > 2) appear to be outliers (refer to Figure \ref{fig:fcnm_rhi_hist} for observed \RHI\ from absorption surveys, and Figure \ref{fig:trends} for the distribution of \RHI\ in training sets). In these instances, the model tends to predict similarly high \RHI\ values, mirroring the trends seen in the training data. It is important to emphasize that, observationally, approximately half of the entire sky is in the optically-thin regime \citep{hi4pi2016}, where \RHI\ $\sim$ 1. Regions with high \RHI\ are typically in directions through the galactic plane or towards giant molecular clouds \citep{Heiles2003a,Lee2015,Nguyen2019}. 

\begin{figure}
\centering
\includegraphics[width=0.5\textwidth]{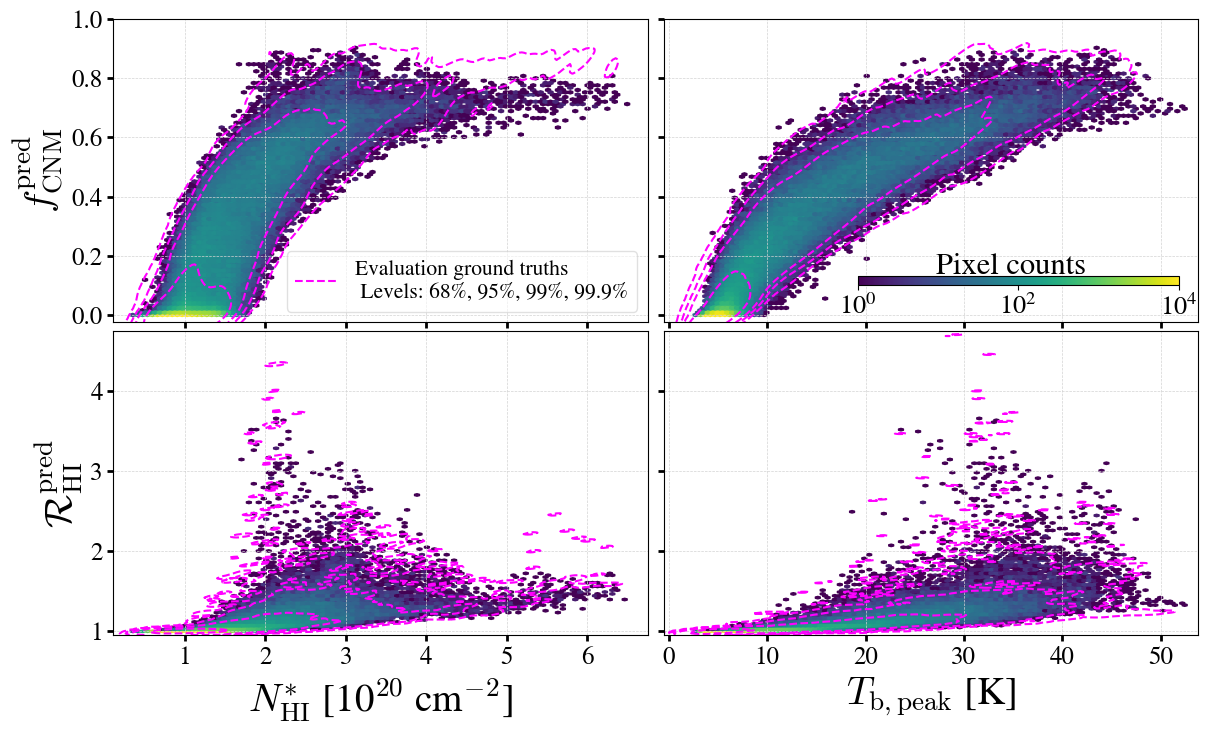}
\caption{Same as Figure \ref{fig:trends} but for \RHI\ and \FCNM\ \ourmod\ predictions (``pred'') using evaluation cube. Four contour levels (68\%, 95\%, 99\%, and 99.9\%) in magenta are generated by evaluation ground truths.}
\label{fig:trends_pred}
\end{figure}

In Figure \ref{fig:trends_pred}, we illustrate the patterns of predicted \FCNM\ and \RHI\ with respect to the optically-thin \hi\ column density \NHIthin\ and peak brightness temperature \TBpeak. Contours at 68\%, 95\%, 99\%, and 99.9\% levels derived from the evaluation ground truths are overlaid for direct comparison. The predicted \FCNM\ and \RHI\ consistently align well with the extent of the four contour levels. Additionally, the model effectively predicts anomalous \RHI\ values at certain \NHIthin\ and \TBpeak\ (albeit slightly higher), akin to those in the ground truths. It is thus evident that \ourmod\ model is capable of effectively reproducing the trends seen in the evaluation cube, specifically showing higher \FCNM\ and \RHI\ with increasing \NHI\ and \TBpeak.

We also do tests to see how different simulations impact the model's performance. When training with datacubes from \cite{Saury2014} and testing with \cite{Seta2022} datacubes, the RMSE for the training phase is $\sim$30\% lower (across all positional encoding techniques). However, the RMSE for predictions on the evaluation set is $\sim$40\% higher. A similar tendency is observed when we swap the training/evaluation sets between the two simulations. In general, we find that models generalize better when trained on combined synthetic datasets from different simulations that include a wider range of astrophysical processes.

\section{Comparison with absorption surveys}
\label{sec:compare_with_absorption}
\begin{figure}
\centering
\includegraphics[width=0.5\textwidth]{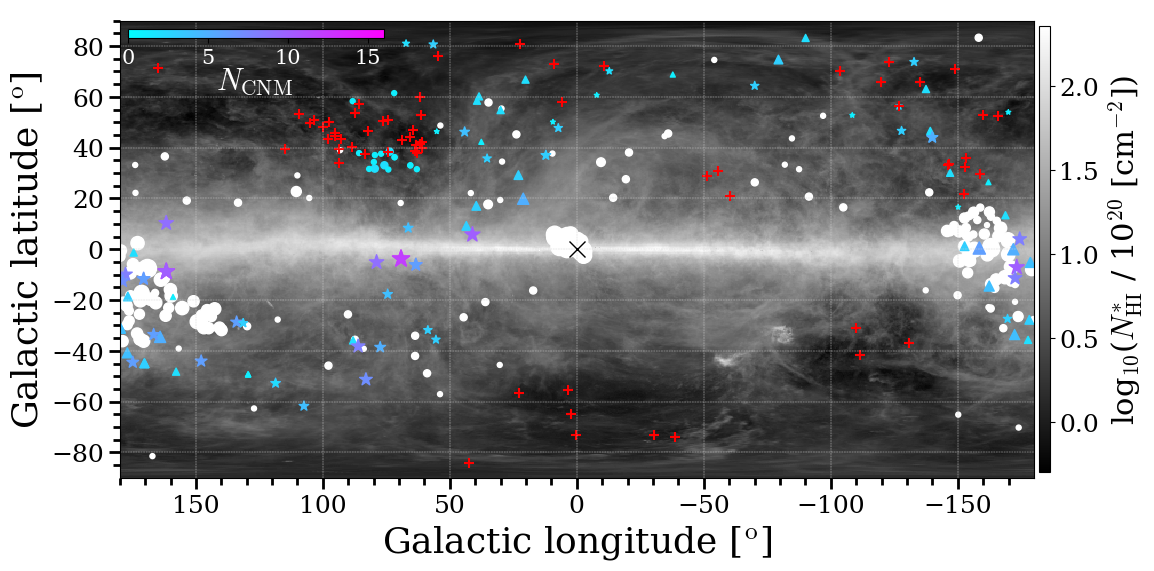}
\caption{Locations of absorption measurements from ``BIGHICAT'' catalog \citep{McClure-Griffiths2023} in the Galactic coordinates, overlaid on the map of HI4PI column density \NHIthin\ \citep{hi4pi2016}. The  ``X'' marker (at $l = 0^{\circ}, b = 0^{\circ}$) labels the Galactic center. The red crosses indicate non-detections. Colored markers (associated with the upper-left color bar) highlight the absorption surveys considered in this work: stars for 21-SPONGE, circles for MACH, and triangles for Millennium Survey. White circles represent other surveys. Both colors and sizes of the markers represent the number of identified absorption Gaussian components $N_\mathrm{CNM}$ (ranging from 1 to 16).}
\label{fig:bighicat_locs}
\end{figure}

\begin{figure*}
\centering
\includegraphics[width=\textwidth]{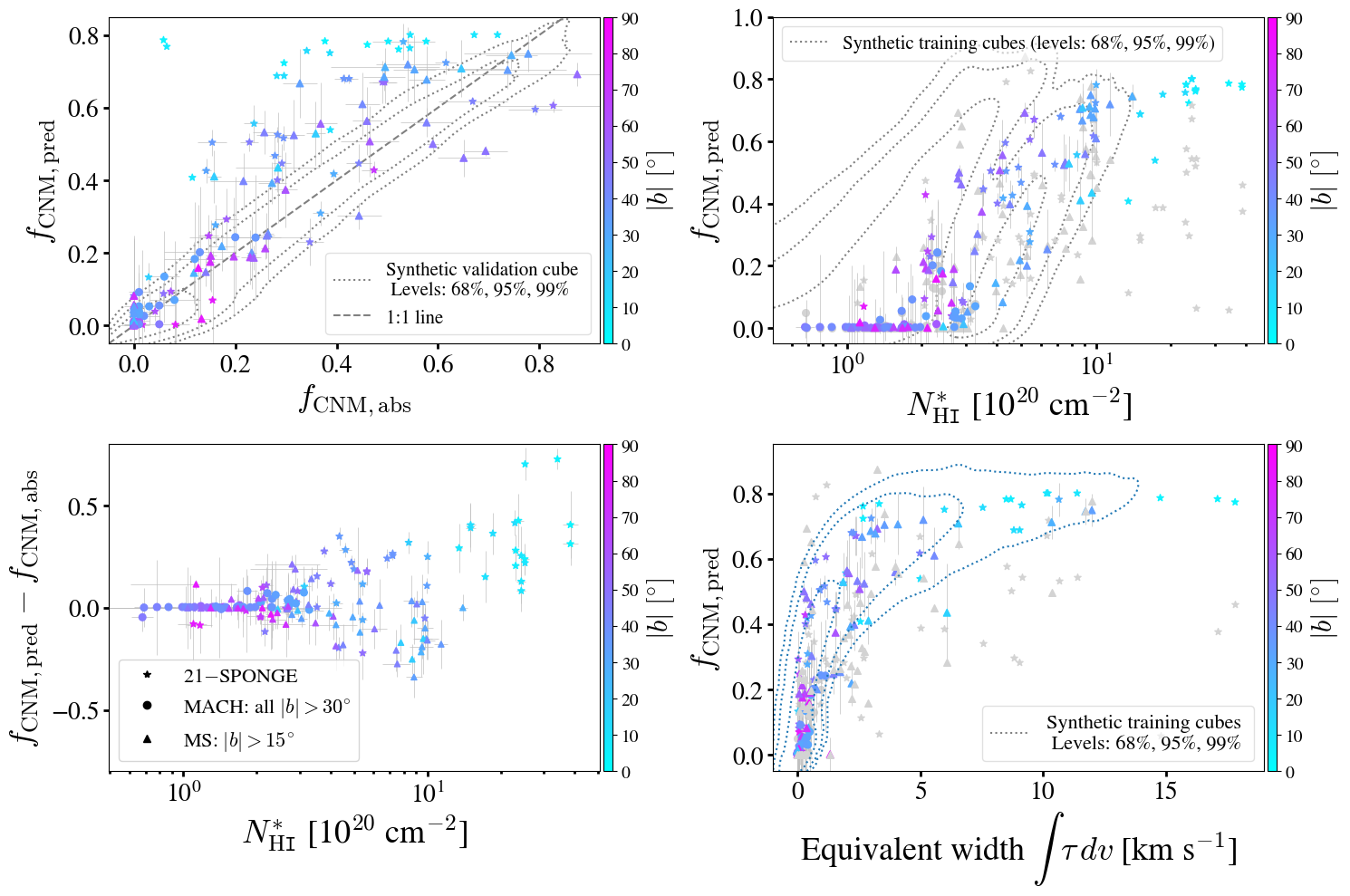}
\caption{Comparison between absorption cold \hi\ gas fraction ($f_\mathrm{CNM,abs}$) and \ourmod\ predictions ($f_\mathrm{CNM,pred}$) using observed data from absorption surveys: stars for 21-SPONGE, circles for MACH, and triangles for Millennium Survey. Color scale indicates the absolute Galactic latitudes $|b|$ in degrees. Upper-left panel: one-to-one comparison ($f_\mathrm{CNM,pred}$ vs $f_\mathrm{CNM,abs}$), with contours indicating the 68\%, 95\%, and 99\% levels of data point densities based on the synthetic evaluation cube. The error bars for $f_\mathrm{CNM,pred}$ are estimated as the standard deviation of the model predictions based on emission spectra around continuum sources. Lower-left panel: cold \hi\ gas fraction difference ($f_\mathrm{CNM,pred} - f_\mathrm{CNM,abs}$) against optically-thin column densities \NHIthin. Subplots on the right column: predicted $f_\mathrm{CNM,pred}$ against \NHIthin\ (upper panel) and equivalent widths $\int$$\tau_v dv$ (lower panel), with contour levels of 68\%, 95\%, and 99\% generated from the synthetic training cubes. Absorption-based $f_\mathrm{CNM,abs}$ is shown in grey without error bars.}
\label{fig:fcnm_vs_abs}
\end{figure*}

\begin{figure*}
\centering
\includegraphics[width=\textwidth]{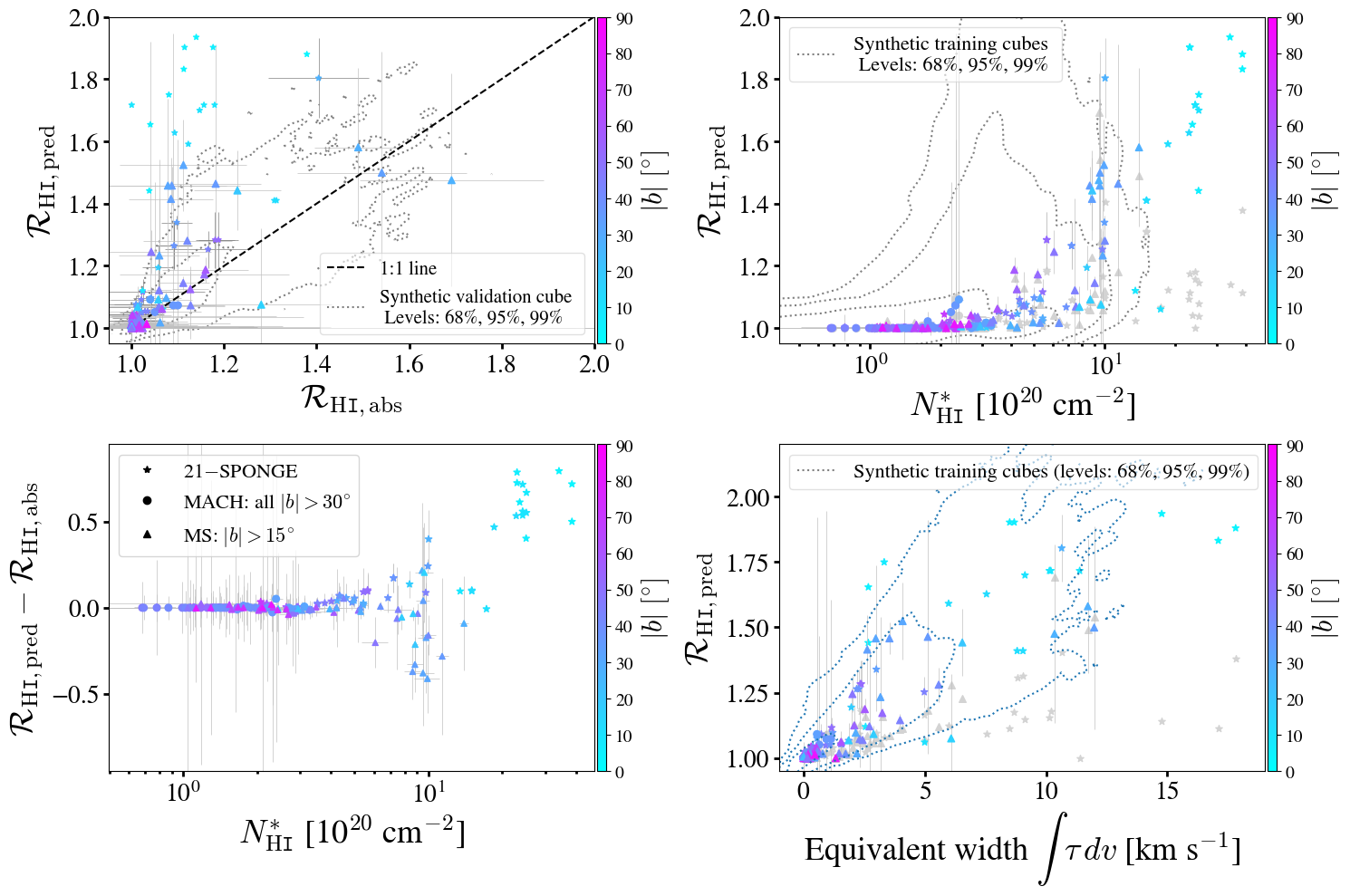}
\caption{Same as Figure \ref{fig:fcnm_vs_abs}, but for the comparison between absorption-based \hi\ opacity correction factor ($\mathcal{R}_\mathrm{HI,abs}$) and \ourmod\ predictions ($\mathcal{R}_\mathrm{HI,pred}$).}
\label{fig:rhi_vs_abs}
\end{figure*}

After training the neural network models, we put them to the test on observed datasets to evaluate their performance and applicability. Here, we compare \ourmod\ predictions with results from absorption surveys that provide $f_\mathrm{CNM,abs}$ and $\mathcal{R}_\mathrm{HI,abs}$ values inferred through Gaussian decompositions of emission-absorption spectra. We specifically focus on absorption surveys towards (1) random sightlines (21-SPONGE and Millennium surveys) or at high galactic latitudes (MACH survey) and (2) directed away from giant molecular clouds. Sightlines with saturated absorption spectra ($e^{-\tau}$ $\sim$ 0) or unusually high $\chi^{2}$ values in the Gaussian fittings, which may impact the accuracy of estimating cold \hi\ gas and optical depth correction, are excluded. Ultimately, we incorporate 157 lines of sight from the selected absorption surveys, as shown in Figure \ref{fig:bighicat_locs}: 44 from MACH \citep[][out of 44 sightlines]{Murray2021}, 56 from 21-SPONGE \citep[][out of 57 sightlines, with one sightline coinciding with the MACH survey]{Murray2018a,Murray2018} and 57 from the Millennium Survey \citep[][out of 79 sightlines]{Heiles2003a}.

To align with our definition of CNM, \FCNM\ along the selected lines of sight is then recomputed with \Ts\ $<$ 500 K for all the CNM components identified by the Gaussian fitting processes. The \RHI\ estimates remain unchanged under the CNM temperature threshold.

For a comparison with absorption $f_\mathrm{CNM,abs}$ and $\mathcal{R}_\mathrm{HI,abs}$, we require emission spectra to apply our trained \ourmods. Before feeding them to the models, the emission spectra must be regridded into 256 channels to align with the required inputs. Since emission measured at the locations of continuum background sources is contaminated by the absorption from the \hi\ gas in the foreground, we thus, following the observational strategy of the absorption surveys, use the off-source emission measurements around a continuum source as the ``expected'' emission spectra (representing the emission profiles we would observe as if the radio continuum source were turned off). For each MACH background continuum source, we utilize its 40 provided emission spectra from the Effelsberg-Bonn All-Northern Sky Survey (angular resolution of 9$^{\prime}$, brightness sensitivity of 90 mK, and 1.3 \kms\ per channel velocity resolution). Similarly, for each 21-SPONGE and Millennium continuum source, we leverage its associated ``expected'' emission spectrum. In line with the MACH survey approach, in addition to the expected spectrum, we extract eight extra emission spectra around each 21-SPONGE/Millennium continuum source from the Arecibo GALFA-\hi\ survey \citep[][angular resolution of 3$^{\prime}$.4, brightness sensitivity of 80 mK, and 0.16 \kms\ per channel velocity resolution]{Peek2011,Peek2018}. In this extraction, we specifically selected one spectrum per Arecibo beam size to minimize correlation within a beam. As a result, the uncertainties of \FCNM\ and \RHI\ are estimated as the standard deviation of the model predictions derived from these emission spectra.

Figure \ref{fig:fcnm_vs_abs} displays the comparison between \ourmod\ predictions ($f_\mathrm{CNM,pred}$) and absorption-based cold gas fraction ($f_\mathrm{CNM,abs}$) as well as their relationships with two observables: optically-thin column density \NHIthin\ and equivalent width $\int$$\tau_v$$dv$. Across all four panels, distinctive markers represent different absorption surveys: stars for 21-SPONGE, circles for MACH, and triangles for Millennium Survey. The color scale indicates the absolute Galactic latitudes $|b|$. Contours (if shown) outline the 68\%, 95\%, and 99\% levels (or 1$\sigma$, 2$\sigma$, and 3$\sigma$, respectively) of data point densities derived from the synthetic PPV cubes. While the contours in the right panels originate from the ground truths of the training cubes, those in the upper-left panel are produced from \ourmod\ predictions on the evaluation cube.

The upper-left panel presents a one-to-one comparison between the predicted cold \hi\ gas fraction $f_\mathrm{CNM,pred}$ and the absorption-based $f_\mathrm{CNM,abs}$. The scatter plot generally reveals a good linear correlation between the two estimates, as evidenced by the majority of data points clustering within the 3$\sigma$ contour level. While the \ourmod\ method demonstrates robust agreement with the MACH absorption survey, which probes the local ISM at low column densities (\NHIthin\ $<$ 2 \nhiUnit), it deviates extensively from absorption values obtained by 21-SPONGE and Millennium surveys as one approaches the Galactic Plane at higher column densities (\NHIthin\ $>$ 5 \nhiUnit), where the observed emission spectra tend to exhibit more complex structural features. This deviation is particularly noticeable in the lower-left panel illustrating the difference $\Delta f_\mathrm{CNM} = (f_\mathrm{CNM,pred} - f_\mathrm{CNM,abs}$) against \NHIthin\ in log scale, with $\Delta f_\mathrm{CNM}$ reaching up to $\sim$0.7. Predictions from \ourmod, based only on emission data, align well with emission-absorption Gaussian fitting methodology in optically-thin regimes \NHIthin\ $\lesssim$ 5 \nhiUnit, nevertheless, at higher column densities, our \ourmod\ model appears to estimate a higher amount of cold gas.

To gain a better understanding of how \ourmod\ predicts cold gas fraction after being trained on training data sets, we plot in the right column the predicted \FCNM\ as a function of the two keys observables: optically-thin column densities \NHIthin\ (upper panel) and equivalent widths (lower panel). Grey markers represent $f_\mathrm{CNM,abs}$ values inferred from absorption surveys for direct comparisons. Overall, the model predictions in both panels closely track the 3$\sigma$ contour, which highlights the trends in training sets (\FCNM\ vs \NHIthin\ and \FCNM\ vs equivalent width, as described in Section \ref{subsec:training_set_trends}). The relationship of \FCNM\ vs \TBpeak, although not shown here, also follows a similar pattern. The absorption-based $f_\mathrm{CNM,abs}$ estimates align well with trends over low column densities (\NHIthin\ $\lesssim$ 5 \nhiUnit) and equivalent widths ($\text{EW} \lesssim$ 5 \kms), but diverge substantially at higher \NHIthin\ and $\text{EW}$ (with significantly lower $f_\mathrm{CNM,abs}$ values falling below the 3$\sigma$ contour).

Toward a few lines of sight pointing through the Galactic Plane, \ourmod\ models predict high $f_\mathrm{CNM,pred}$ ($\sim$0.8) beyond the 3$\sigma$ contours in both panels, but within the 100\% contour levels (not shown) acquired from the whole training sets. Notably, the upper-right panel reveals that sightlines with higher column densities (\NHIthin\ $>$ 3 \nhiUnit) consistently align with the trends seen in \cite{Seta2022} simulation. In contrast, those with lower \NHIthin\ follow the trends observed in both simulations utilized in this study.

Similar conclusions can be derived for \RHI, as depicted in Figure \ref{fig:rhi_vs_abs}. However, the one-to-one comparison between the model prediction ($\mathcal{R}_\mathrm{HI,pred}$) and absorption-based $\mathcal{R}_\mathrm{HI,abs}$ displays greater dispersion compared to the predicted \FCNM. This is also evident in the scatter plot of predicted \RHI\ against equivalent widths. In optically-thin regimes, the \ourmod\ method shows excellent agreement with the emission-absorption Gaussian decomposition approach. Nevertheless, over higher column densities at low Galactic latitudes, the \ourmod\ models predict considerably higher values of opacity correction. It is noteworthy that while \FCNM\ is constrained within both ends (0 $\leq$ \FCNM\ $\leq$ 1), \RHI\ is restricted only at the lower end (\RHI\ $\geq$ 1). Hence, it is likely more feasible for the \ourmod\ models to predict a distribution function of a variable within the range of (0, 1) than for a variable without explicit limits.

Thermal instability is believed to be the primary process in the ISM that leads to the thermal condensation of multiple atomic phases (WNM, UNM, and CNM) in stable pressure equilibrium \citep{Field1965,Field1969,Wolfire1995,Audit2005}. Given that the mass fraction of each phase found in numerical studies strongly depends on the specifics of prescriptions for cooling and heating, turbulence, magnetic fields, and supernova rate \citep{Gazol2001,Audit2005,Audit2010,Kim2013,Saury2014,Hill2018,Seta2022}. The disparities between model predictions and absorption surveys may thus be attributed to two key factors. Firstly, because the models are trained on synthetic datasets generated from numerical simulations where the optically-thick cold \hi\ gas was not transitioned to molecular gas, leaving a large amount of cold neutral gas within the simulated volume. They then appear to learn such patterns to predict high \FCNM\ or \RHI\ at high column densities and high equivalent widths. This tendency is evident in the right panels of Figures \ref{fig:fcnm_vs_abs} and \ref{fig:rhi_vs_abs}, showing predicted \FCNM\ as a function of column densities and equivalent widths. It seems that after being trained, the models transform into approximation (complex) functions, mapping and extrapolating new inputs (emission spectra) to an appropriate distribution of outputs based on the relationships they have learned from synthetic training data sets. More sophisticated HD/MHD simulations with complete heating and cooling prescriptions (including FUV radiation field, turbulence, magnetic field, and feedback from stars and supernovae, etc), are essential for future work to enhance accuracy. Secondly, the discrepancy may arise from the objective in emission-absorption Gaussian fitting, where an underestimated number of CNM components (or an overestimated number of WNM components) can lead to a corresponding underestimation of \FCNM\ and \RHI. Since the Gaussian fitting lacks uniqueness, a more refined approach is required for emission-absorption fitting. This approach should involve simultaneously modeling the emission and absorption spectra using a single loss function and considering spatial coherence. These additional steps could help well constrain the number of CNM and WNM velocity Gaussian components.

As pointed out in Section \ref{subsec:training_set_trends}, we do not observe significant correlations between \FCNM, \RHI, and FWHM line widths in the synthetic training database. We do not, however, rule out the potential impact of line profile complexity on the predicted values. The outliers seen in Figures \ref{fig:fcnm_vs_abs} and \ref{fig:rhi_vs_abs} (data points with significant divergence between the predictions and the absorption-based estimates) correspond to lines of sight through/near the Galactic Plane, where the emission profile structures are broader and contain more peaks compared to high Galactic latitudes. Along these Galactic disk directions, the presence of numerous cold and warm \hi\ clouds results in more complex spectral features, including blended components and diverse line shapes. Since such complex features are likely uncommon in the training set, the models may struggle to learn/recognize them. This can introduce an additional source of uncertainties in the predictions. Training models using synthetic datasets derived from Milky Way-like simulations \citep[e.g.,][]{KimWT2020,Vijayan2023}, with a particular focus on disk areas, could be one possible remedy.

\begin{figure*}
\centering
\includegraphics[width=\textwidth]{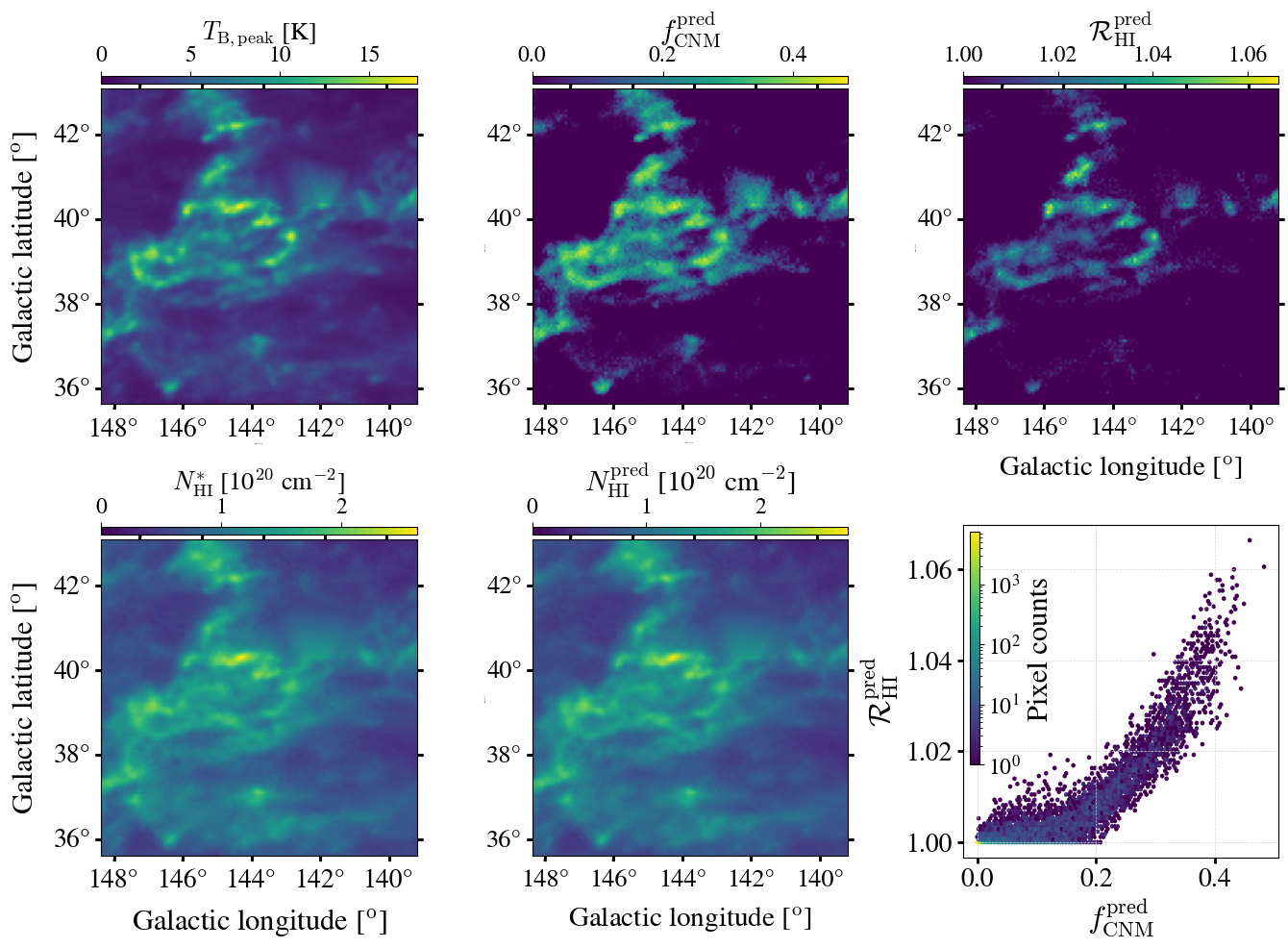}
\caption{Predictions by representation learning \ourmod\ model for the Low-Latitude Intermediate-Velocity Arch 1 (LLIV1) observed by the GHIGLS survey: Maps of peak brightness temperature \TBpeak\ (top left) and optically-thin column density \NHIthin\ (bottom left, in the unit of 10$^{20}$ cm$^{-2}$); \ourmod-predicted $f_\mathrm{CNM,pred}$ (top center) and $\mathcal{R}_\mathrm{HI,pred}$ (top right) maps, with their relationship (bottom right) and the opacity-corrected column density \NHI\ = $\mathcal{R}_\mathrm{HI,pred} \times N^{*}_\mathrm{HI}$ (bottom center).}
\label{fig:uma_replearning_pred}
\end{figure*}  

\section{Application to observed emission data from GHIGLS survey}
\label{sec:application_to_emission}
    We now validate \ourmod\ model using the LLIV1 emission spectral cube observed by the GHIGLS survey. The sensitivity of the GHIGLS survey is approximately $\lesssim$ 100 mK per 0.8 \kms velocity channel, which is sufficient to detect CNM (and parts of the WNM). The left column of Figure \ref{fig:uma_replearning_pred} shows LLIV1 peak brightness temperature (top panel) and optically-thin column density (bottom panel) maps. The peak brightness \TBpeak\ varies from 0.6 K to 17.7 K, while column density \NHIthin\ ranges from (0.25 -- 2.63) \nhiUnit. These values suggest that LLIV1 is an optically-thin region (\NHI\ $<$ 5 $\times$ 10$^{20}$ cm$^{-2}$) with a negligible opacity effect.

We apply the trained models to predict the cold gas fraction \FCNM\ and \hi\ opacity correction \RHI\ in LLIV1. Subsequently, we compare these predictions with publicly available results from the Gaussian decomposition \citep{Vujeva2023}, Fourier Transform \citep{Marchal2024} methods, and \citealt{Murray2020}'s CNN model. The observed cube of the LLIV1 region originally has a native resolution of 1 \kms\ and a channel width of 0.8 \kms. We then regrid the spectra into 101 and 256 channels to align with the required inputs for the models. Predictions from these two regridded spectral cubes are particularly consistent, hence the results presented below, maps of \FCNM\ and \RHI, are averaged from the predicted outputs.

\subsection{\ourmod\ predictions for LLIV1}
\label{subsec:rep_pred}

We represent the predictions made by our representation learning models for the LLIV1 region in Figure \ref{fig:uma_replearning_pred}: $f_\mathrm{CNM,pred}$ (top left) and $\mathcal{R}_\mathrm{HI,pred}$ (top right), along with their relationship (bottom left) and the total opacity-corrected \hi\ column density (bottom right) computed from the predictions, \NHI\ = $\mathcal{R}_\mathrm{HI,pred} \times N^{*}_\mathrm{HI}$. The $f_\mathrm{CNM,pred}$ varies in a wide range from 0 to 0.48 with a mean (median) of 0.17 (0.16), and a standard deviation of 0.1. The predicted $\mathcal{R}_\mathrm{HI,pred}$ is, on the other hand, approximately unity, with a maximum value around 1.07. These opacity correction values aligns with the expectation that the LLIV1 area is optically thin (\NHIthin\ $\approx$ \NHI\ $<$ 5 $\times$ 10$^{20}$ cm$^{-2}$, reflected in the \RHI\ = \NHI/\NHIthin\ $\sim$ 1). Although $\mathcal{R}_\mathrm{HI,pred}$ varies in a narrow range (1 to 1.07), its map captures the main structure akin to the $f_\mathrm{CNM,pred}$ map. Moreover, it increases with increasing predicted $f_\mathrm{CNM,pred}$, indicating that higher \FCNM\ corresponds to higher \RHI, and vice versa.

\subsection{\ourmod\ predictions vs Fourier-transformed method}
\label{subsec:vs_fft}

\begin{figure}
\centering
\includegraphics[width=0.5\textwidth]{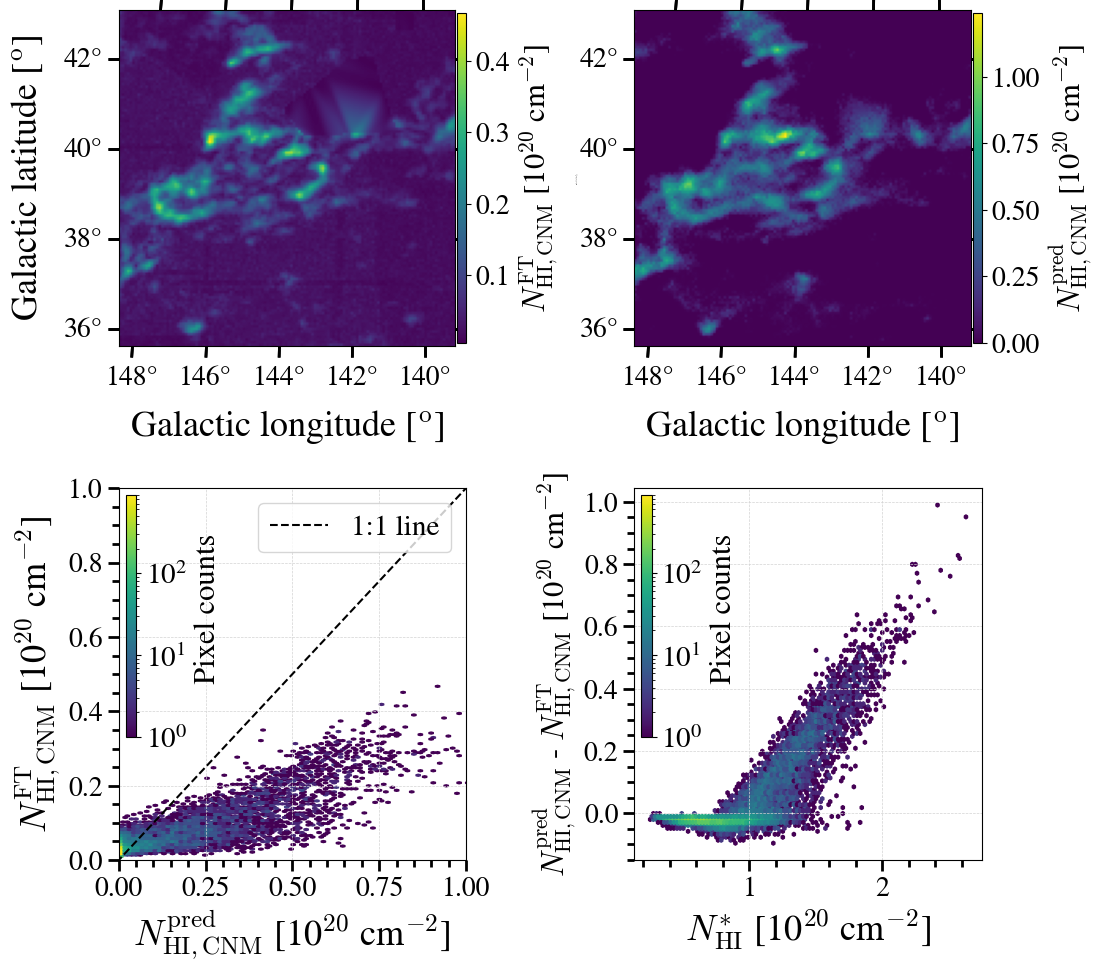}
\caption{Comparison between \ourmod\ predictions and Fourier-transformed method for the Low-Latitude Intermediate-Velocity Arch 1: CNM column density estimated by Fourier transform (top left) and by \ourmod\ model (top right) both in unit of 10$^{20}$ cm$^{-2}$, along with their one-to-one comparison (bottom left) and the dependence of their difference on the optically-thin column density \NHIthin\ (bottom right).}
\label{fig:uma_replearning_vs_fft}
\end{figure}

In Figure \ref{fig:uma_replearning_vs_fft}, we compare the amounts of cold gas obtained from \ourmod\ predictions and Fourier-transformed method using the LLIV1 spectral cube: the Fourier-transformed CNM lower limit column density (top right), CNM column densities predicted by the \ourmod\ models (top right), along with their direct comparison (bottom right). The dependence of their difference on the optically-thin column density \NHIthin\ is also plotted in the bottom right panel.

The Fourier transform operation converts an \hi\ spectrum (brightness temperature as a function of velocity) into a power spectrum in the wave-number domain ($k_v$). In the context of analyzing \hi\ spectra, where the line broadening of the \hi\ features are linked to thermal and non-thermal processes, we can simplify the interpretation of $k_v$ as representing the ``scale'' of patterns/structures within the spectral data. A higher $k_v$ value aligns with higher spectral frequencies, indicating finer colder details (CNM), whereas a lower $k_v$ value corresponds to lower spectral frequencies, reflecting larger-scale warmer structures (WNM).

This interpretation allows to define a threshold wave number $k_v$ to distinguish between CNM and WNM components. Since $k_v$ reflects the scales in spectral widths, a proxy for \hi\ temperatures, it is often chosen finely enough to confidently represent CNM structures. In high density regions of CNM (n $\gtrsim$ 100 cm$^{-3}$, e.g., \citealt{Shaw2017}), collisions (among \hi\ atoms themselves and between \hi\ atoms with electrons, ions) dominate the excitation of \hi\ atoms, spin temperature \Ts\ is thus close to the kinetic temperature \Tk\ \citep[e.g.,][]{Kulkarni1988,Liszt2001}. Observationally, observers generally assume \Ts\ $\approx$ \Tk\ for temperatures up to 1000 K \citep{McClure-Griffiths2023}. 

In order to detect CNM components characterized by a Gaussian with a line-width $\Delta V_\text{FWHM}$ $\sim$ 7 \kms (standard deviation width $\sigma$ = 3 \kms), \citealt{Marchal2024} selected a wave number threshold at $k_{v,\text{thresh}}$ = 0.12 (km s$^{-1}$)$^{-1}$ for their Fourier-transformed method. This threshold corresponds to gas with a maximum kinetic temperature of \Tkmax\ $\sim$ 1000 K, and only rescues a portion $t \sim$ 0.05 of its true total mass. As a result, the CNM fraction estimated from the Fourier transform provides a lower limit while still capturing the CNM spatial structure of large-scale \hi\ emission. This results in a lower CNM column density ($N_\mathrm{HI,CNM, FT}$ = $f_\mathrm{CNM, FT} \times N^{*}_\mathrm{HI}$) than predicted by \ourmod\ model, which takes into account the opacity effect in the calculation of \hi\ column density. This disparity is evident in the bottom panels of Figure \ref{fig:uma_replearning_vs_fft}. While most data points fall below the one-to-one line, there exists a plateau of low but finite $N_\mathrm{HI,CNM, FT}$ ($\sim$ 0.05 \nhiUnit) with a small subset of data points exceeding the equality line, resulting in a negative difference between model predictions and Fourier-transformed $N_\mathrm{HI,CNM}$ (as observed in the bottom right panel). \citealt{Marchal2024} noted that these anomalous lower limit values arise from areas within the LLIV1 map where the Fourier transform yields small but non-zero cold gas fraction $f_\text{CNM,low}$ (typically below 0.05) from broad WNM emission and noise. In contrast, \ourmod\ models predict negligible CNM amounts in these noise areas.

If we exclude these anomalous data points from the LLIV1 map, we notice from the bottom-right panel that at low column density (\NHIthin\ $<$ 1 \nhiUnit), the Fourier transform yields a comparable CNM estimation to the machine learning approach. However, this alignment vanishes as the column density increases. The difference in $N_\mathrm{HI,CNM}$ between the two methods ranges from (0 -- 1) \nhiUnit. This translates to a CNM lower limit of $\sim$95\% below the predicted value by our \ourmods.

\subsection{\ourmod\ predictions vs ROHSA Gaussian Decomposition}
\label{subsec:vs_rohsa}

The ROHSA Gaussian decomposition technique employs a set of Gaussians to model emission spectral data, taking into account the spatial coherence of neighboring spectra. It separates the \hi\ gaseous phases based on the line widths of these Gaussian components, as line width can serve as a proxy for kinetic temperature (see Section \ref{subsec:vs_fft} above). While ROHSA does not incorporate \hi\ optical depth in its fitting process, it is capable of retrieving more accurate CNM column densities compared to the lower limit provided by the Fourier transform. This is because CNM, UNM, and WNM all contribute to the \hi\ emission brightness, and ROHSA methodology enables to better discern these components.

\begin{figure}
\centering
{\label{}\includegraphics[width=1.05\linewidth]{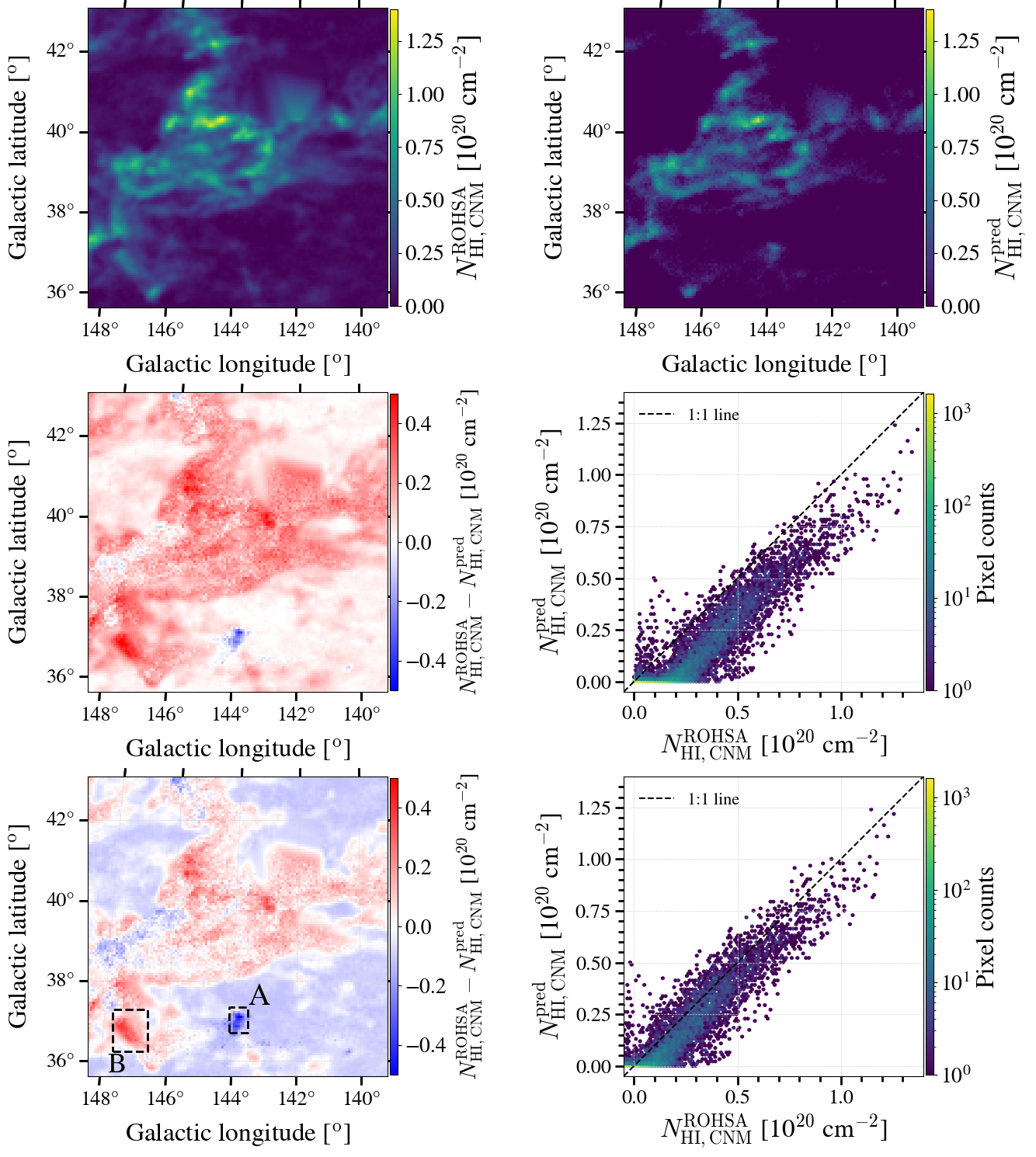}}\\
\caption{Comparison between CNM column densities $N_\mathrm{HI, CNM}$ obtained from \ourmod\ model and ROHSA Gaussian decomposition for the LLIV1 subregion. Top row: ROHSA $N_\mathrm{HI, CNM}$  (left) and predicted $N_\mathrm{HI, CNM}$ maps (right). Middle row: $N_\mathrm{HI, CNM}$ difference between the two methods (left), one-to-one comparison scatter plot (right). Bottom row: Same as the middle row, but an average contribution
of small CNM Gaussian components (amplitudes $T_\mathrm{B} < 2$ K) is subtracted from the ROHSA $N_\mathrm{HI, CNM}$ map (see text). Samples of \hi\ spectra along with their ROHSA decomposed results within the areas A and B (bottom left panel) are shown in Figure \ref{fig:spectra_AB}. Given a distance to the LLIV1 subregion of 0.9$-$1.8 kpc \citep{Wakker2001}, the physical linear sizes of regions A ($\sim$20 pixels) and B ($\sim$10 pixels) are approximately 18$-$36 pc and 9$-$18 pc, respectively.} 
\label{fig:ct_fcnm_rhi_prediction}
\end{figure}

We present, in Figure \ref{fig:ct_fcnm_rhi_prediction}, a comparison of the CNM column densities ($N_\mathrm{HI, CNM}$) predicted from \ourmod\ model with those obtained through ROHSA Gaussian decomposition for the LLIV1 emission cube. The top-left panel illustrates ROHSA CNM column density, while the top-right panel shows the predicted $N_\mathrm{HI, CNM}$ by the model, the middle-left panel depicts their difference, and the middle-right panel presents a scatter plot for a one-to-one comparison. The bottom row mirrors the middle row, with the additional step of subtracting the average contribution of small CNM Gaussian components (amplitudes $T_\mathrm{B} < 2$ K) from the ROHSA $N_\mathrm{HI, CNM}$ map (see below for further clarification).

By accounting for the \hi\ optical depth effect, the CNM column densities derived from \ourmod\ model outputs ($f^\mathrm{pred}_\mathrm{CNM}$, $\mathcal{R}^\mathrm{pred}_\mathrm{HI}$) were computed as:

\begin{equation}
\begin{split}
\frac{N^\mathrm{pred}_\mathrm{HI, CNM}}{[\mathrm{cm^{-2}}]} & = f^\mathrm{pred}_\mathrm{CNM} \times \frac{N_\mathrm{HI}}{[\mathrm{cm^{-2}}]} \\ & = f^\mathrm{pred}_\mathrm{CNM} \times \mathcal{R}^\mathrm{pred}_\mathrm{HI} \times \frac{N^{*}_\mathrm{HI}}{[\mathrm{cm^{-2}}]}
\end{split}
\label{eq:nhicnm_rl}
\end{equation}

\noindent In contrast, the ROHSA CNM column densities were estimated under the optically-thin assumption (see Equation \ref{eq:nhithin} using the total brightness temperature of all (fitted) CNM components. Note that ROHSA identifies cold \hi\ gas components by employing a spectral width threshold (velocity dispersion $\sigma_v \sim 3$ \kms, corresponding to a full width at half maximum $\Delta V_\text{FWHM}$ $\sim 7$ \kms) as a proxy for kinetic temperature ($T_\mathrm{k, max}$ $\sim$ 1000 K).

\begin{figure}
\centering
{\includegraphics[width=0.995\linewidth]{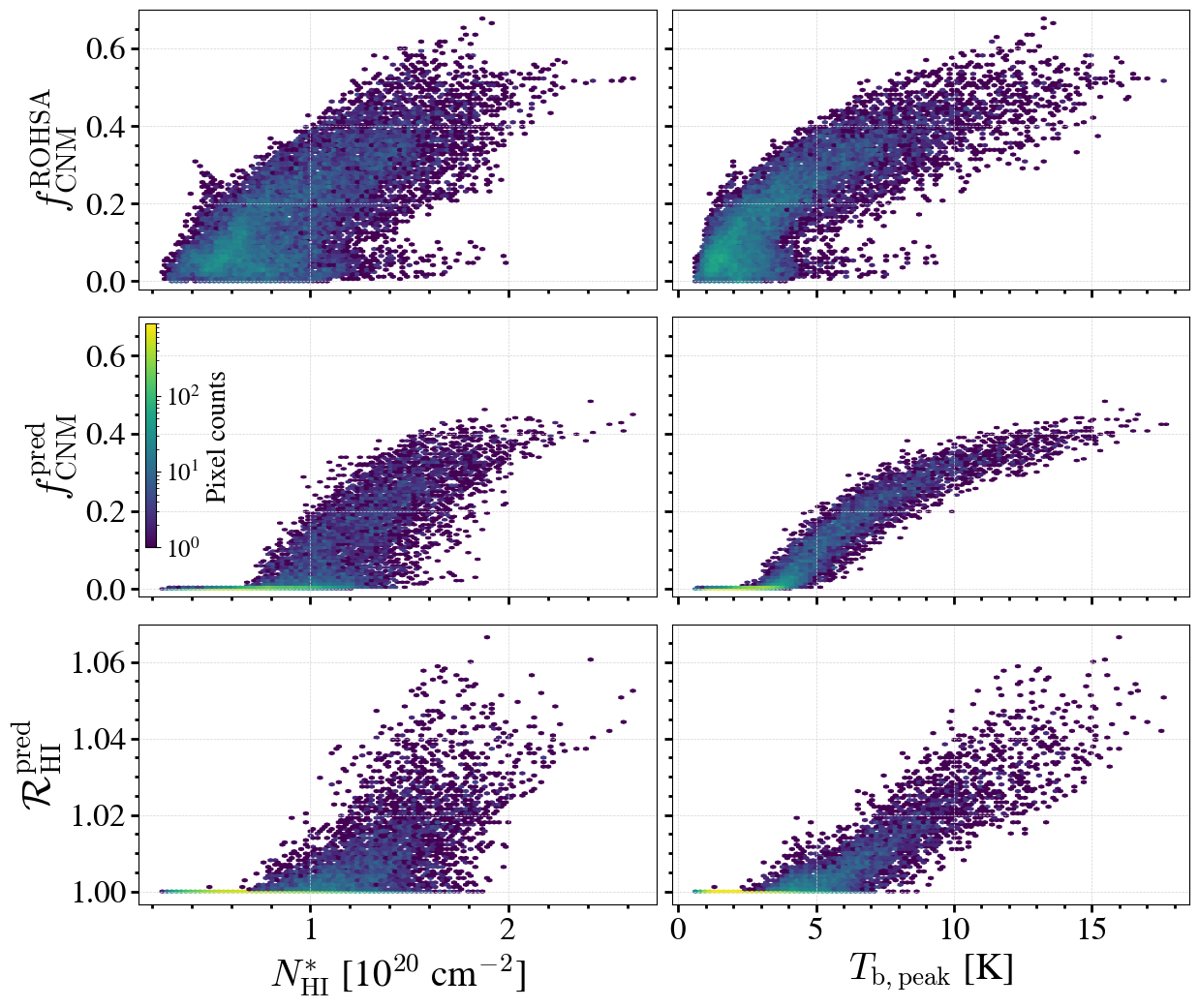}}\\
\caption{The relationships between \FCNM, \RHI\ and \NHIthin\ (left column), \TBpeak\ (right column) in the LLIV1 observed dataset. Top row for \FCNM\ obtained from ROHSA Gaussian decomposition; middle row for \FCNM\ predicted by \ourmod\ model. The y-axis scales in the first two rows are identical for direct visual comparison. Bottom row for opacity correction factor \RHI.} 
\label{fig:fcnm_trends_rohsa_ctrans}
\end{figure}

Figure \ref{fig:fcnm_trends_rohsa_ctrans} (in the first two rows) illustrates the correlation between \FCNM, optically-thin \hi\ column density (left column), and peak brightness temperatures (right column). The top panels display \FCNM\ obtained from ROHSA Gaussian decomposition ($f^\mathrm{ROHSA}_\mathrm{CNM}$ = $N^\mathrm{ROHSA}_\mathrm{HI, CNM} / N^\mathrm{*}_\mathrm{HI}$), the middle panels show \FCNM\ predicted by \ourmod\ model. The \TBpeak\ ranges from 3 -- 18 K, whereas the optically-thin \hi\ column density varies within a narrow range of $\sim$(0.3 -- 2.5) $\times$ 10$^{20}$ cm$^{-2}$. In this optically-thin regime ($N^\mathrm{*}_\mathrm{HI} \lesssim 5 \times 10^{20}$ cm$^{-2}$), both methods demonstrate similar trends for \FCNM: higher \FCNM\ at higher \NHIthin\ and \TBpeak. While \ourmod\ models predict \FCNM\ from 0 to 0.4 without much fluctuation, ROHSA \FCNM\ appears quantitatively higher from 0 to 0.6, with greater scattering.

\begin{figure}
\centering
\includegraphics[width=0.485\textwidth]{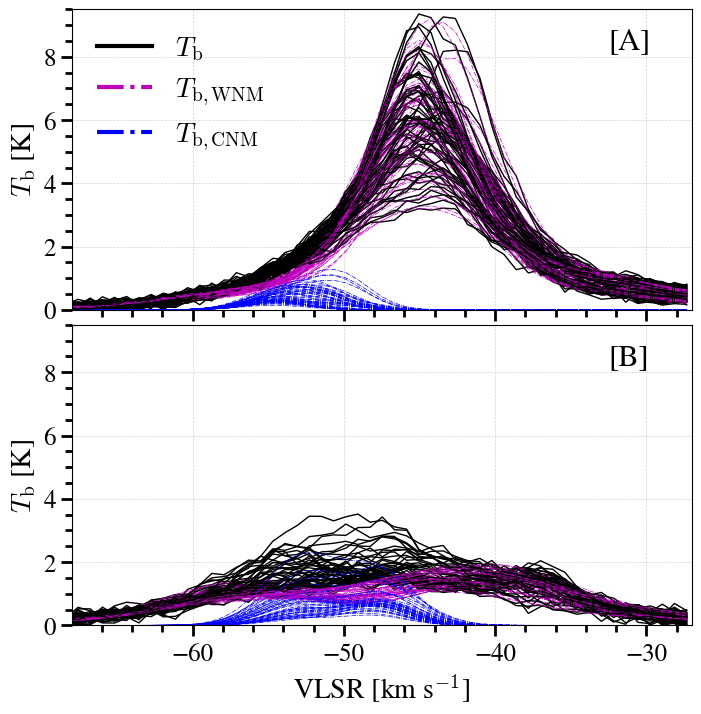}
\caption{Samples of \hi\ spectra and the fitted results from ROHSA Gaussian decomposition towards the areas A (upper panel) and B (lower panel) in Figure \ref{fig:ct_fcnm_rhi_prediction}. Black lines indicate the observed emission spectra and magenta (blue) lines show the total contributions of WNM (CNM) to the emission spectra. The y-axis scales are identical for both panels.}
\label{fig:spectra_AB}
\end{figure}

The prominent disparities observed between the amounts of CNM obtained from the two methods can be attributed to three main factors:

\begin{enumerate}
  \item[1.] Different approaches for estimating CNM column densities (different formulations as given in Equations \ref{eq:nhicnm_rl} and \ref{eq:nhithin}). \ourmod\ model incorporates \hi\ opacity effect, ROHSA uses optically-thin limit on the fitted CNM Gaussian components.
  \item[2.] Distinct CNM kinetic temperature thresholds: \ourmod\ model: $T_{k, \text{CNM}} < 500$ K, ROHSA: $T_{k, \text{CNM}} < 1000$ K ($\sigma_v < 3$ \kms). 
  \item[3.] The most significant distinction, however, lies in the fact that \ourmod\ model predicts negligible CNM fractions ($f_\mathrm{CNM} \sim 0$) to emission spectra with low brightness temperature $T_\mathrm{b, peak} < 2$ K and low column density in optically-thin regime $N^{*}_\mathrm{HI} < 0.7 \times 10^{20}~\mathrm{cm^{-2}}$ (as illustrated in Figure \ref{fig:fcnm_trends_rohsa_ctrans}). On the contrary, ROHSA introduces a considerable number of small narrow CNM Gaussian components (with amplitudes $T_\mathrm{b} < 2$ K, sometimes below the noise level) to fit the emission spectra (see Figure \ref{fig:spectra_AB}, lower panel).

\end{enumerate}

The first factor would lead to a higher CNM fraction from \ourmod\ model compared to ROHSA, the second factor has little effect since most of the CNM Gaussian components have a width of $\sigma_v \lesssim 2$ \kms, associated with a $T_\mathrm{k, max}$ $\sim$ 500 K (see Figure 6 in \citealt{Vujeva2023}), the last factor appears to dominate, leading to a higher CNM column density and, consequently, a higher total CNM fraction. If an average contribution of these small CNM Gaussian components around the noise levels is subtracted from all pixels on the ROHSA $N_\mathrm{HI, CNM}$ map, the CNM column densities from the two methods become more linearly correlated (as indicated in Figure \ref{fig:ct_fcnm_rhi_prediction}, bottom row), and their difference is reduced significantly with a drop in RMSE from 0.138 to 0.088, reflecting an improvement of $\sim$36\%.

There exists a specific small area (box ``A'' in the bottom left panel of Figure \ref{fig:ct_fcnm_rhi_prediction}) where the predicted $N_\mathrm{HI, CNM}$ is greater than ROHSA $N_\mathrm{HI, CNM}$. This difference is due to \ourmod\ model predicting higher \FCNM\ $\sim$ 0.2 for higher values of brightness $T_\mathrm{b}$ ($\sim$8 K) and \hi\ column density $N^{*}_\mathrm{HI}$ ($\sim$1.2 $\times\ 10^{20}~\mathrm{cm^{-2}}$). In these conditions, ROHSA still fits the emission spectra with small, narrow CNM Gaussian components (amplitudes $T_\mathrm{b} < 2$ K), as seen in the upper panel of Figure \ref{fig:spectra_AB}.

Despite the differences revealed by careful comparison above, the cold \hi\ gas column densities generated from the two approaches show a strong linear connection ($R_\mathrm{Pearson} = 0.92$, $p-$value = 0). Intriguingly, even though the \ourmod\ model operates on a spectrum-by-spectrum basis without accounting for spatial coherence as ROHSA does, both methodologies yield comparable 2D CNM structures and morphologies. This suggests that the \ourmod\ model can recognize analogous patterns/features within neighboring \hi\ emission spectra, and subsequently generate spatially consistent output values.

Note that there is no available information about \hi\ optical depth (and no \RHI\ either) from ROHSA Gaussian and Fourier transform methodologies, as these studies do not integrate \hi\ opacity in their analyses, and their results relied instead solely on optically-thin approximations. Direct comparisons with their results are thus not feasible. Nevertheless, we present, in the bottom row of Figure \ref{fig:fcnm_trends_rohsa_ctrans}, the predicted trends of \RHI\ obtained from \ourmod\ model with the observed optically-thin \hi\ column density, peak brightness temperature. The predicted \RHI\ apparently increases with rising \NHIthin, $T_\mathrm{B, peak}$, indicating that at higher \hi\ column density and brightness, one can anticipate higher \RHI.

\subsection{\ourmod\ vs M20 CNN}
\label{subsec:vs_m20cnn}

The development of \ourmod\ has been inspired by the work of M20 CNN. In this Section, we will compare their performances on the observed dataset toward the intermediate velocity LLIV1 region. The application of M20 trained CNN model to the LLIV1 cube produces plain maps of \FCNM\ and \RHI\ with identical values of $f_\mathrm{CNM, M20} \approx 0.025$ and $\mathcal{R}_\mathrm{HI, M20} \approx 1.027$, respectively. These results imply that the CNN model was unable to detect CNM gas in the LLIV1 region.

\cite{Marchal2024} also noted similar results when applying the M20 2-layer CNN model to the HI4PI IVC at high Galactic latitude. The predicted cold gas fraction map is close to 0, exhibiting a structure akin to the lower limit Fourier-transformed $f_\text{CNM,low}$. However, its mean value was notably negative, approximately $-$0.04, which was lower than the background value of 0.01. The authors suggested that this CNN failure is likely attributed to the significant differences between the IVC-dominated spectra and the training set used for M20 CNN. Moreover, they observed that when the IVC spectral peak was shifted to the center of the velocity range, the M20 CNN model could detect CNM gas, yielding an estimated cold gas fraction of $f_\mathrm{CNM} = 0.22$. This improvement occurred partly because the M20 CNN (also our models) was trained on datasets where the \hi\ signal was concentrated in the central velocity channels, with little to no signal at the spectral edges (as discussed in Section \ref{subsec:training_datasets}). This centralization of the training spectra could introduce biases to models without positional encodings, where they are more effective at predicting cold gas over the central spectral channels, but struggle with spectra that deviate from the central channels, such as those observed in IVCs.

\begin{figure}
\centering
\includegraphics[width=0.5\textwidth]{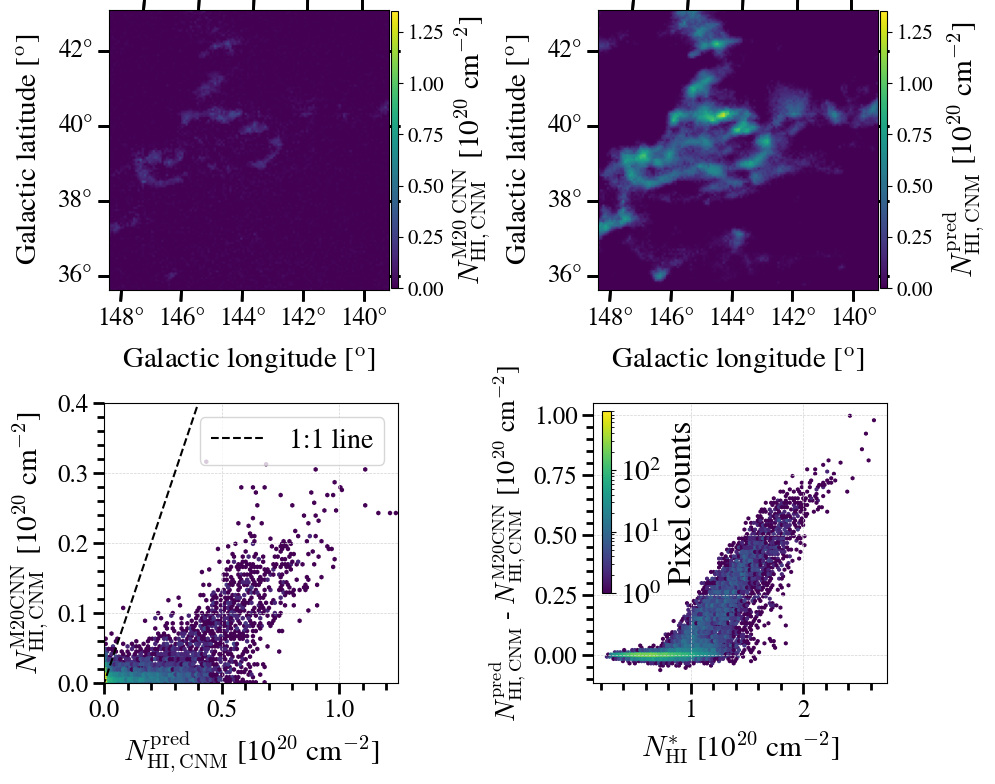}
\caption{Comparison between \ourmod\ predictions and M20 CNN for the LLIV1: CNM column density $N_\mathrm{HI, CNM}$ estimated by \ourmod\ model (top left) and by M20 CNN (top right) both in unit of 10$^{20}$ cm$^{-2}$, along with their one-to-one comparison (bottom left) and the dependence of their absolute difference on the optically-thin column density \NHIthin\ (bottom right).}
\label{fig:m20cnn_prediction_lliv1}
\end{figure}

As in \cite{Marchal2024}, we centralize the spectra of the LLIV1 cube and apply the M20 CNN to the adjusted spectral data. The M20 CNN predicted $f_\mathrm{CNM,M20}$ values ranging from 0 to 0.2, and $\mathcal{R}_\mathrm{HI,M20}$ values between 1.0 to 1.1. Figure \ref{fig:m20cnn_prediction_lliv1} presents a comparison of the CNM column density predictions from both \ourmod\ and the M20 CNN (see Equation \ref{eq:nhicnm_rl}). In the upper-left panel, the M20 CNN predictions are shown, while the upper-right panel displays those from \ourmod. The lower-left panel provides a direct one-to-one comparison between the two models, and the lower-right panel illustrates their absolute differences as a function of optically-thin column density \NHIthin. Although the M20 CNN $N_\mathrm{HI, CNM}$ map somewhat captures the skeletal structure of the LLIV1 cold gas, its values are consistently lower than those predicted by \ourmod\ and ROHSA, as reflected in the lower panels of Figure \ref{fig:m20cnn_prediction_lliv1}. This discrepancy becomes more pronounced at higher \hi\ column densities. Data points where the difference is negligible (lower-right panel) correspond to background noise. Notably, our deep 8-layer CNN models deliver results similar to the M20 CNN. In comparison to \ourmods\ and ROHSA Gaussian fittings, CNN architectures in general are unlikely to effectively depict the overall gas structure of the IVCs.

\begin{figure*}
\centering
{\label{}\includegraphics[width=1.0125\linewidth]{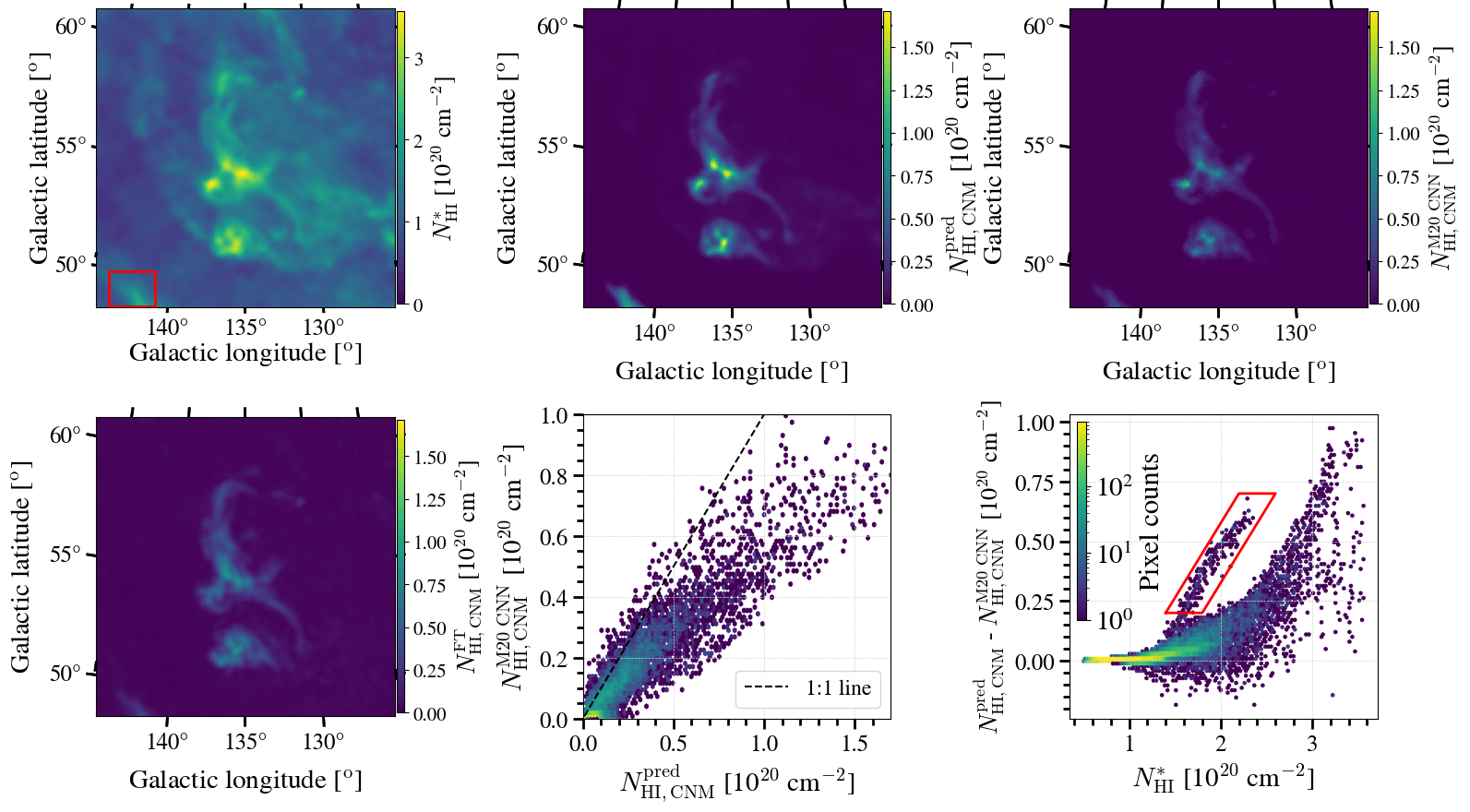}}
\caption{Comparison between \ourmod\ predictions and M20 CNN for the HI4PI IVC at high Galactic latitude, centered at ($l,b$) = ($+$135$^{\circ}$, $+$55$^{\circ}$): CNM column density $N_\mathrm{HI, CNM}$ estimated by \ourmod\ model (top left) and by M20 CNN (top right) both in unit of 10$^{20}$ cm$^{-2}$, along with their one-to-one comparison (bottom left) and the dependence of their absolute difference on the optically-thin column density \NHIthin\ (bottom right). The pixels in the red box on the map (shown in the upper-left panel) are linked to the data points in the bottom-right panel's red box. They are not a part of the IVC since their associated spectra peak at local velocity $v$ $\approx$ $-10$ \kms, whereas the IVC spectra peak at $v \approx -50$ km s$^{-1}$.} 
\label{fig:m20cnn_prediction}
\end{figure*}

Towards this LLIV1 intermediate velocity region, the hybrid \ourmod\ model, which integrates sinusoidal positional encoding, demonstrated improved performance. Namely, it produces predictions comparable to Gaussian fittings and (as expected) significantly higher than the Fourier-transform lower limits. The positional encoding in \ourmods\ likely enhanced its ability to account for spectral variations. This additional information thus appears to mitigate the challenges posed by the IVC spectral cube and helps \ourmod\ to identify the presence of optically-thick cold gas without requiring spectral centralization.

Finally, we apply \ourmod\ to the HI4PI IVC emission cube centered at ($l,b$) = ($+$135$^{\circ}$, $+$55$^{\circ}$) at high Galactic latitude. We also compare its performance with the M20 CNN model, as illustrated in Figure \ref{fig:m20cnn_prediction}. The upper-left panel displays the optically-thin column density, while the upper-middle panel shows the predictions from \ourmod. \citet{Marchal2024} provided the Fourier-transformed lower limit CNM fraction. Following their analysis, we centralize the $T_\mathrm{b}$ peaks in the IVC spectral cube before feeding them into the M20 CNN to predict \FCNM\ and \RHI. We compute the lower limit and M20 CNN-predicted cold \hi\ column densities for the IVC, as shown in the lower-left and upper-right panels, respectively. The lower-middle panel in Figure \ref{fig:m20cnn_prediction} provides a direct one-to-one comparison between the two models, and the lower-right panel illustrates their absolute differences as a function of optically-thin column density \NHIthin. The data points within the red box in the bottom-right panel correspond to the pixels inside the red box on the map (upper-left panel). These points deviate significantly from the main trend between the absolute difference and \NHIthin. They are not part of the IVC, as their spectra peak at a local velocity of $v \approx -10$ km s$^{-1}$, whereas the IVC spectra peak at $v \approx -50$ km s$^{-1}$.

With spectral centralization, the M20 CNN was able to detect CNM in the HI4PI IVC and bring its CNM predictions closer to those of \ourmod, and fairly comparable to the outcomes of the Fourier transforms. In this bright IVC region, the difference in cold \hi\ column densities between \ourmod\ and M20 CNN predictions ranges from 13\% to 80\%, with a mean of 44\%. The difference between \ourmod\ and Fourier transform $N_\mathrm{HI, CNM}$ spans from 10\% to 95\%, with a mean of 50\%. Similar to the LLIV1 case, the CNM amount obtained by \ourmod\ is higher than both Fourier transform and CNN estimates and the discrepancies increase with the rising optically-thin column density \NHIthin.

\section{Conclusions and future work}
\label{sec:conclusions}
    In continuation of \cite{Murray2020} work of CNN models, we introduced a novel deep-learning approach, \ourmod\ models, for \hi\ spectral analysis. These models combine CNN and Transformer architectures and have been trained on synthetic datasets generated from simulations to predict cold \hi\ gas fraction (\FCNM) and \hi\ opacity correction factor (\RHI) from emission spectra. Recognizing the importance of spectral channel order in training the deep-learning models, we experiment with different positional encoding methods. To evaluate the effectiveness of our models, we conduct a comparative analysis between CNN and \ourmod\ architectures (with positional encodings), based on testing RMSE values. Subsequently, we apply these trained models to the evaluation and observed spectral data.

In summary, our key findings are as follows:

-- \ourmod\ outperforms CNNs on the synthetic evaluation set, with an average improvement of 10\% in testing accuracy, convergence speed, and model training stability.

-- \ourmod, trained on synthetic data, generalizes to provide meaningful predictions on observed emission data from both emission and absorption surveys.

-- In comparison with \FCNM\ and \RHI\ inferred from absorption surveys towards high Galactic latitudes and random sightlines, \ourmod\ predictions on emission spectra around continuum radio sources are comparable to the absorption-based values in optically-thin regime of the local Solar neighborhood (\NHIthin\ $<$ 5 \nhiUnit), but deviate significantly in regions closer to the Galactic plane with higher \NHIthin. While the predicted values follow well the relationships (e.g., \FCNM, \RHI\ vs \NHIthin, \TBpeak and equivalent width) learned from the synthetic training database across the entire ranges, absorption-based estimates agree with the trends only at low \NHIthin\ (and low $\tau_\text{HI}$).

-- Applying \ourmod\ model to the observed emission data toward the intermediate velocity regions reveals a good linear correlation between predicted \FCNM\ and \RHI\ with those obtained by ROHSA Gaussian decomposition, despite different approaches for estimating CNM gas amounts. In general, \ourmod\ appears to deliver better performance on observed data sets compared to the CNN architectures.

-- These findings highlight the potential of representation learning architectures in \hi\ spectral analysis. We propose here the \ourmod\ (a hybrid convolutional-Transformer model) and ``\textit{add sinusoidal}'' positional encoding (adding sinusoidal function to the original spectra) for predicting the CNM gas fraction and \hi\ opacity effect as it is the most robust against various perturbations we expect to see in real data (shuffling schemes, convolutional neural network initialization, spectral input length).

-- Using synthetic datasets from diverse simulations enhances our models' generalization ability. Application of the trained models to a synthetic evaluation spectral cube demonstrates \ourmod\ capability to consistently reproduce \FCNM\ and \RHI\ compared to ground truths. This consistency extends to correlations between predicted quantities and \NHI, \TBpeak, equivalent width, and \taupeak. However, to address our concern that the simulations we are using produce an excess of cold optically-thick \hi\ due to the lack of an atomic-molecular transition, we plan to incorporate more recent MHD numerical studies \citep[e.g.,][]{Kim2013,KimOstriker2017,Hu2023,Vijayan2023} that include the phase transition.

In the future, we plan to explore convolutional embedding to improve model performance. This involves tokenizing the spectral input (splitting the input into several segments) and utilizing a convolutional layer to extract features from each token. Our primary focus in the current work lies in predicting \FCNM\ and \RHI\ from individual spectra. Nonetheless, it is important to note that within both synthetic and observed data, the \hi\ emission spectra possess spatial coherence with their neighboring pixels. Future efforts will address the spatial coherence in model training with the use of the Ising model (e.g. Markov Random Field modeling).

In addition to \hi\ spectral line analysis, \ourmods\ can be extended to analyze other molecular lines, such as CO. For CO molecules specifically, we can leverage publicly available data from the MHD TIGRESS simulation framework \citep{KimOstriker2017}, which provides a thorough modeling of three-phase ISM in galaxies. Post-processing TIGRESS simulation data offers the chemistry and temperature of the gas. We then feed these as inputs to the publicly available radiation transfer code RADMC-3D \citep{Dullemond2012} to generate synthetic observations of the CO (1–0) and CO (2–1) line emission \citep[e.g.,][]{Gong2018,Gong2020}. Through this process, we build a synthetic database of CO molecular lines, which can be used to train \ourmods\ to predict molecular gas properties, such as gas density and the CO-to-\h2\ conversion factor (\xco), across various galactic environments.


\section*{Acknowledgements}

This research was partially funded by the Australian Government through an Australian Research Council Australian Laureate Fellowship (project number FL210100039) to NMc-G.

We express our gratitude to A/Prof. Yuan-Sen Ting from the Research School of Astronomy \& Astrophysics, Australian National University for the fruitful discussions regarding the applicability of CNN and Transformer models. We also appreciate his generous offering of supercomputing resources to support our model training processes. Additionally, we acknowledge Dr. Ioana Ciuc\v{a} from the Research School of Astronomy \& Astrophysics, Australian National University, for the insightful discussions on the applications of machine learning techniques in the analysis of radio spectral data. Finally, we thank the anonymous referee for the comments and suggestions that allowed us to improve the quality of our manuscript.

\textit{Software}: Astropy \citep{Astropy2018A}, PyTorch \citep{Pytorch2019}, Matplotlib \citep{MatplotlibHunter2007}, NumPy \citep{vanderWalt2011}, SciPy \citep{Virtanen2020}, Pandas \citep{mckinney2010data}.

\section*{Data Availability}
The model designs, database used in this study, as well as data analysis notebooks, are publicly available at \href{https://doi.org/10.5281/zenodo.14183481}{\textit{DOI 10.5281}}.



\bibliographystyle{mnras}
\bibliography{references} 

\begin{thebibliography}{}
\makeatletter
\relax
\def\mn@urlcharsother{\let\do\@makeother \do\$\do\&\do\#\do\^\do\_\do\%\do\~}
\def\mn@doi{\begingroup\mn@urlcharsother \@ifnextchar [ {\mn@doi@} {\mn@doi@[]}}
\def\mn@doi@[#1]#2{\def\@tempa{#1}\ifx\@tempa\@empty \href {http://dx.doi.org/#2} {doi:#2}\else \href {http://dx.doi.org/#2} {#1}\fi \endgroup}
\def\mn@eprint#1#2{\mn@eprint@#1:#2::\@nil}
\def\mn@eprint@arXiv#1{\href {http://arxiv.org/abs/#1} {{\tt arXiv:#1}}}
\def\mn@eprint@dblp#1{\href {http://dblp.uni-trier.de/rec/bibtex/#1.xml} {dblp:#1}}
\def\mn@eprint@#1:#2:#3:#4\@nil{\def\@tempa {#1}\def\@tempb {#2}\def\@tempc {#3}\ifx \@tempc \@empty \let \@tempc \@tempb \let \@tempb \@tempa \fi \ifx \@tempb \@empty \def\@tempb {arXiv}\fi \@ifundefined {mn@eprint@\@tempb}{\@tempb:\@tempc}{\expandafter \expandafter \csname mn@eprint@\@tempb\endcsname \expandafter{\@tempc}}}

\bibitem[\protect\citeauthoryear{Abdel-Hamid, Mohamed, Jiang, Deng, Penn  \& Yu}{Abdel-Hamid et~al.}{2014}]{AbdelHamid2014}
Abdel-Hamid O.,  Mohamed A.-r.,  Jiang H.,  Deng L.,  Penn G.,   Yu D.,  2014, \mn@doi [IEEE/ACM Transactions on Audio, Speech, and Language Processing] {10.1109/TASLP.2014.2339736}, 22, 1533

\bibitem[\protect\citeauthoryear{{Astropy Collaboration} et~al.,}{{Astropy Collaboration} et~al.}{2018}]{Astropy2018A}
{Astropy Collaboration} et~al., 2018, \mn@doi [\aj] {10.3847/1538-3881/aabc4f}, \href {https://ui.adsabs.harvard.edu/abs/2018AJ....156..123A} {156, 123}

\bibitem[\protect\citeauthoryear{{Audit} \& {Hennebelle}}{{Audit} \& {Hennebelle}}{2005}]{Audit2005}
{Audit} E.,  {Hennebelle} P.,  2005, \mn@doi [\aap] {10.1051/0004-6361:20041474}, \href {http://adsabs.harvard.edu/abs/2005A%26A...433....1A} {433, 1}

\bibitem[\protect\citeauthoryear{{Audit} \& {Hennebelle}}{{Audit} \& {Hennebelle}}{2010}]{Audit2010}
{Audit} E.,  {Hennebelle} P.,  2010, \mn@doi [\aap] {10.1051/0004-6361/200912695}, \href {https://ui.adsabs.harvard.edu/abs/2010A&A...511A..76A} {511, A76}

\bibitem[\protect\citeauthoryear{{Bello}, {Zoph}, {Vaswani}, {Shlens}  \& {Le}}{{Bello} et~al.}{2019}]{Bello2019}
{Bello} I.,  {Zoph} B.,  {Vaswani} A.,  {Shlens} J.,   {Le} Q.~V.,  2019, \mn@doi [arXiv e-prints] {10.48550/arXiv.1904.09925}, \href {https://ui.adsabs.harvard.edu/abs/2019arXiv190409925B} {p. arXiv:1904.09925}

\bibitem[\protect\citeauthoryear{{Bengio}, {Courville}  \& {Vincent}}{{Bengio} et~al.}{2012}]{Bengio2012}
{Bengio} Y.,  {Courville} A.,   {Vincent} P.,  2012, \mn@doi [arXiv e-prints] {10.48550/arXiv.1206.5538}, \href {https://ui.adsabs.harvard.edu/abs/2012arXiv1206.5538B} {p. arXiv:1206.5538}

\bibitem[\protect\citeauthoryear{{Bubeck} et~al.,}{{Bubeck} et~al.}{2023}]{Bubeck2023}
{Bubeck} S.,  et~al., 2023, \mn@doi [arXiv e-prints] {10.48550/arXiv.2303.12712}, \href {https://ui.adsabs.harvard.edu/abs/2023arXiv230312712B} {p. arXiv:2303.12712}

\bibitem[\protect\citeauthoryear{{Carion}, {Massa}, {Synnaeve}, {Usunier}, {Kirillov}  \& {Zagoruyko}}{{Carion} et~al.}{2020}]{Carion2020}
{Carion} N.,  {Massa} F.,  {Synnaeve} G.,  {Usunier} N.,  {Kirillov} A.,   {Zagoruyko} S.,  2020, \mn@doi [arXiv e-prints] {10.48550/arXiv.2005.12872}, \href {https://ui.adsabs.harvard.edu/abs/2020arXiv200512872C} {p. arXiv:2005.12872}

\bibitem[\protect\citeauthoryear{{Caron}, {Misra}, {Mairal}, {Goyal}, {Bojanowski}  \& {Joulin}}{{Caron} et~al.}{2020}]{Caron2020}
{Caron} M.,  {Misra} I.,  {Mairal} J.,  {Goyal} P.,  {Bojanowski} P.,   {Joulin} A.,  2020, \mn@doi [arXiv e-prints] {10.48550/arXiv.2006.09882}, \href {https://ui.adsabs.harvard.edu/abs/2020arXiv200609882C} {p. arXiv:2006.09882}

\bibitem[\protect\citeauthoryear{{Crain} et~al.,}{{Crain} et~al.}{2015}]{Crain2015}
{Crain} R.~A.,  et~al., 2015, \mn@doi [\mnras] {10.1093/mnras/stv725}, \href {https://ui.adsabs.harvard.edu/abs/2015MNRAS.450.1937C} {450, 1937}

\bibitem[\protect\citeauthoryear{{Dav{\'e}}, {Angl{\'e}s-Alc{\'a}zar}, {Narayanan}, {Li}, {Rafieferantsoa}  \& {Appleby}}{{Dav{\'e}} et~al.}{2019}]{Dave2019}
{Dav{\'e}} R.,  {Angl{\'e}s-Alc{\'a}zar} D.,  {Narayanan} D.,  {Li} Q.,  {Rafieferantsoa} M.~H.,   {Appleby} S.,  2019, \mn@doi [\mnras] {10.1093/mnras/stz937}, \href {https://ui.adsabs.harvard.edu/abs/2019MNRAS.486.2827D} {486, 2827}

\bibitem[\protect\citeauthoryear{{Dehghani}, {Gouws}, {Vinyals}, {Uszkoreit}  \& {Kaiser}}{{Dehghani} et~al.}{2018}]{Dehghani2018}
{Dehghani} M.,  {Gouws} S.,  {Vinyals} O.,  {Uszkoreit} J.,   {Kaiser} {\L}.,  2018, \mn@doi [arXiv e-prints] {10.48550/arXiv.1807.03819}, \href {https://ui.adsabs.harvard.edu/abs/2018arXiv180703819D} {p. arXiv:1807.03819}

\bibitem[\protect\citeauthoryear{{D{\'e}nes}, {McClure-Griffiths}, {Dickey}, {Dawson}  \& {Murray}}{{D{\'e}nes} et~al.}{2018}]{Denes2018}
{D{\'e}nes} H.,  {McClure-Griffiths} N.~M.,  {Dickey} J.~M.,  {Dawson} J.~R.,   {Murray} C.~E.,  2018, \mn@doi [\mnras] {10.1093/mnras/sty1384}, \href {http://adsabs.harvard.edu/abs/2018MNRAS.479.1465D} {479, 1465}

\bibitem[\protect\citeauthoryear{{Devlin}, {Chang}, {Lee}  \& {Toutanova}}{{Devlin} et~al.}{2018}]{Devlin2018}
{Devlin} J.,  {Chang} M.-W.,  {Lee} K.,   {Toutanova} K.,  2018, \mn@doi [arXiv e-prints] {10.48550/arXiv.1810.04805}, \href {https://ui.adsabs.harvard.edu/abs/2018arXiv181004805D} {p. arXiv:1810.04805}

\bibitem[\protect\citeauthoryear{Dickey \& Lockman}{Dickey \& Lockman}{1990}]{Dickey1990}
Dickey J.~M.,  Lockman F.~J.,  1990, \mn@doi [Annual Review of Astronomy and Astrophysics] {10.1146/annurev.aa.28.090190.001243}, 28, 215

\bibitem[\protect\citeauthoryear{{Dickey}, {Salpeter}  \& {Terzian}}{{Dickey} et~al.}{1978}]{Dickey1978}
{Dickey} J.~M.,  {Salpeter} E.~E.,   {Terzian} Y.,  1978, \mn@doi [\apjs] {10.1086/190492}, \href {https://ui.adsabs.harvard.edu/abs/1978ApJS...36...77D} {36, 77}

\bibitem[\protect\citeauthoryear{{Dickey}, {McClure-Griffiths}, {Gaensler}  \& {Green}}{{Dickey} et~al.}{2003}]{Dickey2003}
{Dickey} J.~M.,  {McClure-Griffiths} N.~M.,  {Gaensler} B.~M.,   {Green} A.~J.,  2003, \mn@doi [\apj] {10.1086/346081}, \href {http://adsabs.harvard.edu/abs/2003ApJ...585..801D} {585, 801}

\bibitem[\protect\citeauthoryear{{Dickey}, {Strasser}, {Gaensler}, {Haverkorn}, {Kavars}, {McClure-Griffiths}, {Stil}  \& {Taylor}}{{Dickey} et~al.}{2009}]{Dickey2009}
{Dickey} J.~M.,  {Strasser} S.,  {Gaensler} B.~M.,  {Haverkorn} M.,  {Kavars} D.,  {McClure-Griffiths} N.~M.,  {Stil} J.,   {Taylor} A.~R.,  2009, \mn@doi [\apj] {10.1088/0004-637X/693/2/1250}, \href {http://adsabs.harvard.edu/abs/2009ApJ...693.1250D} {693, 1250}

\bibitem[\protect\citeauthoryear{{Dosovitskiy} et~al.,}{{Dosovitskiy} et~al.}{2020}]{Dosovitskiy2020}
{Dosovitskiy} A.,  et~al., 2020, \mn@doi [arXiv e-prints] {10.48550/arXiv.2010.11929}, \href {https://ui.adsabs.harvard.edu/abs/2020arXiv201011929D} {p. arXiv:2010.11929}

\bibitem[\protect\citeauthoryear{{Dufter}, {Schmitt}  \& {Sch{\"u}tze}}{{Dufter} et~al.}{2021}]{Dufter2021}
{Dufter} P.,  {Schmitt} M.,   {Sch{\"u}tze} H.,  2021, \mn@doi [arXiv e-prints] {10.48550/arXiv.2102.11090}, \href {https://ui.adsabs.harvard.edu/abs/2021arXiv210211090D} {p. arXiv:2102.11090}

\bibitem[\protect\citeauthoryear{{Dullemond}, {Juhasz}, {Pohl}, {Sereshti}, {Shetty}, {Peters}, {Commercon}  \& {Flock}}{{Dullemond} et~al.}{2012}]{Dullemond2012}
{Dullemond} C.~P.,  {Juhasz} A.,  {Pohl} A.,  {Sereshti} F.,  {Shetty} R.,  {Peters} T.,  {Commercon} B.,   {Flock} M.,  2012, {RADMC-3D: A multi-purpose radiative transfer tool}, Astrophysics Source Code Library, record ascl:1202.015

\bibitem[\protect\citeauthoryear{{Field}}{{Field}}{1965}]{Field1965}
{Field} G.~B.,  1965, \mn@doi [\apj] {10.1086/148317}, \href {https://ui.adsabs.harvard.edu/abs/1965ApJ...142..531F} {142, 531}

\bibitem[\protect\citeauthoryear{{Field}, {Goldsmith}  \& {Habing}}{{Field} et~al.}{1969}]{Field1969}
{Field} G.~B.,  {Goldsmith} D.~W.,   {Habing} H.~J.,  1969, \mn@doi [\apjl] {10.1086/180324}, \href {http://adsabs.harvard.edu/abs/1969ApJ...155L.149F} {155, L149}

\bibitem[\protect\citeauthoryear{Fukushima}{Fukushima}{1980}]{fukushima1980}
Fukushima K.,  1980, Biological Cybernetics, 36, 193

\bibitem[\protect\citeauthoryear{{Gazol}, {V{\'a}zquez-Semadeni}, {S{\'a}nchez-Salcedo}  \& {Scalo}}{{Gazol} et~al.}{2001}]{Gazol2001}
{Gazol} A.,  {V{\'a}zquez-Semadeni} E.,  {S{\'a}nchez-Salcedo} F.~J.,   {Scalo} J.,  2001, \mn@doi [\apjl] {10.1086/322873}, \href {http://adsabs.harvard.edu/abs/2001ApJ...557L.121G} {557, L121}

\bibitem[\protect\citeauthoryear{{Gehring}, {Auli}, {Grangier}, {Yarats}  \& {Dauphin}}{{Gehring} et~al.}{2017}]{Gehring2017}
{Gehring} J.,  {Auli} M.,  {Grangier} D.,  {Yarats} D.,   {Dauphin} Y.~N.,  2017, \mn@doi [arXiv e-prints] {10.48550/arXiv.1705.03122}, \href {https://ui.adsabs.harvard.edu/abs/2017arXiv170503122G} {p. arXiv:1705.03122}

\bibitem[\protect\citeauthoryear{{Genel} et~al.,}{{Genel} et~al.}{2014}]{Genel2014}
{Genel} S.,  et~al., 2014, \mn@doi [\mnras] {10.1093/mnras/stu1654}, \href {https://ui.adsabs.harvard.edu/abs/2014MNRAS.445..175G} {445, 175}

\bibitem[\protect\citeauthoryear{{Gong}, {Ostriker}  \& {Kim}}{{Gong} et~al.}{2018}]{Gong2018}
{Gong} M.,  {Ostriker} E.~C.,   {Kim} C.-G.,  2018, \mn@doi [\apj] {10.3847/1538-4357/aab9af}, \href {https://ui.adsabs.harvard.edu/abs/2018ApJ...858...16G} {858, 16}

\bibitem[\protect\citeauthoryear{{Gong}, {Ostriker}, {Kim}  \& {Kim}}{{Gong} et~al.}{2020}]{Gong2020}
{Gong} M.,  {Ostriker} E.~C.,  {Kim} C.-G.,   {Kim} J.-G.,  2020, \mn@doi [\apj] {10.3847/1538-4357/abbdab}, \href {https://ui.adsabs.harvard.edu/abs/2020ApJ...903..142G} {903, 142}

\bibitem[\protect\citeauthoryear{Goodfellow, Bengio  \& Courville}{Goodfellow et~al.}{2016}]{Goodfellow2016}
Goodfellow I.,  Bengio Y.,   Courville A.,  2016, Deep Learning.
MIT Press

\bibitem[\protect\citeauthoryear{{Gulati} et~al.,}{{Gulati} et~al.}{2020}]{Gulati2020}
{Gulati} A.,  et~al., 2020, \mn@doi [arXiv e-prints] {10.48550/arXiv.2005.08100}, \href {https://ui.adsabs.harvard.edu/abs/2020arXiv200508100G} {p. arXiv:2005.08100}

\bibitem[\protect\citeauthoryear{{HI4PI Collaboration} et~al.,}{{HI4PI Collaboration} et~al.}{2016}]{hi4pi2016}
{HI4PI Collaboration} et~al., 2016, \mn@doi [\aap] {10.1051/0004-6361/201629178}, \href {http://adsabs.harvard.edu/abs/2016A%26A...594A.116H} {594, A116}

\bibitem[\protect\citeauthoryear{{Han} et~al.,}{{Han} et~al.}{2020}]{Han2020}
{Han} W.,  et~al., 2020, \mn@doi [arXiv e-prints] {10.48550/arXiv.2005.03191}, \href {https://ui.adsabs.harvard.edu/abs/2020arXiv200503191H} {p. arXiv:2005.03191}

\bibitem[\protect\citeauthoryear{{Haud} \& {Kalberla}}{{Haud} \& {Kalberla}}{2007}]{Haud2007}
{Haud} U.,  {Kalberla} P.~M.~W.,  2007, \mn@doi [\aap] {10.1051/0004-6361:20065796}, \href {https://ui.adsabs.harvard.edu/abs/2007A&A...466..555H} {466, 555}

\bibitem[\protect\citeauthoryear{{Heiles} \& {Troland}}{{Heiles} \& {Troland}}{2003a}]{Heiles2003a}
{Heiles} C.,  {Troland} T.~H.,  2003a, \mn@doi [\apjs] {10.1086/367785}, \href {http://adsabs.harvard.edu/abs/2003ApJS..145..329H} {145, 329}

\bibitem[\protect\citeauthoryear{{Heiles} \& {Troland}}{{Heiles} \& {Troland}}{2003b}]{Heiles2003b}
{Heiles} C.,  {Troland} T.~H.,  2003b, \mn@doi [\apj] {10.1086/367828}, \href {http://adsabs.harvard.edu/abs/2003ApJ...586.1067H} {586, 1067}

\bibitem[\protect\citeauthoryear{{Hennebelle} \& {P{\'e}rault}}{{Hennebelle} \& {P{\'e}rault}}{2000}]{Hennebelle2000}
{Hennebelle} P.,  {P{\'e}rault} M.,  2000, \aap, \href {https://ui.adsabs.harvard.edu/abs/2000A&A...359.1124H} {359, 1124}

\bibitem[\protect\citeauthoryear{{Hensley}, {Murray}  \& {Dodici}}{{Hensley} et~al.}{2022}]{Hensley2022}
{Hensley} B.~S.,  {Murray} C.~E.,   {Dodici} M.,  2022, \mn@doi [\apj] {10.3847/1538-4357/ac5cbd}, \href {https://ui.adsabs.harvard.edu/abs/2022ApJ...929...23H} {929, 23}

\bibitem[\protect\citeauthoryear{{Hill}, {Mac Low}, {Gatto}  \& {Ib{\'a}{\~n}ez-Mej{\'{\i}}a}}{{Hill} et~al.}{2018}]{Hill2018}
{Hill} A.~S.,  {Mac Low} M.-M.,  {Gatto} A.,   {Ib{\'a}{\~n}ez-Mej{\'{\i}}a} J.~C.,  2018, \mn@doi [\apj] {10.3847/1538-4357/aacce2}, \href {http://adsabs.harvard.edu/abs/2018ApJ...862...55H} {862, 55}

\bibitem[\protect\citeauthoryear{{Hu}, {Wibking}  \& {Krumholz}}{{Hu} et~al.}{2023}]{Hu2023}
{Hu} Z.,  {Wibking} B.~D.,   {Krumholz} M.~R.,  2023, \mn@doi [\mnras] {10.1093/mnras/stad931}, \href {https://ui.adsabs.harvard.edu/abs/2023MNRAS.521.5604H} {521, 5604}

\bibitem[\protect\citeauthoryear{Hunter}{Hunter}{2007}]{MatplotlibHunter2007}
Hunter J.~D.,  2007, \mn@doi [Computing in Science \& Engineering] {10.1109/MCSE.2007.55}, 9, 90

\bibitem[\protect\citeauthoryear{{Kalberla} \& {Haud}}{{Kalberla} \& {Haud}}{2018}]{Kalberla2018}
{Kalberla} P.~M.~W.,  {Haud} U.,  2018, \mn@doi [\aap] {10.1051/0004-6361/201833146}, \href {http://adsabs.harvard.edu/abs/2018A%26A...619A..58K} {619, A58}

\bibitem[\protect\citeauthoryear{{Ke}, {He}  \& {Liu}}{{Ke} et~al.}{2020}]{Ke2020}
{Ke} G.,  {He} D.,   {Liu} T.-Y.,  2020, \mn@doi [arXiv e-prints] {10.48550/arXiv.2006.15595}, \href {https://ui.adsabs.harvard.edu/abs/2020arXiv200615595K} {p. arXiv:2006.15595}

\bibitem[\protect\citeauthoryear{{Kerp}, {Winkel}, {Ben Bekhti}, {Fl{\"o}er}  \& {Kalberla}}{{Kerp} et~al.}{2011}]{Kerp2011}
{Kerp} J.,  {Winkel} B.,  {Ben Bekhti} N.,  {Fl{\"o}er} L.,   {Kalberla} P.~M.~W.,  2011, \mn@doi [Astronomische Nachrichten] {10.1002/asna.201011548}, \href {https://ui.adsabs.harvard.edu/abs/2011AN....332..637K} {332, 637}

\bibitem[\protect\citeauthoryear{{Khan}, {Naseer}, {Hayat}, {Waqas Zamir}, {Shahbaz Khan}  \& {Shah}}{{Khan} et~al.}{2021}]{Khan2021}
{Khan} S.,  {Naseer} M.,  {Hayat} M.,  {Waqas Zamir} S.,  {Shahbaz Khan} F.,   {Shah} M.,  2021, \mn@doi [arXiv e-prints] {10.48550/arXiv.2101.01169}, \href {https://ui.adsabs.harvard.edu/abs/2021arXiv210101169K} {p. arXiv:2101.01169}

\bibitem[\protect\citeauthoryear{{Kim} \& {Ostriker}}{{Kim} \& {Ostriker}}{2017}]{KimOstriker2017}
{Kim} C.-G.,  {Ostriker} E.~C.,  2017, \mn@doi [\apj] {10.3847/1538-4357/aa8599}, \href {https://ui.adsabs.harvard.edu/abs/2017ApJ...846..133K} {846, 133}

\bibitem[\protect\citeauthoryear{{Kim}, {Ostriker}  \& {Kim}}{{Kim} et~al.}{2013}]{Kim2013}
{Kim} C.-G.,  {Ostriker} E.~C.,   {Kim} W.-T.,  2013, \mn@doi [\apj] {10.1088/0004-637X/776/1/1}, \href {http://adsabs.harvard.edu/abs/2013ApJ...776....1K} {776, 1}

\bibitem[\protect\citeauthoryear{{Kim}, {Kim}  \& {Ostriker}}{{Kim} et~al.}{2020}]{KimWT2020}
{Kim} W.-T.,  {Kim} C.-G.,   {Ostriker} E.~C.,  2020, \mn@doi [\apj] {10.3847/1538-4357/ab9b87}, \href {https://ui.adsabs.harvard.edu/abs/2020ApJ...898...35K} {898, 35}

\bibitem[\protect\citeauthoryear{{Kim}, {Kim}, {Gong}  \& {Ostriker}}{{Kim} et~al.}{2023}]{Kim2023}
{Kim} C.-G.,  {Kim} J.-G.,  {Gong} M.,   {Ostriker} E.~C.,  2023, \mn@doi [\apj] {10.3847/1538-4357/acbd3a}, \href {https://ui.adsabs.harvard.edu/abs/2023ApJ...946....3K} {946, 3}

\bibitem[\protect\citeauthoryear{{Kitaev}, {Kaiser}  \& {Levskaya}}{{Kitaev} et~al.}{2020}]{Kitaev2020}
{Kitaev} N.,  {Kaiser} {\L}.,   {Levskaya} A.,  2020, \mn@doi [arXiv e-prints] {10.48550/arXiv.2001.04451}, \href {https://ui.adsabs.harvard.edu/abs/2020arXiv200104451K} {p. arXiv:2001.04451}

\bibitem[\protect\citeauthoryear{{Kriman} et~al.,}{{Kriman} et~al.}{2019}]{Kriman2019}
{Kriman} S.,  et~al., 2019, \mn@doi [arXiv e-prints] {10.48550/arXiv.1910.10261}, \href {https://ui.adsabs.harvard.edu/abs/2019arXiv191010261K} {p. arXiv:1910.10261}

\bibitem[\protect\citeauthoryear{{Krumholz}, {McKee}  \& {Tumlinson}}{{Krumholz} et~al.}{2009}]{Krumholz2009}
{Krumholz} M.~R.,  {McKee} C.~F.,   {Tumlinson} J.,  2009, \mn@doi [\apj] {10.1088/0004-637X/693/1/216}, \href {http://adsabs.harvard.edu/abs/2009ApJ...693..216K} {693, 216}

\bibitem[\protect\citeauthoryear{{Kulkarni} \& {Heiles}}{{Kulkarni} \& {Heiles}}{1988}]{Kulkarni1988}
{Kulkarni} S.~R.,  {Heiles} C.,  1988, {Neutral hydrogen and the diffuse interstellar medium}.
Springer-Verlag, pp 95--153

\bibitem[\protect\citeauthoryear{{Lan}, {Chen}, {Goodman}, {Gimpel}, {Sharma}  \& {Soricut}}{{Lan} et~al.}{2019}]{Lan2019}
{Lan} Z.,  {Chen} M.,  {Goodman} S.,  {Gimpel} K.,  {Sharma} P.,   {Soricut} R.,  2019, \mn@doi [arXiv e-prints] {10.48550/arXiv.1909.11942}, \href {https://ui.adsabs.harvard.edu/abs/2019arXiv190911942L} {p. arXiv:1909.11942}

\bibitem[\protect\citeauthoryear{Lecun, Bottou, Bengio  \& Haffner}{Lecun et~al.}{1998}]{Lecun1998}
Lecun Y.,  Bottou L.,  Bengio Y.,   Haffner P.,  1998, \mn@doi [Proceedings of the IEEE] {10.1109/5.726791}, 86, 2278

\bibitem[\protect\citeauthoryear{{Lee}, {Stanimirovi{\'c}}, {Murray}, {Heiles}  \& {Miller}}{{Lee} et~al.}{2015}]{Lee2015}
{Lee} M.-Y.,  {Stanimirovi{\'c}} S.,  {Murray} C.~E.,  {Heiles} C.,   {Miller} J.,  2015, \mn@doi [\apj] {10.1088/0004-637X/809/1/56}, \href {http://adsabs.harvard.edu/abs/2015ApJ...809...56L} {809, 56}

\bibitem[\protect\citeauthoryear{{Lei} \& {Clark}}{{Lei} \& {Clark}}{2023}]{Lei2023}
{Lei} M.,  {Clark} S.~E.,  2023, \mn@doi [arXiv e-prints] {10.48550/arXiv.2312.03846}, \href {https://ui.adsabs.harvard.edu/abs/2023arXiv231203846L} {p. arXiv:2312.03846}

\bibitem[\protect\citeauthoryear{{Li}, {Lavrukhin}, {Ginsburg}, {Leary}, {Kuchaiev}, {Cohen}, {Nguyen}  \& {Teja Gadde}}{{Li} et~al.}{2019a}]{Li2019CNN}
{Li} J.,  {Lavrukhin} V.,  {Ginsburg} B.,  {Leary} R.,  {Kuchaiev} O.,  {Cohen} J.~M.,  {Nguyen} H.,   {Teja Gadde} R.,  2019a, \mn@doi [arXiv e-prints] {10.48550/arXiv.1904.03288}, \href {https://ui.adsabs.harvard.edu/abs/2019arXiv190403288L} {p. arXiv:1904.03288}

\bibitem[\protect\citeauthoryear{{Li}, {Wang}, {Liu}, {Tang}, {Lei}  \& {Li}}{{Li} et~al.}{2019b}]{Li2019Trans}
{Li} H.,  {Wang} A. Y.~C.,  {Liu} Y.,  {Tang} D.,  {Lei} Z.,   {Li} W.,  2019b, \mn@doi [arXiv e-prints] {10.48550/arXiv.1910.13634}, \href {https://ui.adsabs.harvard.edu/abs/2019arXiv191013634L} {p. arXiv:1910.13634}

\bibitem[\protect\citeauthoryear{{Liszt}}{{Liszt}}{2001}]{Liszt2001}
{Liszt} H.,  2001, \mn@doi [\aap] {10.1051/0004-6361:20010395}, \href {http://adsabs.harvard.edu/abs/2001A%26A...371..698L} {371, 698}

\bibitem[\protect\citeauthoryear{{Liu} et~al.,}{{Liu} et~al.}{2019}]{Liu2019}
{Liu} Y.,  et~al., 2019, \mn@doi [arXiv e-prints] {10.48550/arXiv.1907.11692}, \href {https://ui.adsabs.harvard.edu/abs/2019arXiv190711692L} {p. arXiv:1907.11692}

\bibitem[\protect\citeauthoryear{{Lu}, {Li}, {He}, {Sun}, {Dong}, {Qin}, {Wang}  \& {Liu}}{{Lu} et~al.}{2019}]{Lu2019Trans}
{Lu} Y.,  {Li} Z.,  {He} D.,  {Sun} Z.,  {Dong} B.,  {Qin} T.,  {Wang} L.,   {Liu} T.-Y.,  2019, \mn@doi [arXiv e-prints] {10.48550/arXiv.1906.02762}, \href {https://ui.adsabs.harvard.edu/abs/2019arXiv190602762L} {p. arXiv:1906.02762}

\bibitem[\protect\citeauthoryear{{Marchal}, {Miville-Desch{\^e}nes}, {Orieux}, {Gac}, {Soussen}, {Lesot}, {d'Allonnes}  \& {Salom{\'e}}}{{Marchal} et~al.}{2019}]{Marchal2019}
{Marchal} A.,  {Miville-Desch{\^e}nes} M.-A.,  {Orieux} F.,  {Gac} N.,  {Soussen} C.,  {Lesot} M.-J.,  {d'Allonnes} A.~R.,   {Salom{\'e}} Q.,  2019, \mn@doi [\aap] {10.1051/0004-6361/201935335}, \href {https://ui.adsabs.harvard.edu/abs/2019A&A...626A.101M} {626, A101}

\bibitem[\protect\citeauthoryear{{Marchal}, {Martin}, {Miville-Desch{\^e}nes}, {McClure-Griffiths}, {Lynn}, {Bracco}  \& {Vujeva}}{{Marchal} et~al.}{2024}]{Marchal2024}
{Marchal} A.,  {Martin} P.~G.,  {Miville-Desch{\^e}nes} M.-A.,  {McClure-Griffiths} N.~M.,  {Lynn} C.,  {Bracco} A.,   {Vujeva} L.,  2024, \mn@doi [\apj] {10.3847/1538-4357/ad0f21}, \href {https://ui.adsabs.harvard.edu/abs/2024ApJ...961..161M} {961, 161}

\bibitem[\protect\citeauthoryear{{Martin}, {Blagrave}, {Lockman}, {Pinheiro Gon{{c}}alves}, {Boothroyd}, {Joncas}, {Miville-Desch{\^e}nes}  \& {Stephan}}{{Martin} et~al.}{2015}]{Martin2015}
{Martin} P.~G.,  {Blagrave} K.~P.~M.,  {Lockman} F.~J.,  {Pinheiro Gon{{c}}alves} D.,  {Boothroyd} A.~I.,  {Joncas} G.,  {Miville-Desch{\^e}nes} M.~A.,   {Stephan} G.,  2015, \mn@doi [\apj] {10.1088/0004-637X/809/2/153}, \href {https://ui.adsabs.harvard.edu/abs/2015ApJ...809..153M} {809, 153}

\bibitem[\protect\citeauthoryear{{McClure-Griffiths} et~al.,}{{McClure-Griffiths} et~al.}{2009}]{McClureGriffiths2009}
{McClure-Griffiths} N.~M.,  et~al., 2009, \mn@doi [\apjs] {10.1088/0067-0049/181/2/398}, \href {https://ui.adsabs.harvard.edu/abs/2009ApJS..181..398M} {181, 398}

\bibitem[\protect\citeauthoryear{{McClure-Griffiths} et~al.,}{{McClure-Griffiths} et~al.}{2015}]{McClureGriffiths2015aska}
{McClure-Griffiths} N.~M.,  et~al., 2015, in Advancing Astrophysics with the Square Kilometre Array (AASKA14). p.~130 (\mn@eprint {arXiv} {1501.01130}), \mn@doi{10.22323/1.215.0130}

\bibitem[\protect\citeauthoryear{{McClure-Griffiths}, {Stanimirovi{\'c}}  \& {Rybarczyk}}{{McClure-Griffiths} et~al.}{2023}]{McClure-Griffiths2023}
{McClure-Griffiths} N.~M.,  {Stanimirovi{\'c}} S.,   {Rybarczyk} D.~R.,  2023, \mn@doi [\araa] {10.1146/annurev-astro-052920-104851}, \href {https://ui.adsabs.harvard.edu/abs/2023ARA&A..61...19M} {61, 19}

\bibitem[\protect\citeauthoryear{{McKee} \& {Ostriker}}{{McKee} \& {Ostriker}}{1977}]{McKee1977}
{McKee} C.~F.,  {Ostriker} J.~P.,  1977, \mn@doi [\apj] {10.1086/155667}, \href {http://adsabs.harvard.edu/abs/1977ApJ...218..148M} {218, 148}

\bibitem[\protect\citeauthoryear{McKinney et~al.}{McKinney et~al.}{2010}]{mckinney2010data}
McKinney W.,  et~al., 2010, in Proceedings of the 9th Python in Science Conference. pp 51--56

\bibitem[\protect\citeauthoryear{{Mebold}, {Winnberg}, {Kalberla}  \& {Goss}}{{Mebold} et~al.}{1982}]{Mebold1982}
{Mebold} U.,  {Winnberg} A.,  {Kalberla} P.~M.~W.,   {Goss} W.~M.,  1982, \aap, \href {http://adsabs.harvard.edu/abs/1982A%26A...115..223M} {115, 223}

\bibitem[\protect\citeauthoryear{{Mohan}, {Dwarakanath}  \& {Srinivasan}}{{Mohan} et~al.}{2004}]{Mohan2004}
{Mohan} R.,  {Dwarakanath} K.~S.,   {Srinivasan} G.,  2004, \mn@doi [Journal of Astrophysics and Astronomy] {10.1007/BF02702370}, \href {https://ui.adsabs.harvard.edu/abs/2004JApA...25..143M} {25, 143}

\bibitem[\protect\citeauthoryear{{Murray} et~al.,}{{Murray} et~al.}{2015}]{Murray2015}
{Murray} C.~E.,  et~al., 2015, \mn@doi [\apj] {10.1088/0004-637X/804/2/89}, \href {http://adsabs.harvard.edu/abs/2015ApJ...804...89M} {804, 89}

\bibitem[\protect\citeauthoryear{{Murray}, {Stanimirovi{\'c}}, {Goss}, {Heiles}, {Dickey}, {Babler}  \& {Kim}}{{Murray} et~al.}{2018a}]{Murray2018}
{Murray} C.~E.,  {Stanimirovi{\'c}} S.,  {Goss} W.~M.,  {Heiles} C.,  {Dickey} J.~M.,  {Babler} B.,   {Kim} C.-G.,  2018a, \mn@doi [\apjs] {10.3847/1538-4365/aad81a}, \href {http://adsabs.harvard.edu/abs/2018ApJS..238...14M} {238, 14}

\bibitem[\protect\citeauthoryear{{Murray}, {Peek}, {Lee}  \& {Stanimirovi{\'c}}}{{Murray} et~al.}{2018b}]{Murray2018a}
{Murray} C.~E.,  {Peek} J.~E.~G.,  {Lee} M.-Y.,   {Stanimirovi{\'c}} S.,  2018b, \mn@doi [\apj] {10.3847/1538-4357/aaccfe}, \href {http://adsabs.harvard.edu/abs/2018ApJ...862..131M} {862, 131}

\bibitem[\protect\citeauthoryear{{Murray}, {Peek}  \& {Kim}}{{Murray} et~al.}{2020}]{Murray2020}
{Murray} C.~E.,  {Peek} J.~E.~G.,   {Kim} C.-G.,  2020, \mn@doi [\apj] {10.3847/1538-4357/aba19b}, \href {https://ui.adsabs.harvard.edu/abs/2020ApJ...899...15M} {899, 15}

\bibitem[\protect\citeauthoryear{{Murray}, {Stanimirovi{\'c}}, {Heiles}, {Dickey}, {McClure-Griffiths}, {Lee}, {M. Goss}  \& {Killerby-Smith}}{{Murray} et~al.}{2021}]{Murray2021}
{Murray} C.~E.,  {Stanimirovi{\'c}} S.,  {Heiles} C.,  {Dickey} J.~M.,  {McClure-Griffiths} N.~M.,  {Lee} M.~Y.,  {M. Goss} W.,   {Killerby-Smith} N.,  2021, \mn@doi [\apjs] {10.3847/1538-4365/ac0f0b}, \href {https://ui.adsabs.harvard.edu/abs/2021ApJS..256...37M} {256, 37}

\bibitem[\protect\citeauthoryear{{Nguyen} et~al.,}{{Nguyen} et~al.}{2018}]{Nguyen2018}
{Nguyen} H.,  et~al., 2018, \mn@doi [\apj] {10.3847/1538-4357/aac82b}, \href {http://adsabs.harvard.edu/abs/2018ApJ...862...49N} {862, 49}

\bibitem[\protect\citeauthoryear{{Nguyen}, {Dawson}, {Lee}, {Murray}, {Stanimirovi{\'c}}, {Heiles}, {Miville-Desch{\^e}nes}  \& {Petzler}}{{Nguyen} et~al.}{2019}]{Nguyen2019}
{Nguyen} H.,  {Dawson} J.~R.,  {Lee} M.-Y.,  {Murray} C.~E.,  {Stanimirovi{\'c}} S.,  {Heiles} C.,  {Miville-Desch{\^e}nes} M.~A.,   {Petzler} A.,  2019, \mn@doi [\apj] {10.3847/1538-4357/ab2b9f}, \href {https://ui.adsabs.harvard.edu/abs/2019ApJ...880..141N} {880, 141}

\bibitem[\protect\citeauthoryear{{Nguyen} et~al.,}{{Nguyen} et~al.}{2024}]{Nguyen2024}
{Nguyen} H.,  et~al., 2024, \mn@doi [\mnras] {10.1093/mnras/stae2274}, \href {https://ui.adsabs.harvard.edu/abs/2024MNRAS.534.3478N} {534, 3478}

\bibitem[\protect\citeauthoryear{{Pan}, {Ting}  \& {Yu}}{{Pan} et~al.}{2022}]{Pan2022}
{Pan} J.,  {Ting} Y.-S.,   {Yu} J.,  2022, in Machine Learning for Astrophysics. p.~10 (\mn@eprint {arXiv} {2207.02787}), \mn@doi{10.48550/arXiv.2207.02787}

\bibitem[\protect\citeauthoryear{{Paszke} et~al.,}{{Paszke} et~al.}{2019}]{Pytorch2019}
{Paszke} A.,  et~al., 2019, \mn@doi [arXiv e-prints] {10.48550/arXiv.1912.01703}, \href {https://ui.adsabs.harvard.edu/abs/2019arXiv191201703P} {p. arXiv:1912.01703}

\bibitem[\protect\citeauthoryear{{Peek} et~al.,}{{Peek} et~al.}{2011}]{Peek2011}
{Peek} J.~E.~G.,  et~al., 2011, \mn@doi [\apjs] {10.1088/0067-0049/194/2/20}, \href {http://adsabs.harvard.edu/abs/2011ApJS..194...20P} {194, 20}

\bibitem[\protect\citeauthoryear{{Peek} et~al.,}{{Peek} et~al.}{2018}]{Peek2018}
{Peek} J.~E.~G.,  et~al., 2018, \mn@doi [\apjs] {10.3847/1538-4365/aa91d3}, \href {http://adsabs.harvard.edu/abs/2018ApJS..234....2P} {234, 2}

\bibitem[\protect\citeauthoryear{{Radford}, {Wu}, {Child}, {Luan}, {Amodei}  \& {Sutskever}}{{Radford} et~al.}{2018}]{Radford2019}
{Radford} A.,  {Wu} J.,  {Child} R.,  {Luan} D.,  {Amodei} D.,   {Sutskever} I.,  2018, \href {https://d4mucfpksywv.cloudfront.net/better-language-models/language_models_are_unsupervised_multitask_learners.pdf} {}

\bibitem[\protect\citeauthoryear{{Raffel} et~al.,}{{Raffel} et~al.}{2019}]{Raffel2019}
{Raffel} C.,  et~al., 2019, \mn@doi [arXiv e-prints] {10.48550/arXiv.1910.10683}, \href {https://ui.adsabs.harvard.edu/abs/2019arXiv191010683R} {p. arXiv:1910.10683}

\bibitem[\protect\citeauthoryear{{Roy}, {Kanekar}, {Braun}  \& {Chengalur}}{{Roy} et~al.}{2013a}]{Roy2013}
{Roy} N.,  {Kanekar} N.,  {Braun} R.,   {Chengalur} J.~N.,  2013a, \mn@doi [\mnras] {10.1093/mnras/stt1743}, \href {http://adsabs.harvard.edu/abs/2013MNRAS.436.2352R} {436, 2352}

\bibitem[\protect\citeauthoryear{{Roy}, {Kanekar}  \& {Chengalur}}{{Roy} et~al.}{2013b}]{Roy2013b}
{Roy} N.,  {Kanekar} N.,   {Chengalur} J.~N.,  2013b, \mn@doi [\mnras] {10.1093/mnras/stt1746}, \href {https://ui.adsabs.harvard.edu/abs/2013MNRAS.436.2366R} {436, 2366}

\bibitem[\protect\citeauthoryear{{R{\'o}{\.z}a{\'n}ski}, {Ting}  \& {Jab{\l}o{\'n}ska}}{{R{\'o}{\.z}a{\'n}ski} et~al.}{2023}]{Rozanski2023}
{R{\'o}{\.z}a{\'n}ski} T.,  {Ting} Y.-S.,   {Jab{\l}o{\'n}ska} M.,  2023, \mn@doi [arXiv e-prints] {10.48550/arXiv.2306.15703}, \href {https://ui.adsabs.harvard.edu/abs/2023arXiv230615703R} {p. arXiv:2306.15703}

\bibitem[\protect\citeauthoryear{Sainath, Mohamed, Kingsbury  \& Ramabhadran}{Sainath et~al.}{2013}]{Sainath2013}
Sainath T.~N.,  Mohamed A.-r.,  Kingsbury B.,   Ramabhadran B.,  2013, in 2013 IEEE International Conference on Acoustics, Speech and Signal Processing. pp 8614--8618, \mn@doi{10.1109/ICASSP.2013.6639347}

\bibitem[\protect\citeauthoryear{{Saury}, {Miville-Desch{\^e}nes}, {Hennebelle}, {Audit}  \& {Schmidt}}{{Saury} et~al.}{2014}]{Saury2014}
{Saury} E.,  {Miville-Desch{\^e}nes} M.-A.,  {Hennebelle} P.,  {Audit} E.,   {Schmidt} W.,  2014, \mn@doi [\aap] {10.1051/0004-6361/201321113}, \href {http://adsabs.harvard.edu/abs/2014A%26A...567A..16S} {567, A16}

\bibitem[\protect\citeauthoryear{{Schaye} et~al.,}{{Schaye} et~al.}{2015}]{Schaye2015}
{Schaye} J.,  et~al., 2015, \mn@doi [\mnras] {10.1093/mnras/stu2058}, \href {https://ui.adsabs.harvard.edu/abs/2015MNRAS.446..521S} {446, 521}

\bibitem[\protect\citeauthoryear{{Seta} \& {Federrath}}{{Seta} \& {Federrath}}{2022}]{Seta2022}
{Seta} A.,  {Federrath} C.,  2022, \mn@doi [\mnras] {10.1093/mnras/stac1400}, \href {https://ui.adsabs.harvard.edu/abs/2022MNRAS.514..957S} {514, 957}

\bibitem[\protect\citeauthoryear{{Shaw}, {Ferland}  \& {Hubeny}}{{Shaw} et~al.}{2017}]{Shaw2017}
{Shaw} G.,  {Ferland} G.~J.,   {Hubeny} I.,  2017, \mn@doi [\apj] {10.3847/1538-4357/aa7747}, \href {http://adsabs.harvard.edu/abs/2017ApJ...843..149S} {843, 149}

\bibitem[\protect\citeauthoryear{{Springel} et~al.,}{{Springel} et~al.}{2018}]{Springel2018}
{Springel} V.,  et~al., 2018, \mn@doi [\mnras] {10.1093/mnras/stx3304}, \href {https://ui.adsabs.harvard.edu/abs/2018MNRAS.475..676S} {475, 676}

\bibitem[\protect\citeauthoryear{{Stanimirovi{\'c}}, {Murray}, {Lee}, {Heiles}  \& {Miller}}{{Stanimirovi{\'c}} et~al.}{2014}]{Stanimirovic2014}
{Stanimirovi{\'c}} S.,  {Murray} C.~E.,  {Lee} M.-Y.,  {Heiles} C.,   {Miller} J.,  2014, \mn@doi [\apj] {10.1088/0004-637X/793/2/132}, \href {http://adsabs.harvard.edu/abs/2014ApJ...793..132S} {793, 132}

\bibitem[\protect\citeauthoryear{{Strasser} et~al.,}{{Strasser} et~al.}{2007}]{Strasser2007}
{Strasser} S.~T.,  et~al., 2007, \mn@doi [\aj] {10.1086/522794}, \href {https://ui.adsabs.harvard.edu/abs/2007AJ....134.2252S} {134, 2252}

\bibitem[\protect\citeauthoryear{Szegedy et~al.,}{Szegedy et~al.}{2015a}]{Szegedy2015}
Szegedy C.,  et~al., 2015a, in 2015 IEEE Conference on Computer Vision and Pattern Recognition (CVPR). pp~1--9, \mn@doi{10.1109/CVPR.2015.7298594}

\bibitem[\protect\citeauthoryear{{Szegedy}, {Vanhoucke}, {Ioffe}, {Shlens}  \& {Wojna}}{{Szegedy} et~al.}{2015b}]{Szegedy2016}
{Szegedy} C.,  {Vanhoucke} V.,  {Ioffe} S.,  {Shlens} J.,   {Wojna} Z.,  2015b, \mn@doi [arXiv e-prints] {10.48550/arXiv.1512.00567}, \href {https://ui.adsabs.harvard.edu/abs/2015arXiv151200567S} {p. arXiv:1512.00567}

\bibitem[\protect\citeauthoryear{{Uelwer}, {Robine}, {Sylvius Wagner}, {H{\"o}ftmann}, {Upschulte}, {Konietzny}, {Behrendt}  \& {Harmeling}}{{Uelwer} et~al.}{2023}]{Uelwer2023}
{Uelwer} T.,  {Robine} J.,  {Sylvius Wagner} S.,  {H{\"o}ftmann} M.,  {Upschulte} E.,  {Konietzny} S.,  {Behrendt} M.,   {Harmeling} S.,  2023, \mn@doi [arXiv e-prints] {10.48550/arXiv.2308.11455}, \href {https://ui.adsabs.harvard.edu/abs/2023arXiv230811455U} {p. arXiv:2308.11455}

\bibitem[\protect\citeauthoryear{{Vaswani}, {Shazeer}, {Parmar}, {Uszkoreit}, {Jones}, {Gomez}, {Kaiser}  \& {Polosukhin}}{{Vaswani} et~al.}{2017}]{Vaswani2017}
{Vaswani} A.,  {Shazeer} N.,  {Parmar} N.,  {Uszkoreit} J.,  {Jones} L.,  {Gomez} A.~N.,  {Kaiser} L.,   {Polosukhin} I.,  2017, \mn@doi [arXiv e-prints] {10.48550/arXiv.1706.03762}, \href {https://ui.adsabs.harvard.edu/abs/2017arXiv170603762V} {p. arXiv:1706.03762}

\bibitem[\protect\citeauthoryear{{Vijayan}, {Krumholz}  \& {Wibking}}{{Vijayan} et~al.}{2023}]{Vijayan2023}
{Vijayan} A.,  {Krumholz} M.~R.,   {Wibking} B.~D.,  2023, \mn@doi [arXiv e-prints] {10.48550/arXiv.2309.07955}, \href {https://ui.adsabs.harvard.edu/abs/2023arXiv230907955V} {p. arXiv:2309.07955}

\bibitem[\protect\citeauthoryear{{Virtanen} et~al.,}{{Virtanen} et~al.}{2020}]{Virtanen2020}
{Virtanen} P.,  et~al., 2020, \mn@doi [Nature Methods] {10.1038/s41592-019-0686-2}, \href {https://ui.adsabs.harvard.edu/abs/2020NatMe..17..261V} {17, 261}

\bibitem[\protect\citeauthoryear{{Vogelsberger} et~al.,}{{Vogelsberger} et~al.}{2014}]{Vogelsberger2014}
{Vogelsberger} M.,  et~al., 2014, \mn@doi [\mnras] {10.1093/mnras/stu1536}, \href {https://ui.adsabs.harvard.edu/abs/2014MNRAS.444.1518V} {444, 1518}

\bibitem[\protect\citeauthoryear{{Vujeva}, {Marchal}, {Martin}  \& {Taank}}{{Vujeva} et~al.}{2023}]{Vujeva2023}
{Vujeva} L.,  {Marchal} A.,  {Martin} P.~G.,   {Taank} M.,  2023, \mn@doi [\apj] {10.3847/1538-4357/acd340}, \href {https://ui.adsabs.harvard.edu/abs/2023ApJ...951..120V} {951, 120}

\bibitem[\protect\citeauthoryear{{Wakker}}{{Wakker}}{2001}]{Wakker2001}
{Wakker} B.~P.,  2001, \mn@doi [\apjs] {10.1086/321783}, \href {https://ui.adsabs.harvard.edu/abs/2001ApJS..136..463W} {136, 463}

\bibitem[\protect\citeauthoryear{{Winkel}, {Kalberla}, {Kerp}  \& {Fl{\"o}er}}{{Winkel} et~al.}{2010}]{Winkel2010}
{Winkel} B.,  {Kalberla} P.~M.~W.,  {Kerp} J.,   {Fl{\"o}er} L.,  2010, \mn@doi [\apjs] {10.1088/0067-0049/188/2/488}, \href {https://ui.adsabs.harvard.edu/abs/2010ApJS..188..488W} {188, 488}

\bibitem[\protect\citeauthoryear{{Winkel}, {Kerp}, {Fl{\"o}er}, {Kalberla}, {Ben Bekhti}, {Keller}  \& {Lenz}}{{Winkel} et~al.}{2016}]{Winkel2016}
{Winkel} B.,  {Kerp} J.,  {Fl{\"o}er} L.,  {Kalberla} P.~M.~W.,  {Ben Bekhti} N.,  {Keller} R.,   {Lenz} D.,  2016, \mn@doi [\aap] {10.1051/0004-6361/201527007}, \href {https://ui.adsabs.harvard.edu/abs/2016A&A...585A..41W} {585, A41}

\bibitem[\protect\citeauthoryear{{Wolfire}, {Hollenbach}, {McKee}, {Tielens}  \& {Bakes}}{{Wolfire} et~al.}{1995}]{Wolfire1995}
{Wolfire} M.~G.,  {Hollenbach} D.,  {McKee} C.~F.,  {Tielens} A.~G.~G.~M.,   {Bakes} E.~L.~O.,  1995, \mn@doi [\apj] {10.1086/175510}, \href {http://adsabs.harvard.edu/abs/1995ApJ...443..152W} {443, 152}

\bibitem[\protect\citeauthoryear{{Wolfire}, {McKee}, {Hollenbach}  \& {Tielens}}{{Wolfire} et~al.}{2003}]{Wolfire2003}
{Wolfire} M.~G.,  {McKee} C.~F.,  {Hollenbach} D.,   {Tielens} A.~G.~G.~M.,  2003, \mn@doi [\apj] {10.1086/368016}, \href {http://adsabs.harvard.edu/abs/2003ApJ...587..278W} {587, 278}

\bibitem[\protect\citeauthoryear{{Wu}, {Liu}, {Lin}, {Lin}  \& {Han}}{{Wu} et~al.}{2020}]{Wu2020Trans}
{Wu} Z.,  {Liu} Z.,  {Lin} J.,  {Lin} Y.,   {Han} S.,  2020, \mn@doi [arXiv e-prints] {10.48550/arXiv.2004.11886}, \href {https://ui.adsabs.harvard.edu/abs/2020arXiv200411886W} {p. arXiv:2004.11886}

\bibitem[\protect\citeauthoryear{{Yang}, {Wang}, {Wong}, {Chao}  \& {Tu}}{{Yang} et~al.}{2019a}]{Yang2019CNNTRANS}
{Yang} B.,  {Wang} L.,  {Wong} D.,  {Chao} L.~S.,   {Tu} Z.,  2019a, \mn@doi [arXiv e-prints] {10.48550/arXiv.1904.03107}, \href {https://ui.adsabs.harvard.edu/abs/2019arXiv190403107Y} {p. arXiv:1904.03107}

\bibitem[\protect\citeauthoryear{{Yang}, {Dai}, {Yang}, {Carbonell}, {Salakhutdinov}  \& {Le}}{{Yang} et~al.}{2019b}]{Yang2019}
{Yang} Z.,  {Dai} Z.,  {Yang} Y.,  {Carbonell} J.,  {Salakhutdinov} R.,   {Le} Q.~V.,  2019b, \mn@doi [arXiv e-prints] {10.48550/arXiv.1906.08237}, \href {https://ui.adsabs.harvard.edu/abs/2019arXiv190608237Y} {p. arXiv:1906.08237}

\bibitem[\protect\citeauthoryear{{Yu}, {Dohan}, {Luong}, {Zhao}, {Chen}, {Norouzi}  \& {Le}}{{Yu} et~al.}{2018}]{Yu2018CNNTrans}
{Yu} A.~W.,  {Dohan} D.,  {Luong} M.-T.,  {Zhao} R.,  {Chen} K.,  {Norouzi} M.,   {Le} Q.~V.,  2018, \mn@doi [arXiv e-prints] {10.48550/arXiv.1804.09541}, \href {https://ui.adsabs.harvard.edu/abs/2018arXiv180409541Y} {p. arXiv:1804.09541}

\bibitem[\protect\citeauthoryear{{Zhang}, {Lu}, {Sak}, {Tripathi}, {McDermott}, {Koo}  \& {Kumar}}{{Zhang} et~al.}{2020}]{Zhang2020}
{Zhang} Q.,  {Lu} H.,  {Sak} H.,  {Tripathi} A.,  {McDermott} E.,  {Koo} S.,   {Kumar} S.,  2020, \mn@doi [arXiv e-prints] {10.48550/arXiv.2002.02562}, \href {https://ui.adsabs.harvard.edu/abs/2020arXiv200202562Z} {p. arXiv:2002.02562}

\bibitem[\protect\citeauthoryear{{van der Walt}, {Colbert}  \& {Varoquaux}}{{van der Walt} et~al.}{2011}]{vanderWalt2011}
{van der Walt} S.,  {Colbert} S.~C.,   {Varoquaux} G.,  2011, \mn@doi [Computing in Science and Engineering] {10.1109/MCSE.2011.37}, \href {https://ui.adsabs.harvard.edu/abs/2011CSE....13b..22V} {13, 22}

\makeatother
\end{thebibliography}




\appendix

\section{Model designs}
\label{sec:models}
    Following the standard mathematical conventions and notations, we represent scalars with lowercase letters $x \in \mathbb{R}$, vectors of $N$ elements with boldface $\mathbf{x} \in \mathbb{R}^N$, and matrices of $M$ rows and $N$ columns with boldface uppercase letters $\mathbf{X} \in \mathbb{R}^{M \times N}$. Vector and matrix indexing is denoted respectively as follows: $(x_i)_{i=1,2,...,N} = \mathbf{x}$, $(X_{ij})_{i=1,2,...,M,j=1,2,...,N} = \mathbf{X}$.

\subsection{CNN}
\label{subsec:cnn}

Convolutional Neural Networks \citep[CNNs,][]{Lecun1998,fukushima1980} are a type of neural network models widely employed in deep learning, specifically tailored to process grid-like data such as images, audio spectrograms, and time series data. These networks operate in a feed-forward manner, in which the output from one layer serves as the input to the next layer. The key idea behind CNNs is to learn hierarchical representations of features directly from the input data.

A typical CNN architecture consists of multiple layers starting with an input layer, followed by hidden layers, fully connected layers, and an output layer.

The input layer receives the raw data, such as images or sequences, and passes it to the subsequent layers. In image processing, the pixel values from a small rectangular region (or patch) within an image are referred to as the \textit{receptive field}. Similarly, in the context of a 1D CNN applied to spectral data, the receptive field represents a segment of input values that affect the output of a specific neuron in the network.

Hidden layers comprise convolutional, activation, and (optional) pooling layers. Convolutional layers apply convolution operations to the input data, where small filter(s) (a set of kernels with the same size) slide over the input grid and compute dot products to extract hierarchical features. These layers are followed by
non-linear activation functions (e.g., \textit{tanh}, \textit{sigmoid}, \textit{softmax}, or \textit{rectified linear unit}) to introduce non-linearity into the network and facilitate the learning of complex relationships in the data. Optionally, pooling layers may follow to perform vector-to-scalar transformation on local regions of the input, aiding in feature extraction, reducing computational cost, and preventing overfitting.

Convolutional layers are essentially made up of neurons. In each convolutional layer, neurons receive inputs from neurons in the previous layer. Each input ($x$) is assigned a random weight ($w$), and a dot product operation is performed between these weights and the corresponding input values, followed by an addition of a bias term ($b$). The resulting scalar output is passed through a non-linear activation function ($f$). Thus, the output of a single neuron ($z$) is a result of a non-linear transformation and can be mathematically expressed as:

\begin{equation}
z = f(w \ast x) = f(\mathbf{w^\mathrm{T}} \mathbf{x} + b) = f \left( \sum_{i=1}^{n} w_i \cdot x_i + b \right)
\label{eq:activ}
\end{equation}

\noindent where ``$\ast$'' represents the convolution operator, $\mathbf{w}$ is the weight vector associated with the input vector $\mathbf{x}$ for a receptive field, and $n$ represents kernel size.

Similar to the convolutional layers, fully connected layers (also known as dense layers) consist of neurons. These layers connect every neuron in one layer to every neuron in the next layer. One or more fully connected layers may be used, towards the end of the network, to perform high-level reasoning and classification based on the learned features.

Finally, the output layer produces the final predictions or classifications by using an appropriate activation function  (e.g., \textit{softmax} for classification or \textit{linear} for regression).

In a supervised machine learning task, the network must be trained using a labeled training database, where the ground-truth outputs corresponding to the inputs are known, to determine the optimal model parameters (i.e., weights and biases) at each neuron's input across layers. The training process of a CNN model typically involves several key steps. First, the network is initialized with random values for the weights and biases. Then, input data is fed through the network, resulting in a predicted output. This predicted output is compared to the ground truth using a predefined loss function to compute the loss. Through back-propagation, the gradients of the loss function with respect to the model parameters are computed. These gradients are then used in gradient descent to adjust the weights and biases, aiming to minimize the loss and improve model performance. This iterative process continues with different batches (small subsets) of training data until convergence or a predefined stopping criterion is met.

\begin{figure}
\centering
\includegraphics[width=.475\textwidth]{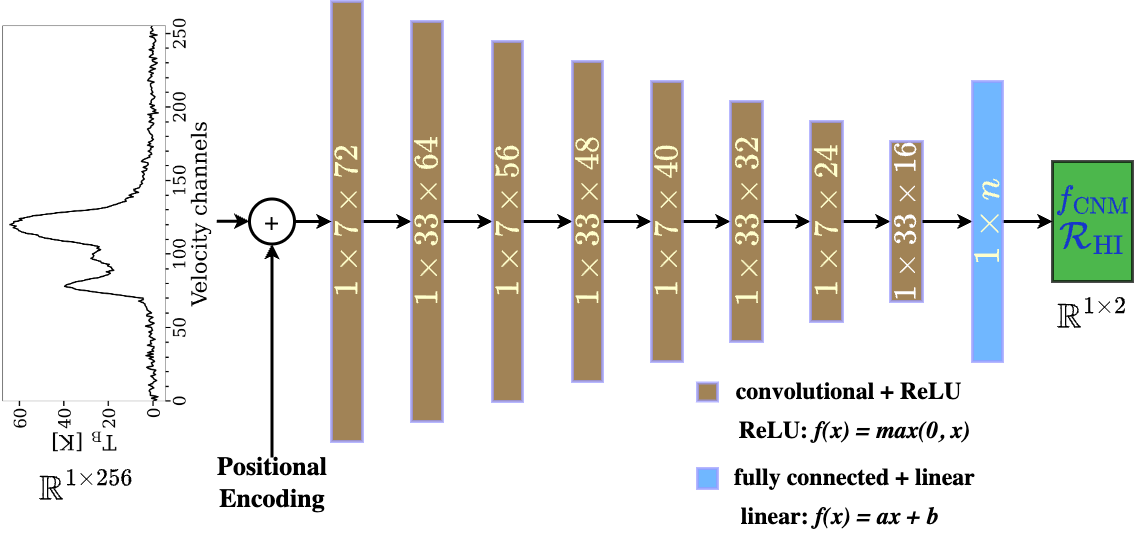}
\caption{Our CNN architecture with 8 convolutional layers using Rectified Linear Unit (ReLU) activation function: $f(x) = \text{max}(0,x)$. The raw spectrum input is position encoded before being passed to the CNN. Two kernel sizes, (1$\times$7) and (1$\times$33), are iteratively employed, with the number of filters reduced by 8 units after each convolutional layer. The model outputs two desired parameters: cold \hi\ gas fraction \FCNM\ and opacity correction factor \RHI.}
\label{fig:1dcnn}
\end{figure}

Following \cite{Murray2020}'s CNN for \hi\ emission analysis, we adopt a typical CNN design with input, convolutional, activation, fully-connected, and output layers. Our CNN architecture is illustrated in Figure \ref{fig:1dcnn}. The input layer accepts either the raw synthetic brightness spectrum or the position-encoded spectrum (see Appendix \ref{sec:data_repr} for details on positional encoding techniques). The width (number of neurons) dynamically adjusts for each convolutional layer based on the total number of layers in the model, with each subsequent layer having fewer neurons than the preceding one. In our approach, we iteratively employ two kernel sizes, 7 and 33, for the convolutional layers to reduce input dimensionality. This architectural choice, combined with varying kernel sizes, is expected to enable the network to learn features across various spectral scales, such as those of CNM (typical width FWHM $\lesssim$ 3 \kms) and WNM (FWHM $\sim$ 10 \kms). Instead of utilizing pooling layers, we opt for the ``Batch Normalization'' operation, after the convolutional layers and before the activation function, to stabilize the training process of deep neural networks. Batch normalization normalizes the inputs of a layer by subtracting the batch mean and dividing by the batch standard deviation.

At the final stage, the outputs from hidden layers are passed through the output layer, which is a fully connected layer equipped with a linear activation function, to predict the \FCNM\ and \RHI\ values.

We determine the optimal CNN configuration through experiments involving varying numbers of hidden layers. The selection process relied on comparing training, validation, and testing RMSE, as indicated in Figure \ref{fig:cnn_nlayers}, Section \ref{subsec:exp_settings}.

\subsection{Transformer}
\label{subsec:trans}

\begin{figure}
\centering
\includegraphics[width=.5\textwidth]{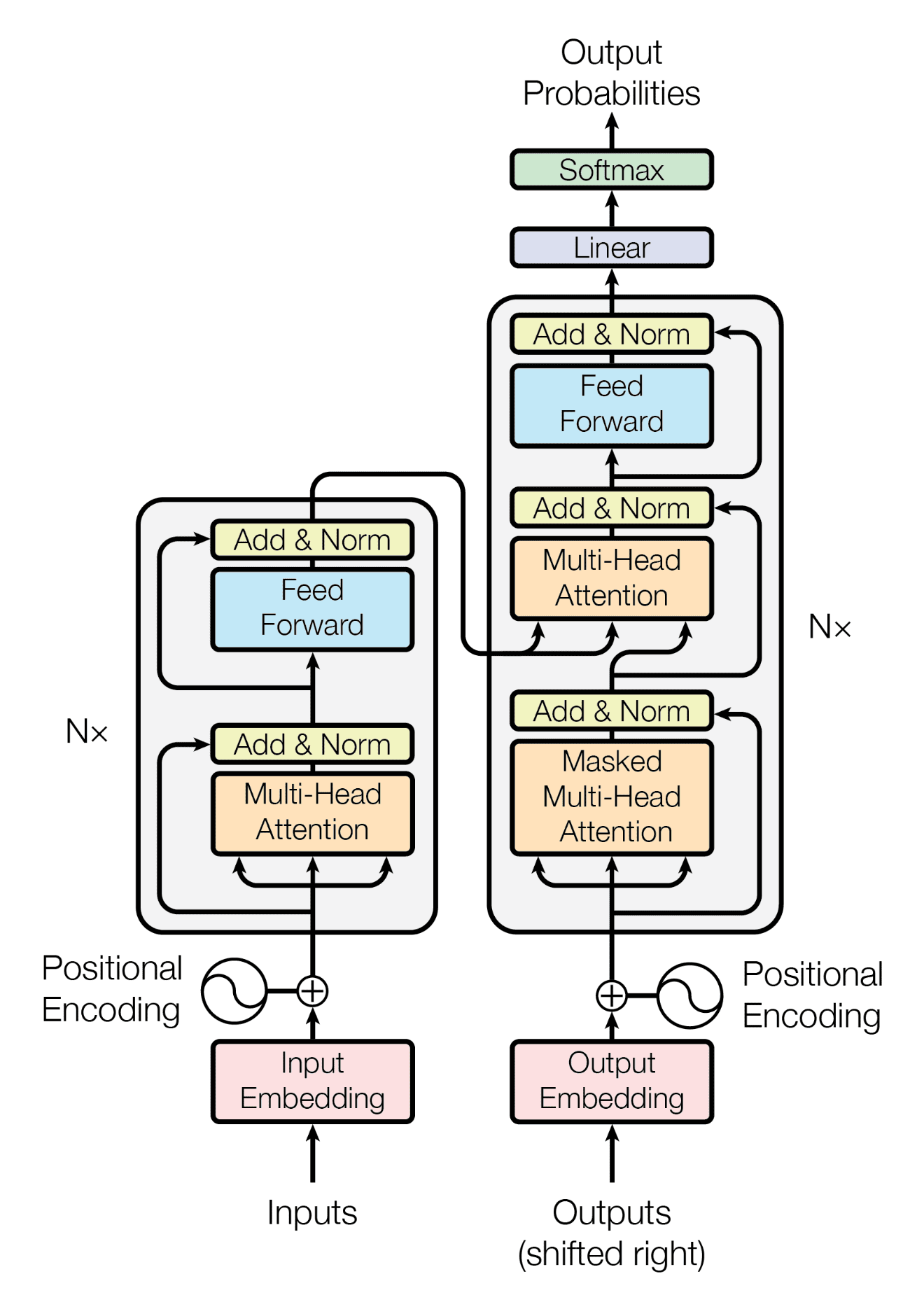}
\caption{A typical architecture of the Transformer deep learning model designed for natural language processing \citep[taken from][]{Vaswani2017}. The architecture comprises two main components: encoder stack (Nx identical encoder layers), and decoder stack (Nx identical decoder layers), incorporating embedding, multi-head self-attention, positional encoding, feedforward, and ``add \& norm'' (residual
connections and layer normalization) layers.}
\label{fig:trans_encoder_decoder}
\end{figure}

The Transformer is a deep learning architecture introduced by \cite{Vaswani2017} used for sequence modeling tasks, such as natural language processing (NLP) and time series analysis. Unlike traditional recurrent neural networks (RNNs) and convolutional neural networks (CNNs), Transformers rely on self-attention mechanisms to capture long-range dependencies within input sequences efficiently \citep[e.g.,][]{Zhang2020,Gulati2020,Pan2022}. A typical Transformer architecture is illustrated in Figure \ref{fig:trans_encoder_decoder} and will be briefly discussed below, though refer to the original paper by \cite{Vaswani2017} for more details.

In the context of natural language processing, sequence data is composed of \textit{tokens}, which are its fundamental units of input. Tokens often represent words or subwords that form a sentence or a piece of text. However, tokens can also denote other types of entities or elements in different types of sequences, such as time series data or spectral data.

At the core of the Transformer architecture is the self-attention mechanism, which allows the model to weigh the importance of different tokens within a sequence when making predictions. The input sequence is first embedded into a continuous vector space using learned embedding matrices, which convert each token into a high-dimensional representation vector (with a length of \textit{D} features). To account for the sequential order of tokens, positional encoding is added to the input embeddings (data matrix $\mathbf{X} \in \mathbb{R}^{N \times D}$ of \textit{N} tokens and \textit{D} features) before they are fed into the self-attention mechanism. This positional encoding typically involves adding directly sinusoidal functions of different frequencies to the embeddings. The combination of token embedding and positional encoding provides the Transformer model with a comprehensive understanding of both the content and the sequential order of the input sequence.

The self-attention mechanism in the Transformer operates on three key matrices: the query matrix \textbf{Q} $\in \mathbb{R}^{N \times D}$, the key matrix \textbf{K} $\in \mathbb{R}^{N \times D}$, and the value matrix \textbf{V} $\in \mathbb{R}^{N \times D}$. These matrices are derived from the input embeddings ($\mathbf{X}$) via the weight matrices:

\begin{equation}
\begin{split}
\mathbf{Q} = \mathbf{X}\mathbf{W}^{(q)} \\
\mathbf{K} = \mathbf{X}\mathbf{W}^{(k)} \\
\mathbf{V} = \mathbf{X}\mathbf{W}^{(v)}
\label{}
\end{split}
\end{equation}

\noindent where the matrices $\mathbf{W}^{(q)}$, $\mathbf{W}^{(q)}$ and $\mathbf{W}^{(q)}$ represent the parameters that are learned during the training. The dimensionalities of the matrices $\mathbf{W}^{(q)}$, $\mathbf{W}^{(k)}$, and $\mathbf{W}^{(v)}$ are ($D \times D_q$), ($D \times D_k$), and ($D \times D_v$), respectively. The two matrices $\mathbf{W}^{(q)}$ and $\mathbf{W}^{(k)}$ must have the same dimensionalities to form dot products between query and key vectors, hence $D_q = D_k$. The \textbf{Q}, \textbf{K}, and \textbf{V} matrices are used to calculate attention scores between each pair of tokens in the sequence. The attention scores determine how much each token attends to other tokens in the sequence. The attention scores are then normalized using the \textit{softmax} function to obtain attention weights, which are applied to the value vectors to compute the output \textbf{Y} of the self-attention mechanism:

\begin{equation}
\mathbf{Y} = \text{attention}(\mathbf{Q}, \mathbf{K}, \mathbf{V}) = \text{softmax}\left[ \frac{\mathbf{Q}\mathbf{K}^{\text{T}}}{\sqrt{D_\mathrm{k}}}\right] \mathbf{V}
\label{}
\end{equation}

The Transformer also employs multi-head self-attention (MHSA), where the self-attention mechanism is applied multiple times in parallel, each with its own set of learned parameters. This allows the model to attend to different aspects of the input sequence simultaneously, capturing both local and global dependencies more effectively.

In addition to self-attention layers, the Transformer architecture includes feedforward neural networks and ``add \& norm'' layers. The feedforward layers apply linear transformations followed by non-linear activation functions to the output of the self-attention layers, enabling the model to capture complex patterns and relationships within the sequence. The ``add \& norm'' layers then perform residual connections and layer normalization, stabilizing the training process and improving the flow of information through the network.

In a typical Transformer architecture, the input sequence is processed by an encoder-decoder structure, with each component containing multiple stacks of MHSA, feedforward, and ``add \& norm'' layers. The encoder encodes the input sequence into a set of context-rich representations. The decoder then uses these representations, along with additional self-attention layers and cross-attention layers, to generate the output sequence token by token.

\subsection{Representation learning models}
\label{subsec:cnn_transformer}

Representation learning in machine learning, also referred to as feature learning, involves algorithms extracting meaningful patterns from raw data to generate representations (or features) that are easier to process. Representation learning serves as a powerful tool for gaining deeper insights into the inherent patterns, trends, and structures within the data \citep[e.g.,][]{Bengio2012,Goodfellow2016,Caron2020,Uelwer2023}.

In this study, we introduce, \ourmod, a representation learning neural network, tailored for processing \hi\ emission analysis. This model draws inspiration from both convolution and Transformer architectures. Models that use either convolution or Transformer self-attention individually have their own limitations. While convolutional neural networks excel in capturing short-range features within the input data via local receptive fields \citep{Sainath2013,AbdelHamid2014,Li2019CNN,Kriman2019,Han2020}, they need many more layers or parameters to extract global information. The Transformer self-attention, on the other hand, specializes in revealing the intricate long-range interactions; yet, they may struggle with extracting fine-grained local feature patterns \citep[e.g.,][]{Zhang2020,Gulati2020}.

Recent research has demonstrated that combining convolution and self-attention enhances performance over utilizing them independently. With this combined method, the model can learn both position-wise local characteristics and global interactions \citep{Bello2019,Lu2019Trans,Yang2019CNNTRANS,Yu2018CNNTrans,Wu2020Trans,Pan2022,Rozanski2023}. Since both global and local interactions within frequency/velocity spectral channels are crucial for model generalization, given their relation to the kinematical scales of \hi\ phases, we design \ourmod\ via a combination of CNN (representation learning block) and Transformer self-attention mechanism (prediction block). CNNs use convolutional layers to extract compact features within the emission spectra. These features are then transformed into an embedding matrix input before being passed to the multi-head attention of a Transformer. The intuition is that the deep CNN learns patterns from \hi\ spectra, and the Transformer's attention mechanism enables long-range interactions between different parts of the spectra to predict the cold gas fraction \FCNM\ and opacity correction \RHI. Our investigation also encompasses an evaluation of the influence of positional encodings on the models' performance.

\begin{figure}
\centering
\includegraphics[width=.5\textwidth]{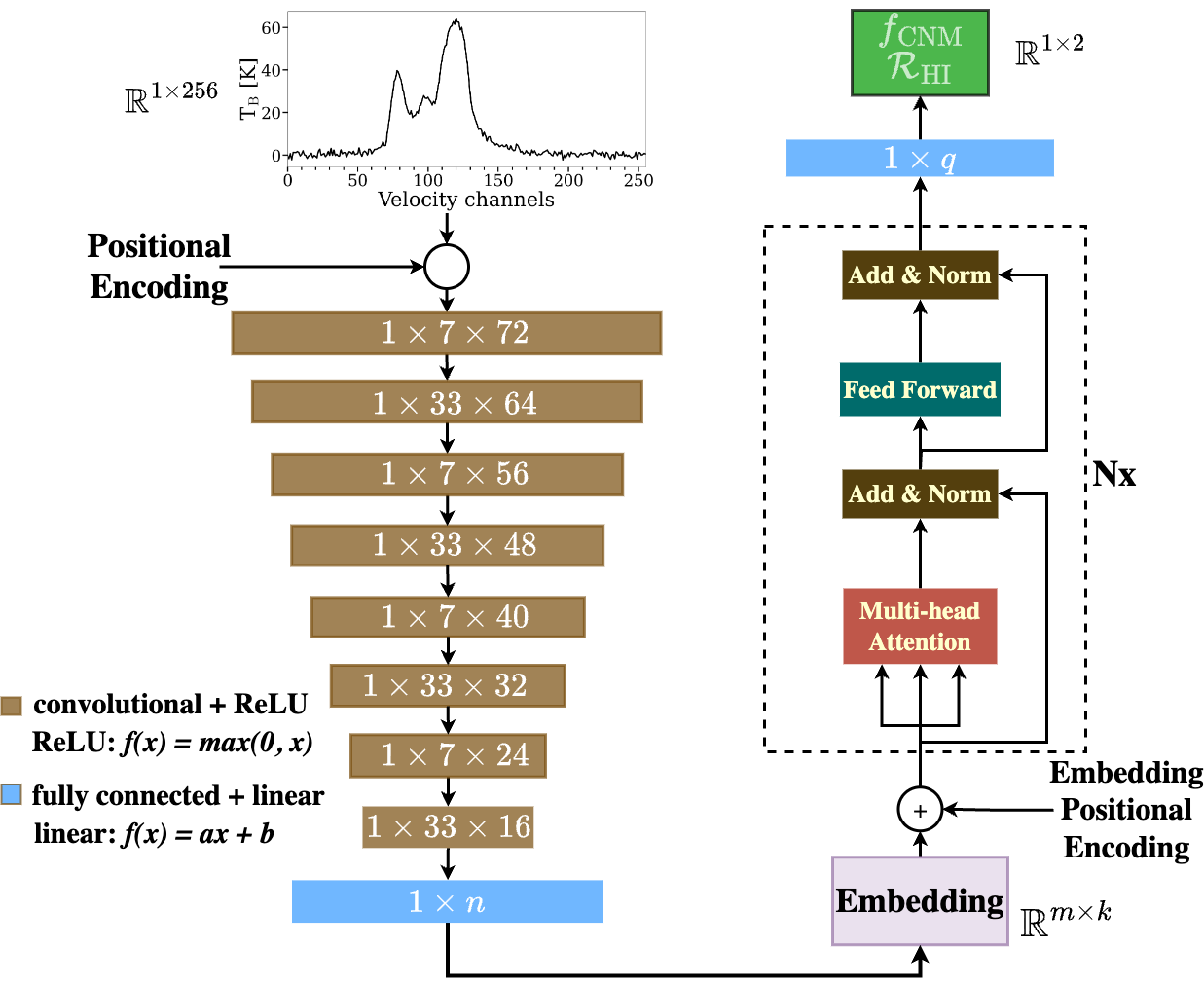}
\caption{A basic architecture of our \ourmod\ model: CNN on the left branch and a stack of Nx identical Transformer decoder layers on the right. The feature vector with length $n$ from the CNN output is converted into a $m \times k$ embedding matrix (containing \textit{m} tokens, each of length \textit{k}, hence $n = m \times k$).}
\label{fig:cnn_transformer}
\end{figure}

Figure \ref{fig:cnn_transformer} illustrates the basic design of our \ourmod. The initial input to the CNN is either the original spectrum or a position-encoded spectrum. The last fully-connected layer of the CNN outputs a feature vector with a length $n$, which is then transformed into a matrix with dimensions of $m \times k$ to mimic an embedding matrix (with $m$ tokens, each with a length of $k$, so $n = m \times k$). We experiment with different shapes for the embedding matrix and yield comparable RMSEs with a range of $m$ and $k$ values. This is likely because the embedding matrix, regardless of its size, contains the same information as the 1D feature vector. We then opt for an ``arbitrary'' length of $k = 9$ for our Transformer's embedding input.

Next, each token (or row) in the embedding matrix is position-encoded using a series of sine and cosine functions with gradually increasing wavelength before being given to the Transformer decoder (see \cite{Vaswani2017} for the choice of embedding sinusoidal positional encoding). Note that the embedding positional encoding is applied to the embedding matrix input of the Transformer. This positional embedding is different from the initial positional encodings applied to the original spectra. Finally, the Transformer decoder will predict the two \hi\ quantities of interest: \FCNM\ and \RHI.

\section{Experimental settings and hyperparameter optimization}
\label{sec:hyperparams}
    \subsection{Experimental settings and finding hyperparameters}
\label{subsec:exp_settings}
In the context of our supervised regression task, where we simultaneously predict two quantities: cold gas fraction and opacity correction factor, we opt for the Root Mean Square Error (RMSE) as the model evaluation metric throughout this work. RMSE quantifies the difference between ground truths (\FCNM\ and \RHI) and their corresponding predictions ($\hat{f}_{\text{CNM}}$ and $\hat{\mathcal{R}}_{\text{HI}}$) values, providing an error measurement in the same units as the target variables.
The overall RMSE for the two quantities is calculated as:
\begin{small}
\begin{equation}
\text{RMSE} = \sqrt{\frac{1}{2K} \left(\sum_{i=1}^{K} (f_{\text{CNM},i} - \hat{f}_{\text{CNM},i})^2 + \sum_{i=1}^{K} (\mathcal{R}_{\text{HI},i} - \hat{\mathcal{R}}_{\text{HI},i})^2 \right)}
\label{eq:rmse}
\end{equation}
\end{small}
\noindent where $K$ is the sample size, and the subscripts \textit{i} index individual data points.

We use four spectral cubes in the training database for training and reserve the fifth for the final evaluation phase. To facilitate our assessment, we split the training set of 1,048,576 spectra into three subsets: 60\% for training (629,146 spectra), 20\% for validation (209,715), and 20\% for testing (209,715). For the training split above,
we experiment with ten random initial weights for each neural network architecture,
and train for 60 epochs.

Our model design is such that it operates on a spectrum-by-spectrum basis. The initial step is to rearrange the PPV spectral cubes into an array of individual spectra. Positional encodings are then applied to the original spectra. Subsequently, these position-encoded spectra arrays, along with their associated \FCNM\ and \RHI, proceed to a data-loading procedure before being fed into the models as batches of samples (i.e., subsets of the training set). During the training phase, we use a batch size of 256 samples, RMSE as the loss function to minimize the difference between the predictions and ground truths, and \textit{Adam} optimizer to adjust model parameters (weights and biases). Additionally, we employ the early stopping technique to continuously assess the models' performance on a validation dataset throughout the training process, aiming to avert overfitting. Whenever the performance on the validation dataset stops improving or starts to degrade, early stopping interrupts training to prevent the model from learning the noise in the training data, which may lead to overfitting. The weights of the model achieving the best performance on the validation sets are then saved for the testing phase.

\begin{figure}
\centering
\includegraphics[width=0.475\textwidth]{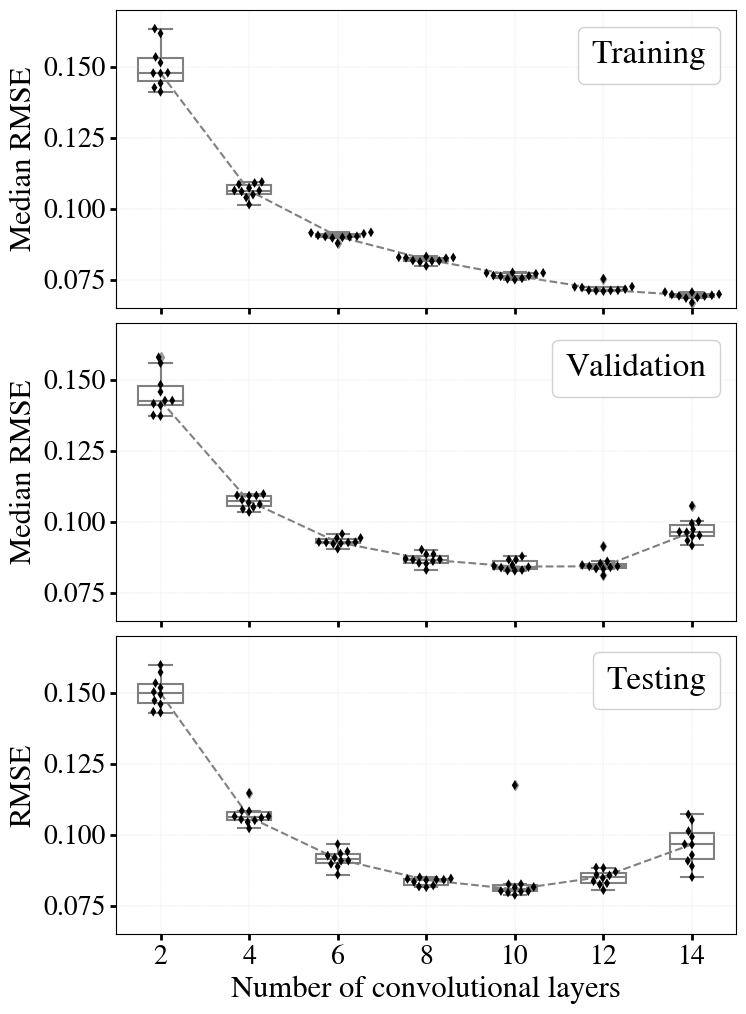}
\caption{Swarm-box plots for training (top), validation (middle), and testing (bottom) RMSE values from ten-trial training sessions of CNN models with different numbers of convolutional layers (using the original spectral data, i.e. without positional encoding). Median training/validation RMSE values are computed over 60 epochs for each trial, while testing RMSE values are obtained individually from each trial. The y-axis scales in all panels are identical for better direct comparisons.}
\label{fig:cnn_nlayers}
\end{figure}

In the first experiment, we seek the optimal parameters of CNNs for further processing. Figure \ref{fig:cnn_nlayers} presents swarm-box plots of the testing and validation RMSE obtained from ten random initial weights for the CNN models with varying numbers of layers. A clear understanding emerges regarding the observed trends in the relationship between the number of convolutional layers and model performance. Although the RMSE values consistently decrease in the training phase with the addition of more convolutional layers, both testing and validation RMSEs show no substantial improvement beyond eight layers. This pattern strongly indicates overfitting in complex CNN architectures, especially with a higher number of convolutional layers ($N > 10$ layers) and underfitting with shallower CNN architectures ($N < 6$ layers). Moreover, the validation/testing RMSE values in cases of overfitting and underfitting show considerable scatter. Once the number of layers surpasses eight, adding more convolutional layers does not significantly yield any further enhancements in model performance. In addition, the swarm plots indicate a significant increase in RMSE standard deviations across different initial conditions after ten convolutional layers.

It is worth noting that increasing the number of convolutional layers in a CNN greatly extends training time. A 14-layer CNN, for example, requires approximately three times the training time of a 10-layer CNN, and about six times the training time of an 8-layer CNN. Considering a balance between model complexity, training efficiency, and accuracy, we will utilize an 8-layer CNN for processing our spectral data.

We then focus on the learning rate -- a hyperparameter that determines how much a neural network weight is adjusted using backpropagation with respect to the amount of RMSE between the predicted values and labels. After experimenting with multiple learning rate values on training sets with 256 (101) velocity channels, we determine the optimal learning rates to be $5 \times 10^{-3}$ ($5 \times 10^{-2}$) for deep CNNs and $5 \times 10^{-3}$ ($9 \times 10^{-3}$) for \ourmods. We will consistently use these learning rates and an 8-layer CNN architecture throughout all subsequent experiments.

\section{Positional encodings (PE)}
\label{sec:data_repr}
    \begin{table*}
\fontsize{10}{9}\selectfont
\centering
\renewcommand{\arraystretch}{1.25}
\begin{tabular}{l| l| l} 
 \hline
 Abbreviation  & Representation & Input shape \\ [0.5ex] 
 \hline\hline
 1. original & Original spectrum & ($1 \times N$) \\ 
 2. concat\_index & Append index array alongside original spectrum & ($2 \times N$) \\
 3. add\_index & Add index array to original spectrum & ($1 \times N$) \\ 
 4. concat\_sinusoidal & Append index sinusoidal function alongside original spectrum & ($2 \times N$) \\
 5. add\_sinusoidal & Add index sinusoidal function to original spectrum & ($1 \times N$) \\ 
 6. concat\_polynomial & Append index 2$^\mathrm{nd}$ polynomial function alongside original spectrum & ($2 \times N$) \\
 7. learnable$^{*}$ & Add learnable positional encoding to original spectrum & ($1 \times N$) \\[1ex] 
 \hline
\end{tabular}
\caption{Positional encodings for 1D spectral data of $N$ channels. ~~~~\\\\ {$^{(*)}$} Learnable positional encoding is only used in \ourmods.}
\label{table:pe}
\end{table*}

In the analysis of sequence-like data, such as spectral data, the order of spectral channels is indeed crucial. Each position in the spectrum corresponds to a signal intensity value at a specific frequency (i.e., wavelength, or velocity channel), and the arrangement of these positions conveys meaningful information about the underlying processes. Providing information about the order of elements in a sequence using positional encodings empowers the model to capture dependencies and patterns associated with the order of the data, resulting in enhanced performance. Positional encodings have proven effective in sequence-to-sequence models and natural language processing, particularly in contexts where the order of words has a great impact on semantics \cite[][]{Vaswani2017,Gehring2017,Devlin2018,Radford2019,Raffel2019, Lan2019, Yang2019,Liu2019,Ke2020,Dufter2021}. In this work, we use and evaluate different positional encoding variants.

For a given original input vector \textbf{x} (e.g., a spectrum), we will construct its corresponding positional encoding vector \textbf{r}. Then, we will combine the vector \textbf{r} with the input vector \textbf{x} element-wise. We can either add (``\textit{add}'') or concatenate (``\textit{concat}'') the positional encoding vector to the original input vector. However, it is important to note that concatenating will increase the dimensionality of the input space, potentially leading to a significant increase in computational cost.

This work employs a total of seven positional encoding techniques: ``original spectrum'', ``concat index'', ``add index'', ``concat sinusoidal'', ``add sinusoidal'', ``concat second--order polynomial'' and ``(add) learnable''. These techniques are all drawn from absolute positional encodings which provide information about the absolute position or order of elements in a sequence. Table \ref{table:pe} provides an overview of the designed positional encodings, and each technique is further elaborated below.

\subsection{Original spectrum}

This data representation uses the original spectrum (i.e. without positional encoding), which represents a 1D spectrum characterized by the brightness temperature $T_\mathrm{b}$ (measured in Kelvin) as a function of velocity $v$ (in $km~s^{-1}$). In this context, the model input has a size of ($1 \times N$), comprising one row and $N$ columns, where $N$ is the number of spectral channels (either 256 or 101 in this work). Serving as the foundational representation, the ``original spectrum'' requires no data preprocessing. The model input is given by:

\begin{equation}
\mathbf{x_\mathrm{original}} = (x_1, x_2, \ldots, x_N) \in \mathbb{R}^N
\label{pe:original}
\end{equation}
\noindent where $x_{i}$ corresponds to the brightness temperature in the $i^{th}$ velocity channel.

\subsection{Concatenating index PE to original spectrum}

In this positional encoding technique, a positional encoding vector (PEV) is generated to match the size of the original vector ($1 \times N$). Each element in the positional encoding vector is calculated as the integer index number divided by $N$, resulting in values within the range of 0 to 1.

\begin{equation}
\mathbf{x_\mathrm{index, c}} = (c_1, c_2, \ldots, c_N) = (\frac{0}{N}, \frac{1}{N}, \ldots, \frac{N}{N}) \in \mathbb{R}^N
\label{pe:indexc}
\end{equation}

This approach introduces a fixed absolute positional encoding, where the values of the PEV directly represent the position of each element in the original spectrum. The absolute positional encoding provides the model with explicit information about the absolute order of elements in the sequence.

After generating the positional encoding vector, it is concatenated with the original vector along the velocity axis. The resulting matrix has a size of ($2 \times N$), where the first row represents the original spectrum $\mathbf{x_\mathrm{original}} = (x_1, x_2, \ldots, x_N)$, and the second row contains the PEV values $\mathbf{x_\mathrm{index,c}} = (c_1, c_2, \ldots, c_N)$. The model input is given by the following matrix:

\begin{equation}
\mathbf{X}_\mathrm{index,c} =
\begin{bmatrix}
\mathbf{x_\mathrm{original}} \\
\mathbf{x_\mathrm{index,c}}
\end{bmatrix}
=
\begin{bmatrix}
    x_{1} & x_{2} & \ldots & x_{N} \\
    0/N & 1/N & \ldots & N/N
\end{bmatrix}
\in \mathbb{R}^{2 \times N}
\end{equation}

\subsection{Adding index PE to original spectrum}

Following the same approach as the ``concatenating index PE'' technique, a positional encoding vector is also created with the same size as the original vector ($1 \times N$).

Subsequently, the positional encoding vector is directly element--wise added to the original spectrum $\mathbf{x_\mathrm{original}} = (x_1, x_2, \ldots, x_N)$, resulting in a new vector with the same size as the original spectrum ($1 \times N$). This method enhances the original spectrum with positional information for improved model understanding. The final model input is represented by the following vector:

\begin{equation}
  \mathbf{x_\mathrm{index,a}} = (x_1 + \frac{0}{N}, x_2 + \frac{1}{N}, \ldots, x_N + \frac{N}{N}) \in \mathbb{R}^N
\end{equation}

\subsection{Concatenating sinusoidal PE to original spectrum}
Presenting the ``concatenating sinusoidal PE to original spectrum'' technique, inspired by Transformer positional encodings \citep{Vaswani2017,Dehghani2018,Li2019Trans}, which utilize \textit{sinusoidal} and \textit{cosine} functions for generating the positional encoding matrix.

In the context of spectral data, unlike the Transformer's input, which consists of multiple vectors representing words, our input format involves a single spectral vector. To adapt, we simplify the positional encoding formula expression:

\begin{equation}
\mathbf{x_\mathrm{sin, c}} = sin(0, 1, \ldots, N) \in \mathbb{R}^N
\label{pe:indexc}
\end{equation}

After being generated, the positional encoding vector $\mathbf{x_\mathrm{sin, c}}$ is concatenated with the original vector along the velocity axis. The resulting matrix has a size of ($2 \times N$), with the first row representing the original spectrum $\mathbf{x_\mathrm{original}} = (x_1, x_2, \ldots, x_N)$, and the second row containing the positional encoding vector values $\mathbf{x_\mathrm{sin, c}} = sin(0, 1, \ldots, N)$. The final model input is thus represented by the following matrix:

\begin{equation}
\mathbf{X}_\mathrm{sin,c} =
\begin{bmatrix}
\mathbf{x_\mathrm{original}} \\
\mathbf{x_\mathrm{sin,c}}
\end{bmatrix}
=
\begin{bmatrix}
    x_{1} & x_{2} & \ldots & x_{N} \\
    sin(0) & sin(1) & \ldots & sin(N)
\end{bmatrix}
\in \mathbb{R}^{2 \times N}
\end{equation}

\subsection{Adding sinusoidal PE to original spectrum}

Introducing the ``adding sinusoidal PE to original spectrum'' technique, which adopts the same approach as the ``concatenating sinusoidal PE'' technique for generating the positional encoding vector. However, in this PE technique, we then perform element-wise addition of the positional encoding vector to the original spectrum $\mathbf{x_\mathrm{original}} = (x_1, x_2, \ldots, x_N)$. This results in a new vector of the same size as the original spectrum ($1 \times N$). The ultimate input to be fed into the model is represented by the following vector:

\begin{equation}
  \mathbf{x_\mathrm{sin,a}} = (x_1 + sin(0), x_2 + sin(1), \ldots, x_N + sin(N)) \in \mathbb{R}^N
\end{equation}

\subsection{Concatenating polynomial PE to original spectrum}

We present the ``concatenating polynomial'' positional encoding technique, where we explore polynomial representation and evaluate the model performance for a second-order polynomial feature ($\alpha = 2$). The generation of the polynomial feature vector from the original spectrum is expressed as:

\begin{equation}
\mathbf{x_\mathrm{poly, c}} = \mathbf{x_\mathrm{original}} \odot \mathbf{x_\mathrm{original}} = (x^2_1, x^2_2, \ldots, x^2_N) \in \mathbb{R}^N
\end{equation}

The resulting polynomial positional encoding vector, $\mathbf{x_\mathrm{poly, c}}$, is then concatenated with the original vector along the velocity axis. This produces a matrix of size ($2 \times N$), where the first row represents the original spectrum $\mathbf{x_\mathrm{original}} = (x_1, x_2, \ldots, x_N)$, and the second row contains the polynomial positional encoding vector $\mathbf{x_\mathrm{poly, c}} = (x^2_1, x^2_2, \ldots, x^2_N)$. The final model input is represented by the ensuing matrix:

\begin{equation}
\mathbf{X}_\mathrm{poly,c} =
\begin{bmatrix}
\mathbf{x_\mathrm{original}} \\
\mathbf{x_\mathrm{poly,c}}
\end{bmatrix}
=
\begin{bmatrix}
    x_{1} & x_{2} & \ldots & x_{N} \\
    x^2_1 & x^2_2 & \ldots & x^2_N
\end{bmatrix}
\in \mathbb{R}^{2 \times N}
\end{equation}

\subsection{Adding learnable PE to original spectrum.}

The learnable positional encoding (LPE) allows deep learning models to learn positional information during training. Instead of using fixed predefined positional encodings, we use learnable encoding for each position in the spectral input. The learnable vector is treated as additional parameters, and their values are optimized along with the rest of the model parameters during the training process. While LPE introduces more flexibility, it also increases the number of parameters in the models for training, which can affect training and generalization. Therefore, we only assess the learnable positional encoding to our \ourmod.

In terms of implementation, the learnable positional encoding is often added to the input. The LPE parameters are initialized randomly from a random distribution, such as a normal distribution $\mathcal{N}(0, \sigma^2)$ with a mean $\mu = 0$ and a small standard deviation $\sigma$.

Mathematically, the learnable positional encoding can be represented as a vector:
\begin{equation}
\mathbf{p} = (p_1, p_2, \ldots, p_N) \in \mathbb{R}^N
\label{pe:lpe}
\end{equation}
\noindent with each $p_i \sim \mathcal{N}(0, \sigma^2)$. Subsequently, this learnable positional encoding vector is added element--wise to the original spectrum vector $\mathbf{x_\mathrm{original}} = (x_1, x_2, \ldots, x_N)$ to form the final model input vector:

\begin{equation}
\begin{aligned}
  \mathbf{x_\mathrm{lpe}} & = \mathbf{x_\mathrm{original}} + \mathbf{p}\\
  & = (x_1 + p_1, x_2 + p_2, \ldots, x_N + p_N) \in \mathbb{R}^N
  \end{aligned}
\end{equation}

\section{Importance of positional encoding}
\label{sec:pes}
    We train our CNN and \ourmod\ models on synthetic spectral cubes, experimenting with diverse data representations by combining original data with positional encodings as the model inputs. While evaluating our CNN performance with six \textit{absolute} positional encoding techniques: ``\textit{original spectrum}'', ``\textit{concat index}'', ``\textit{add index}'', ``\textit{concat sinusoidal}'', ``\textit{add sinusoidal}'' and ``\textit{concat second-order polynomial}'', we are keen to examine \ourmod\ models with an additional data representation: ``\textit{learnable}'' positional encoding. The parameters of LPE can be learned during the training phase. Yet, it adds more free parameters to the models, thus requiring more computational resources and training time. The models might opt to fine-tune the parameters of the positional vector to reduce the loss function instead of optimizing the parameters in the neural network. This may cause a reduction in performance.

Absolute positional encodings provide information about the absolute position or order of elements in a sequence \cite[e.g.,][]{Vaswani2017,Devlin2018,Kitaev2020,Liu2019,Dufter2021}. These encodings are typically incorporated into models either by addition (``\textit{add index}'', ``\textit{add sinusoidal}'', ``\textit{learnable}'') or concatenation (``\textit{concat index}'', ``\textit{concat sinusoidal}'', ``\textit{concat second-order polynomial}'') with the original data. This integration enables the models to distinguish between different positions in the sequence. In natural language processing, for example, when processing a sentence, absolute positional encodings help the models understand the order of words in a sentence.

Refer to Appendix \ref{sec:data_repr} for more details on the positional encoding techniques employed in this work.

\begin{figure}
\centering
\includegraphics[width=0.5\textwidth]{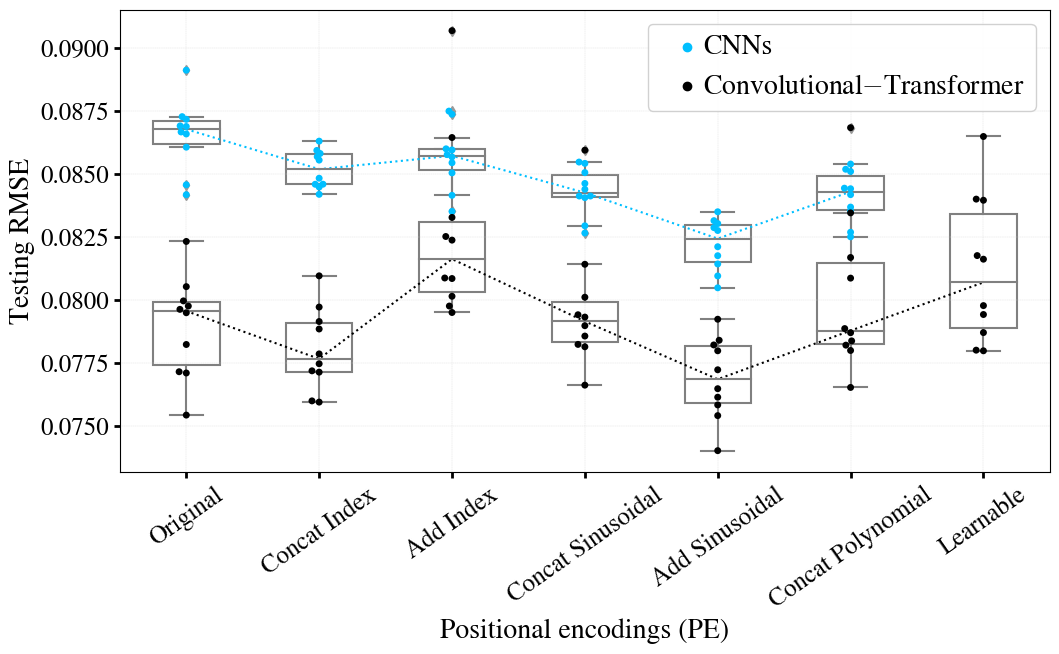}
\caption{Swarm-box plots testing RMSE values obtained from ten-trial training sessions of CNN (cyan) and \ourmod\ (black) models. The models utilize different positional encoding techniques on spectral PPV cubes.}
\label{fig:mse_vt_cnn_cnntrans_cube}
\end{figure}

\begin{figure*}
\centering
\includegraphics[width=1.\textwidth]{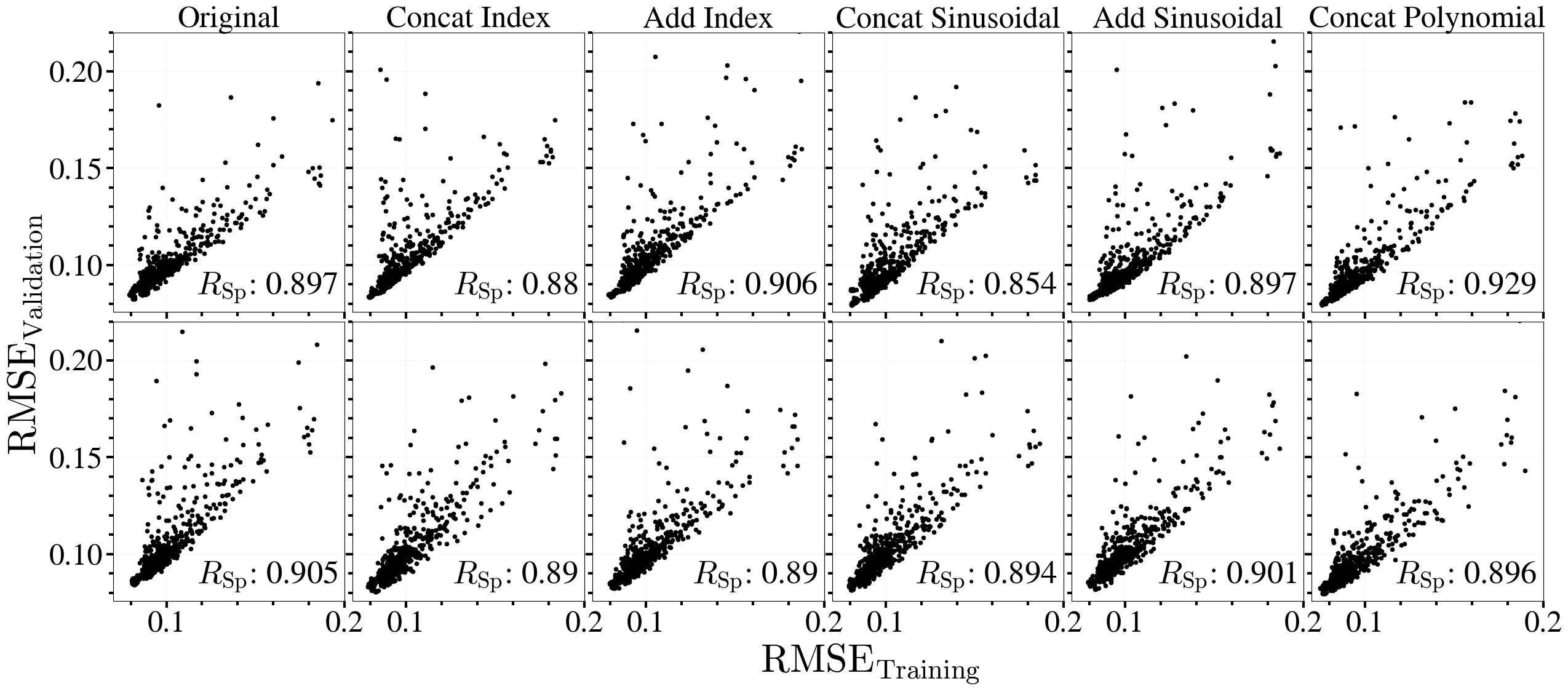}
\caption{Correlations between validation and training RMSE values obtained from CNN models across positional encodings, employing the same training/validation/testing (top panels) and different shuffling setups (bottom panels). Each panel includes Spearman's correlation coefficient $R_\mathrm{Sp}$ (all \textit{p-values} $< 10^{-10}$). The axis scales in all panels are identical.}
\label{fig:cnn_train_valid_rmse_corr_split_1split}
\end{figure*}

\subsection{\ourmod\ with added sinusoidal positions is the best model }
\label{subsec:comparison}

We compare the performances of \ourmod\ with deep CNNs using their testing RMSEs. To emphasize the importance of the representation learning method (in extracting the compact features of \hi\ emission spectra), we ensured that the models have nearly identical numbers of learnable parameters and similar hyperparameters.
Figure \ref{fig:mse_vt_cnn_cnntrans_cube} summarises the testing RMSEs from ten-training trials for the two deep learning architectures: CNN in cyan and \ourmod\ in black. Dotted lines connect the median RMSEs from each positional encoding method. It is observed that self-attention-based models consistently achieve ($\sim$8\%) lower testing RMSEs than CNNs across all positional encodings, although \ourmod\ RMSEs slightly vary over a wider range.

Among the CNN models alone, those incorporating positional encodings outperform the one (labeled as ``original'') without positional encoding. However, in the case of \ourmod\ models, RMSEs between various data representations do not show a clear trend. This implies that the \ourmod\ models seem able to mitigate the influence of positional encodings on their performance.

To assess how sensitive the positional encodings are to weight initialization and shuffling configurations. We explore two experimental setups. In the first setup, we maintain the same training/validation/testing shuffle but different initial neural network weights, ensuring that the way we shuffle and partition our entire dataset into training, validation, and testing sets remains identical throughout. In the second setup, we randomly shuffle the entire dataset differently before splitting it into training, validation, and test sets in each run. The percentages for each split remain constant (60\% for training, 20\% for validating, and 20\% for testing), but the specific data points assigned to each split change due to the random shuffling. 

We then estimate Spearman's correlation coefficients between validation and training RMSE values. These coefficients assess how well the relationship between two variables can be described using a monotonic function (the closer the coefficient is to $\pm1$, the stronger the monotonic relationship). The lack of correlation between the training and validation RMSEs implies suboptimal models and positional encodings.

In Figures \ref{fig:cnn_train_valid_rmse_corr_split_1split}, we display the correlation between validation and training RMSE values obtained from CNN models utilizing various positional encodings. The upper panels show the experiments using the same training/validation/testing shuffle, while the lower panels experiment with different shuffle procedures. All Spearman's correlation coefficients fall in the range of 0.85 to 0.93, indicating robust monotonic relationships between validation and training RMSE values. In particular, Spearman correlation coefficients of each positional encoding remain comparable between the two splitting settings. However, the slight scatterings in the correlations suggest that the positional encoding methods used with CNN models are merely sensitive to neural network weight initialization and training/validation/testing splitting configurations.

\begin{table*}
\fontsize{10}{9}\selectfont
\centering
\begin{tabular}{l | l | c | c }
 \hline
 Model & Positional Encoding  & $R_\mathrm{Spearman}$ & $R_\mathrm{Spearman}$ \\
 \  & \   & (same shuffles) & (different shuffles) \\ [0.5ex] 
 \hline\hline
 
CNN & Concat Polynomial & 0.929 & 0.896 \\
CNN & Add Index & 0.906 & 0.890 \\
CNN & Original & 0.897 & 0.905 \\
CNN & Add Sinusoidal & 0.897 & 0.901 \\
CNN & Concat Index & 0.880 & 0.890 \\
CNN & Concat Sinusoidal & 0.854 & 0.894 \\
\hline
\ourmod\ & Concat Index & 0.970 & 0.931 \\
\ourmod\ & Concat Sinusoidal & 0.968 & 0.943 \\
\ourmod\ & Learnable & 0.964 & 0.951 \\
\ourmod\ & Original & 0.962 & 0.938 \\
\ourmod\ & Add Sinusoidal & 0.960 & 0.934 \\
\ourmod\ & Concat Polynomial & 0.958 & 0.922 \\
\ourmod\ & Add Index & 0.953 & 0.910 \\
\hline\hline

\end{tabular}
\caption{Spearman coefficients ($R_\mathrm{Spearman}$) of correlations between validation and training RMSEs obtained from CNN and \ourmod\ across positional encodings, using the same
training/validation/testing (``same shuffle'') and different shuffling (``different shuffle'') setups. Within each model type, the Spearman coefficients are sorted by $R_\mathrm{Spearman}$ obtained from the same shuffle scheme (third column). All corresponding \textit{p--values} $< 10^{-10}$.}
\label{table:rspearman}
\end{table*}

\begin{figure*}
\centering
\includegraphics[width=1.\textwidth]{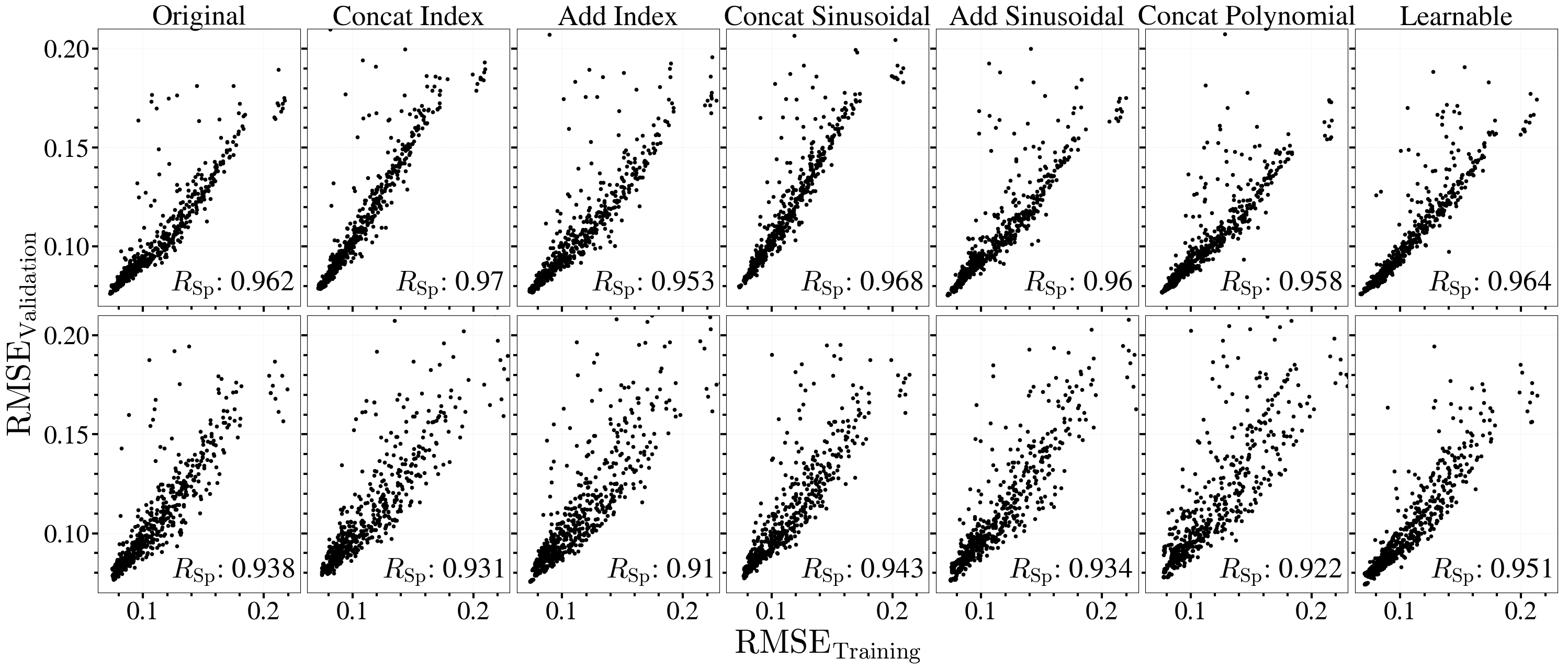}
\caption{Correlations between validation and training RMSE values obtained from \ourmod\ models across positional encodings, employing the same training/validation/testing (upper panels) and different shuffling setups (bottom panels). Each panel includes Spearman's correlation coefficient $R_\mathrm{Sp}$ (all \textit{p--values} $< 10^{-10}$).}
\label{fig:r_spearman_cnntrans_1split_split}
\end{figure*}

In Figure \ref{fig:r_spearman_cnntrans_1split_split}, we describe the correlation between validation and training RMSE values obtained from \ourmod\ models employing various positional encodings. Similar to Figure \ref{fig:mse_vt_cnn_cnntrans_cube}, in the top row, we follow the same training/validation/testing shuffle, while we explore different shuffling setups in the bottom row. The correlations between training and validation RMSE values for \ourmod\ models appear concentrated, with Spearman correlation coefficients ranging from 0.91 to 0.97. This points to strong monotonic relationships between the validation and training phases. The training/validation RMSEs from ten-training trials are also summarized in Table \ref{table:rspearman}.

Of particular note, the Spearman coefficients obtained by \ourmod\ models are consistently higher than those obtained by CNNs. This trend still holds whether using the same training/validation/testing shuffling setup or different shuffling strategies. As a result, the positional encodings used in \ourmod\ are ($\sim$10\%) less sensitive to neural network weight initializations and shuffle configurations compared to CNN models.

\begin{figure}
\centering
\includegraphics[width=0.475\textwidth]{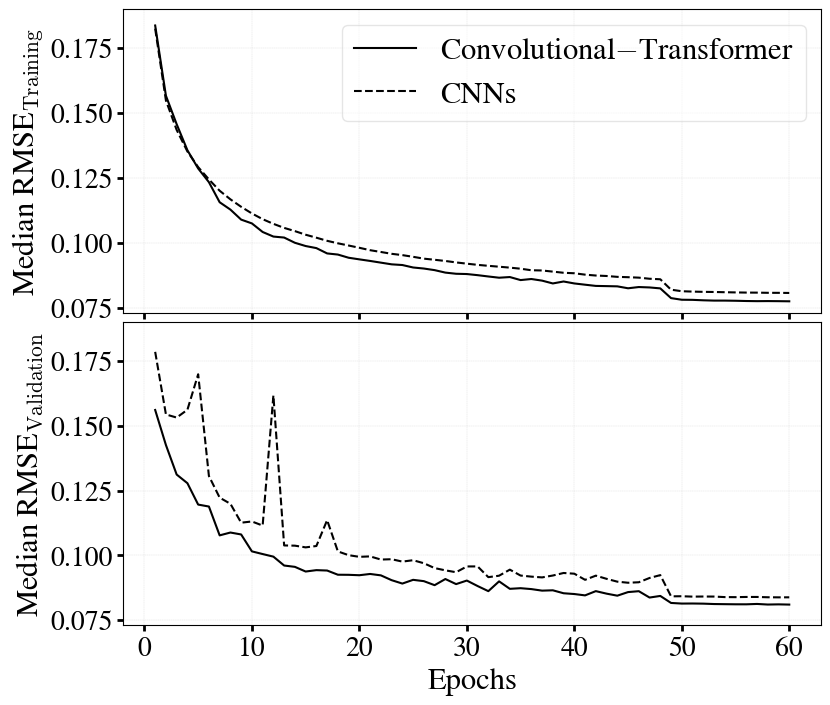}
\caption{Convergence speed comparison for CNN (dashed line) and \ourmod\ (solid line) models achieved from the ten-trial training sessions. The y-axes show the median RMSE for training (top) and validation (bottom) sessions through all positional encodings.
}
\label{fig:converge_speed}
\end{figure}

The \ourmod\ models outperform the CNN models in terms of convergence speed. In both training and validating, the \ourmod\ converges more quickly than the CNNs as illustrated in Figure \ref{fig:converge_speed} where we show the median RMSE for training (top panel) and validation (bottom panel)
sessions through all positional encodings. Furthermore, the \ourmod\ attains lower final RMSEs in the last training epochs than the CNNs. Additionally, the hybrid models' training processes are more stable than the CNNs. In particular, CNNs' training processes display significant fluctuations with peaks in validation occurring at early training epochs (before epoch 20), which contrasts with a smoother training behavior of the \ourmod. In terms of overall accuracy, as measured by RMSE, \ourmod\ surpasses deep CNNs by an average of 10\%.

Taking convolutional weight initialization/shuffling sensitivity, and the fluctuations in testing RMSEs into account, we consider \ourmod\ with ``\textit{add sinusoidal}'' positional encoding (adding a sinusoidal positional function to the original spectrum) as the most robust model. We thus employ this model for analyses in the present paper.

\subsection{Impact of kernel sizes on CNN performance}
\label{subsec:kernel_width}

In a CNN, the convolutional layers indeed play a crucial role in extracting features of the input data. These layers use filters (i.e., convolutional kernels) of various sizes to convolve over the input and capture features at different scales \citep[e.g.,][]{Szegedy2015,Szegedy2016}. Small kernel sizes (e.g., 3$\times$3, 5$\times$5 for 2D image processing) are commonly used for capturing fine-grained features in the input data, effectively learning local patterns. On the other hand, large kernel sizes (e.g., 9$\times$9, 11$\times$11) are employed to capture larger patterns and global features in the input data.

\begin{figure}
\centering
\includegraphics[width=0.5\textwidth]{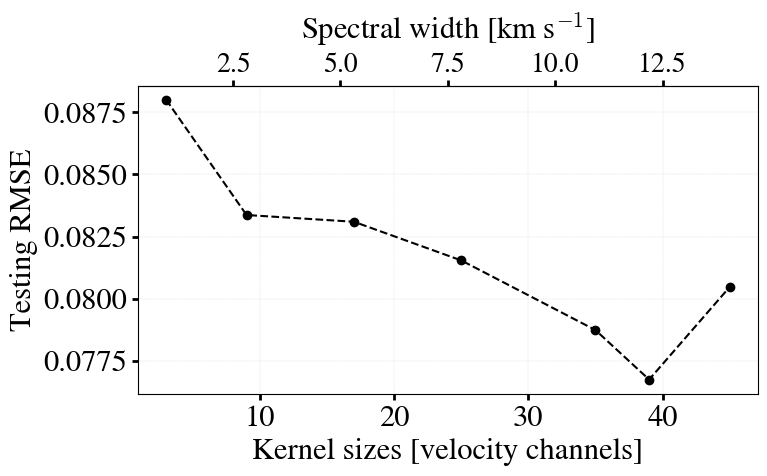}
\caption{Testing RMSEs obtained from our deep CNN (without positional encoding) vs kernel sizes (3, 9, 17, 25, 35, 39, 45 spectral channels). The corresponding spectral widths
are illustrated on the top axis. The channel width is 0.3125 \kms.}
\label{fig:rmse_kernel_size}
\end{figure}

Here, we conduct a simple experiment to examine the impact of kernel sizes on the performance of our convolutional network. Namely, we train an 8-layer CNN model without positional encoding using varying kernel sizes (3, 9, 17, 25, 35, 39, and 45 velocity channels) in the first convolutional layer, while keeping the parameters of the other layers unchanged. The RMSE error for each kernel size is shown in Figure \ref{fig:rmse_kernel_size}.

Initially, the RMSE decreases with increasing kernel size, reaching a minimum at 39 spectral channels, equivalent to a 12\% improvement. However, beyond this point, as kernel sizes continue to widen, RMSE starts to increase. The spectral width of 39 channels is about $\sim$10 \kms, the typical width of a WNM component, and approximately four times the typical width of a CNM component. This observation suggests that both very narrow and overly wide spectral kernels may struggle to adequately capture local/global features within the input. While narrow spectral filters (e.g., 3, 5, or 7 velocity channels) might cause CNNs to learn primarily the fluctuation of noise features, wide kernel windows may lead to potential oversmoothing. It is important to note that wider kernel sizes require significantly more time for training. For instance, training a CNN model with a kernel size of 35 channels takes ten times longer than with a kernel size of nine channels, yet with $\sim$6\% improvement in RMSE error. Considering training time and model performance, we design our models to iteratively use two kernel sizes, (1$\times$7) and (1$\times$33), for convolutional layers to capture \hi\ features at various spectral scales.

When shallow CNNs (without positional encoding) are applied to an observed dataset where the detected signal is not centralized in the center of the spectral velocity ranges, they may malfunction, resulting in blank maps with all pixel values (\FCNM\ or \RHI) set to zero, as reported by \cite{Marchal2024} (see their Section 8). In contrast, \ourmod\ models demonstrate the capability to produce results comparable to those obtained from the ROHSA Gaussian decomposition (see Section \ref{sec:application_to_emission}). This emphasizes the important roles played by the combination of positional encodings, CNN-based feature extraction, and the Transformer self-attention mechanism in spectral data analysis. According to recent research by \cite{Gulati2020}, \cite{Pan2022}, and \cite{Rozanski2023}, CNNs are crucial to capturing local spectral features and extracting compact patterns from the input data, whereas the Transformer self-attention focuses on the global spectral patterns. In particular, \cite{Pan2022} used \textit{Astroconformer}, a convolution-augmented Transformer model, to analyze stellar spectroscopic data, highlighting the important function of Transformer blocks in capturing long-range information embedded within spectra. Through comprehensive ablation studies, they found that CNNs do not outperform their baseline Multilayer Perceptron counterparts in analyzing the intrinsic complexity of stellar spectra.

With the use of the hybrid convolutional-Transformer model, this may provide a potential improvement of the high latitude \FCNM\ map produced by \cite{Hensley2022} using M20 CNN and recently adopted by \cite{Lei2023}.

\subsection{Impacts of spectral resolutions and noise levels on CNM fraction prediction}
\label{subsec:spectral_res}

\begin{figure}
\centering
\includegraphics[width=0.5\textwidth]{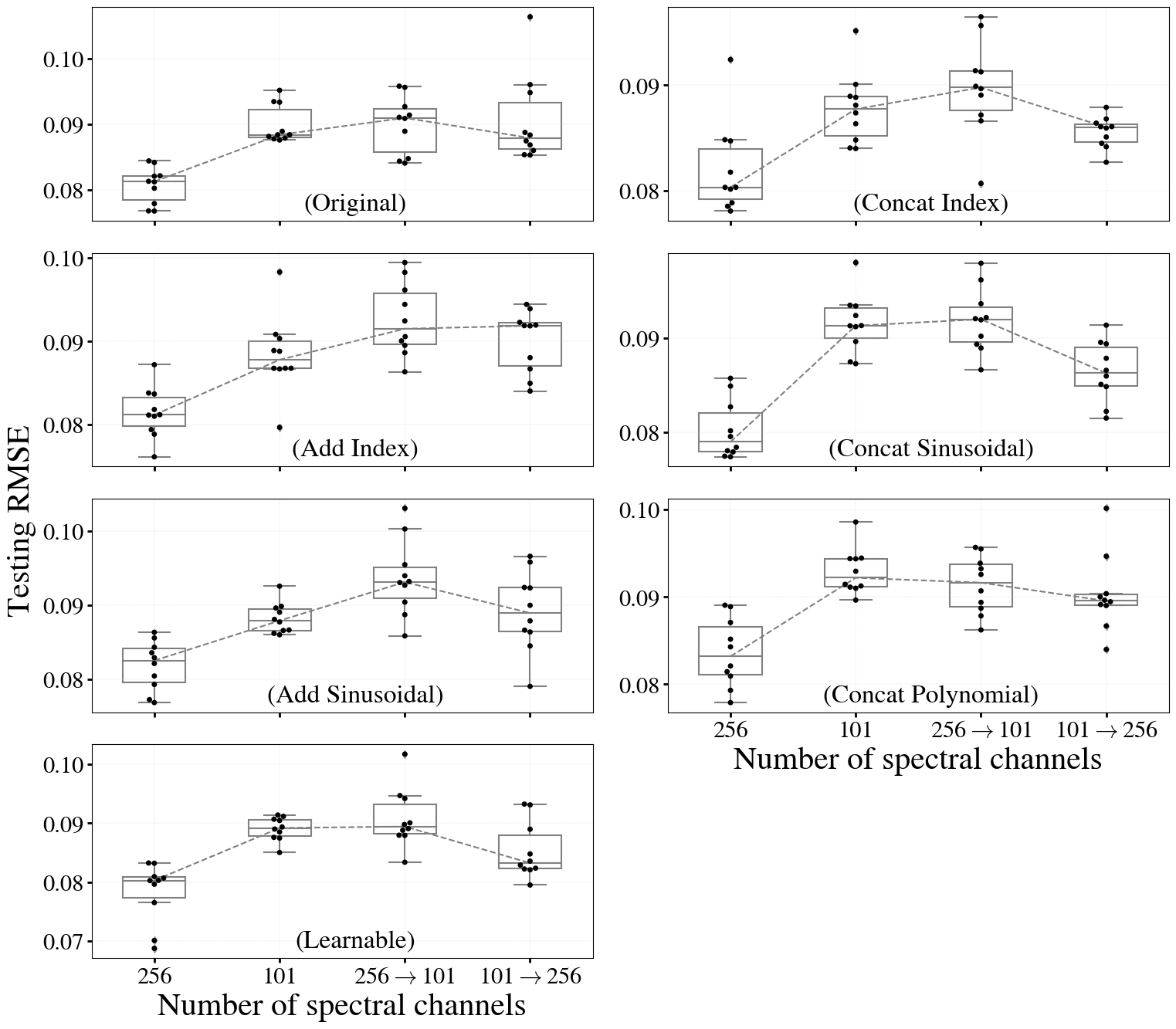}
\caption{Swarm-box plots for the testing RMSEs from \ourmod\ models employing various positional encoding techniques across different numbers of spectral channels. The spectral resolutions include 101 channels with 0.8 \kms\ spectral width and 256 channels with 0.3125 \kms\ spectral width, using PPV spectral datacubes. The annotations ``256$\rightarrow$101'' and ``101$\rightarrow$256'' refer to regridding from 256 channels to 101 channels and vice versa.}
\label{fig:spectral_res}
\end{figure}

Throughout this paper, we consistently utilize training spectral datacubes consisting of 256 spectral channels, corresponding to a spectral resolution of 0.3125 \kms, to train the \ourmods. However, we further investigate the impact of spectral resolutions on the performance of the models in this subsection. In Figure \ref{fig:spectral_res}, swarm-box plots display test RMSE values for various positional encoding techniques across different numbers of spectral channels. The spectral resolutions under examination include 0.8 \kms\ with 101 channels and 0.3125 \kms\ with 256 channels. The annotations ``256$\rightarrow$101'' and ``101$\rightarrow$256'' denote regridding from 256 channels to 101 channels and vice versa (to mimic a fabricated spectral resolution downgrade/upgrade, respectively).

Across all cases of positional encodings, higher spectral resolution with 256 velocity channels results in lower RMSE error compared to 101 velocity channels. In addition, regridding from 101 channels to 256 channels or vice versa does not enhance the models' performances. This emphasizes the significance of genuine increased spectral resolution as inputs to models. Nonetheless, \hi\ spectral data with high spectral resolutions typically need a greater number of spectral channels (e.g., 512, 1024, or 2048 channels), unavoidably leading to more complex model architectures and, as a result, requiring more time for training and testing processes. Hence, when utilizing the \textit{``tano signal''} package, which can automatically construct an appropriate \ourmod\ model based on spectral inputs, users are advised to carefully select a spectral resolution and spectral length that align with the scientific requirements.

In practice, it would be ideal to systematically explore various parameters -- such as beam size, noise level, and spatial resolution -- and train deep learning models on a synthetic database with parameters closely matching those of the observations. However, due to limitations in high-performance computing resources, instead of investigating a broad range of observing parameters, we introduced a relatively high noise level of 0.5 -- 0.8 K and smoothed the training data cubes with a beam size of 45 arcsec (as described in Section \ref{subsec:training_datasets}). Despite these constraints, the models have demonstrated promising performance when applied to different observed datasets with varying noise levels and spatial resolutions. The results remain reasonable when compared to other methods (see Sections \ref{sec:compare_with_absorption} and \ref{sec:application_to_emission} for details), indicating that the model can generalize well across different observational conditions. This robustness aligns with the principles of 21-cm radiative transfer, where the contributions of CNM gas -- characterized by low temperatures and high optical depth -- to emission profiles are pronounced and primarily dominate the brightness temperature peaks \citep[see ][]{Murray2015,Murray2018,Marchal2024,Nguyen2024}. For instance, an \hi\ parcel with a temperature of \Ts\ = 50 K and an optical depth $\tau$ = 0.1 would have a brightness temperature $T_\mathrm{b}$ $\sim$ 5 K, which is above the 3$\sigma$ detection limit of existing \hi\ surveys.

Furthermore, during the development of the Bayesian version of \ourmod\ (Muller et al., in preparation), we observed that noise levels barely affected the CNM fraction prediction. Namely, increased noise led to slightly more scatter in the one-to-one comparisons between ground truths and predictions. On the other hand, predictions for UNM and WNM fractions were marginally more sensitive to noise (and the physical scales that the models are trained on), due to the typically lower $T_\mathrm{b}$ values of these components. For instance, a UNM cloud with a temperature of \Ts\ = 1000 K and an optical depth $\tau$ = 10$^{-3}$ would have a brightness temperature $T_\mathrm{b}$ $\sim$ 1 K. If the noise level exceeds 1 K per velocity channel, it would introduce more uncertainty into the predictions for UNM and WNM fractions.   



\bsp	
\label{lastpage}
\end{document}